\documentclass[a4paper,11pt]{article}
\pdfoutput=1

\usepackage{cancel}
\usepackage{jheppub}
\usepackage[T1]{fontenc}
\usepackage{overpic}

\usepackage{xcolor}
\usepackage{transparent}
\usepackage{calc}
\usepackage{amsthm}
\usepackage{graphicx}
\usepackage{slashed}
\usepackage{amssymb}
\usepackage{xspace}            % for correct spacing after commands
\usepackage{comment}
\usepackage{cleveref}
\crefformat{footnote}{#2\footnotemark[#1]#3}

\def\slashchar#1{\setbox0=\hbox{$#1$}           % set a box for #1
   \dimen0=\wd0                                 % and get its size
   \setbox1=\hbox{/} \dimen1=\wd1               % get size of /
   \ifdim\dimen0>\dimen1                        % #1 is bigger
      \rlap{\hbox to \dimen0{\hfil/\hfil}}#1
   \else                                        % / is bigger
      \rlap{\hbox to \dimen1{\hfil$#1$\hfil}}/                                    \fi}

\usepackage{tikz}

\title{Fuzzy Jets}

\author{Lester Mackey,${}^a$}
\author{Benjamin Nachman,${}^{b,c}$}
\author{Ariel Schwartzman,${}^c$ and}
\author{Conrad Stansbury${}^b$}

\affiliation{$^{a}$Department of Statistics, Stanford University, Stanford, CA 94305, USA}

\affiliation{$^{b}$Department of Physics, Stanford University, Stanford, CA 94305, USA}

\affiliation{$^{c}$SLAC National Accelerator Laboratory, Stanford University, 2575 Sand Hill Rd, Menlo Park,
  CA 94025, U.S.A.}

\emailAdd{lmackey@stanford.edu, bnachman@cern.ch, sch@slac.stanford.edu, chstan@stanford.edu}

\abstract{
Collimated streams of particles produced in high energy physics experiments are organized using clustering algorithms to form {\it jets}.   To construct jets, the experimental collaborations based at the Large Hadron Collider (LHC) primarily use agglomerative hierarchical clustering schemes known as sequential recombination.  We propose a new class of algorithms for clustering jets that use infrared and collinear safe mixture models.  These new algorithms, known as {\it fuzzy jets}, are clustered using maximum likelihood techniques and can dynamically determine various properties of jets like their size.  We show that the fuzzy jet size adds additional information to conventional jet tagging variables.  Furthermore, we study the impact of pileup and show that with some slight modifications to the algorithm, fuzzy jets can be stable up to high pileup interaction multiplicities.
}

\begin{document}
\maketitle
\flushbottom

\section{Introduction}

As the result of a proton-proton collision at a hadron collider, hundreds of particles are created and detected~\cite{Aad:2010ac,Khachatryan:2010nk}.  While some particles can be identified by their type, such as electrons~\cite{Khachatryan:2015hwa,Aad:2014nim} and muons~\cite{Chatrchyan:2012xi,Aad:2014rra}, most of the detected particles are light hadrons produced in collimated sprays called {\it jets}.  Jets are the consequence of high energy quarks or gluons fragmenting into colorless hadrons.  Experimentally, jets are defined by clustering schemes which group together measured calorimeter energy deposits or reconstructed charged particle tracks.  A jet algorithm is a clustering scheme that connects the measured objects with theoretical quantities that can be calculated and simulated.  At a hadron collider, the natural coordinates for describing particles are $p_T$, $y$, and $\phi$, where $p_T$ is the magnitude of the momentum transverse to the proton beam, $y$ is the rapidity, and $\phi$ is the azimuthal angle.  Particles or calorimeter energy deposits are clustered using jet algorithms based on distance metrics on their coordinates in $(p_T,\vec{\rho})=(p_T,y,\phi)$.   In order for a jet algorithm to be useful to experimentalists and theorists, the collection of jets should be IRC safe in the following sense:

\begin{enumerate}
\item Infrared safe (IR): if a particle $i$ is added with $|p_T|\rightarrow 0$, the jets are unaffected.
\item Collinear safe (C): if a particle $i$ with momentum $p_i$ is replaced with two particles $j$ and $k$ with momenta $p_j+p_k=p_i$ such that $|\vec{\rho}_i-\vec{\rho}_j|= 0$, then the jets are unaffected.
\end{enumerate}

\noindent The jet algorithms most widely used at hadron colliders fall into a class of schemes known as {\it sequential recombination}~\cite{Ellis:1993tq}.  These IRC safe schemes require metrics $d$ on momenta $d_{ij}=d(p_i,p_j):(p_i,p_j)\rightarrow\mathbb{R}^+,d_{iB}=d(p_i):p_i \rightarrow \mathbb{R}^+$ and proceed as follows:

\begin{enumerate}
\item Assign each particle as a proto-jet.
\item Repeat until there are no proto-jets left: Let $(k,\ell)=\text{argmin}_{i,j}d(p_i,p_j)$ and without loss of generality, $d_{kB}<d_{\ell B}$.  If $d_{kB} < d_{k\ell}$, declare proto-jet $k$ a jet and remove it from the list.  Otherwise, combine proto-jets $k$ and $\ell$ into a new proto-jet with momentum $p_\text{new}=p_\ell+p_k$.
\end{enumerate}

\noindent One common prescription is called the Cambridge-Aachen (C/A) algorithm~\cite{Dokshitzer:1997in,Wobisch:1998wt}, which uses $d_{ij}=|\vec{\rho}_i-\vec{\rho}_j|^2/R^2$ and $d_{iB}=1$. The fixed quantity $R$ is roughly the size of the jet in $(y,\phi)$.  By far, the most ubiquitous jet algorithm used at the Large Hadron Collider (LHC) is the anti-$k_t$ algorithm~\cite{Cacciari:2008gp} with $d_{ij}=\min(p_{T,i}^{-2},p_{T,j}^{-2})|\vec{\rho}_i-\vec{\rho}_j|^2/R^2$ and $d_{iB}=p_{T,i}^{-2}$.

The purpose of this paper is to introduce a new paradigm for jet clustering, called {\it fuzzy jets}, based on probabilistic mixture modeling.  Section~\ref{sec:stats} introduces the statistical concept of a mixture model and describes the necessary modification to make the procedure IRC safe.  Section~\ref{sec:EMalgorithm} gives one efficient method for clustering fuzzy jets based on the Expectation-Maximization (EM) algorithm.  Section~\ref{sec:tagging} contains several examples comparing fuzzy jets with sequential recombination and Sec.~\ref{sec:pileup} describes how one might mitigate the impact of overlapping proton-proton collisions (pileup).  We conclude in Sec.~\ref{sec:conclusions} with some summary remarks and outlook for the future.

\section{Mixture Model Jets}
\label{sec:stats}

\begin{figure}[h!]
\begin{center}
\includegraphics[width=0.45\textwidth]{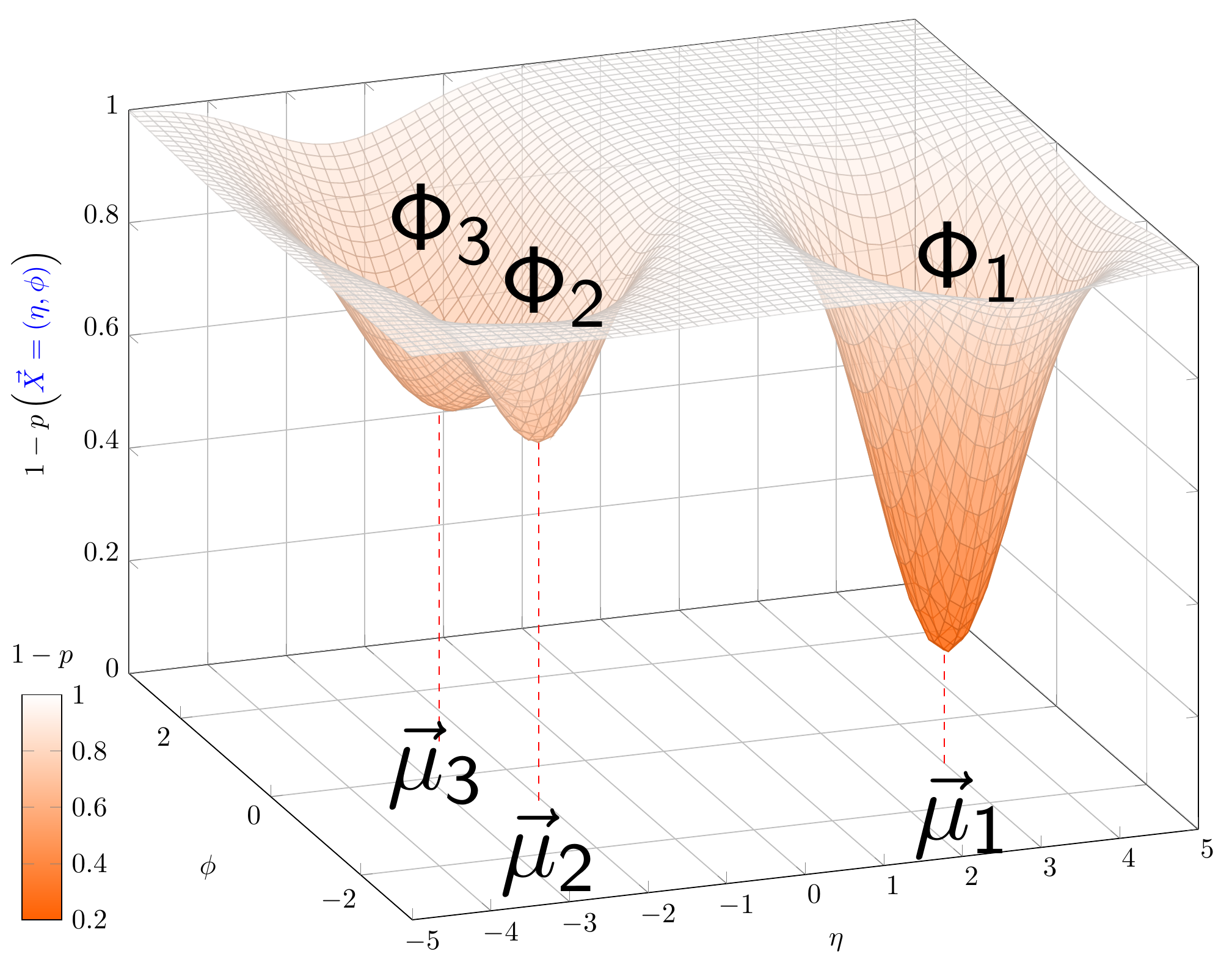}
\caption{An example of the learned per-particle probability density specified in Eq.~(\ref{eq:mm}) with $k=3$ and Gaussian $f=\Phi$ in $n=2$ dimensions.  One cluster is associated with each component density $\Phi_i=\Phi(\cdot \mid \mu_i,\Sigma_i)$, where the dot $\cdot$ is a placeholder for the function argument.}
\label{fig:densitymap}
\end{center}
\end{figure}

Mixture models~\cite{opac-b1097397} are a statistical tool for clustering which postulate a particular class of probability densities for the data to be clustered.   Generically, for grouping $m$ $n$-dimensional data points into $k$ clusters, the mixture model density is

\begin{align}
\label{eq:mm}
p(x_1,...,x_m|\theta)=\prod_{i=1}^m\left(\sum_{j=1}^k \pi_j f(x_i|\theta_j)\right),
\end{align}

\noindent where $\pi_j$ is the unknown weight of cluster $j$ such that $\sum_j \pi_j=1$ and $f(x_i|\theta_j)$ is a probability density on $n$-dimensions with unknown parameters $\theta_j$ to be learned from the data.  A common choice for $f$ is the normal density $\Phi$ with $\theta_j=(\mu_j,\Sigma_j)$ for $\mu_j$ the $n$-dimensional mean and $\Sigma_j$ the $n\times n$ covariance matrix.   In the mixture model paradigm, the $\theta_j$ are the cluster properties; in the Gaussian case, $\mu_j$ is the location of cluster $j$ and $\Sigma_j$ describes its shape in the $n$-dimensional space.  When clustering with a finite mixture, the number of clusters $k$ must be specified ahead of time\footnote{There is a wealth of literature on the subject of choosing $k$, for a survey of methods, see~\cite{determiningk}.  The likelihood monotonically increases with $k$; as alternatives to maximum likelihood, one can for instance look for kinks in the likelihood as a function of $k$~\cite{gap}.}, which is dual to the usual use of sequential recombination\footnote{It is similar to the exclusive form of the $k_T$ sequential recombination scheme~\cite{Catani:1993hr}.  The exclusive nature of the algorithm (and the minimization procedure used to find the jets) is similar to the XCone algorithm~\cite{Stewart:2015waa,Thaler:2015xaa} that became public as this manuscript was in its final preparation.} in which $k$ is learned and the size of jets is specified ahead of time.  The standard objective in (frequentist) mixture modeling is to select the parameters $\theta_j$ which maximize the likelihood (Eq.~\ref{eq:mm}) of the observed dataset.  Figure~\ref{fig:densitymap} illustrates what the learned event density might look like for $k=3$ and Gaussian $f=\Phi$ in $n=2$ dimensions.  

An equivalent way of approaching mixture modeling is to view Eq.~(\ref{eq:mm}) as the density used to generate the data.  We view the data as having been drawn randomly from the density specified in Eq.~(\ref{eq:mm}), with the following setup:

\begin{enumerate}
\item Throw $n$ independent and identical $k$-sided dice with probability $\pi_j$ to land on side $j=1,...,k$ and label the outcomes $\lambda_1,...,\lambda_n$.  %Symbolically, this means we let\footnote{In the form given here, the data are drawn independently from the mixture.  There are other schemes, such as Hidden Markov Models~\cite{baum1966} which allow for correlations between particles and which might be useful in describing the tree structure of the parton shower.}  $\lambda_i\stackrel{\text{i.i.d.}}{\sim} \text{Multinomial}(\pi,1)$. The variable $\lambda_i$ identifies to which cluster the data point $i$ belongs and $\pi\in[0,1]^k$ with $\sum_i \pi_i=1$ is the prior probability of belonging to any one of the $k$ clusters.
\item Independent of the others, data point $i\in\{1,...,n\}$ is drawn randomly from $f(\cdot\mid\theta_{\lambda_i})$.%, i.e. $x_i|\lambda_i\stackrel{\text{i.i.d.}}{\sim} f(\cdot\mid\theta_{\lambda_i})$.
\end{enumerate}

% and $x_i|\lambda_i\stackrel{\text{i.i.d.}}{\sim} f(\cdot\mid\theta_{\lambda_i})$. The variable $\lambda_i$ identifies to which cluster the data point $i$ belongs and $\pi\in[0,1]^k$ with $\sum_i \pi_i=1$ is the prior probability of belonging to any one of the $k$ clusters.  The procedure for generating data under Eq.~(\ref{eq:mm}) is then given by the following algorithm:

%A multinomial distribution for each data point $j$ means that it has probability $\pi_i$ of belonging to cluster $i$.  Given that one knows that data point $j$ came from cluster $i$, it follows the distribution $f(\cdot\mid\theta_{\lambda_i})$.  

Once $\theta$ and $\pi$ are learned by minimizing Eq.~(\ref{eq:mm}), we can compute $q_{ij} = \Pr(\lambda_i = j \mid x_i)$, the posterior probability that $x_i$ was generated by $f(\cdot\mid\theta_j)$ or, intuitively, the posterior probability that $x_i$ belongs to cluster $j$.  The $q_{ij}$ are the {\it soft assignments} of particles $i$ to jet $j$ and will play an important role in Sec.~\ref{sec:EMalgorithm} when we show how to maximize the likelihood in Eq.~(\ref{eq:mm}).  Jets produced with mixture modeling are called {\it fuzzy jets} because of the soft memberships - every particle can belong to every jet with some probability\footnote{Soft assignments for jets during clustering was studied in the context of the ``optimal jet finder''~\cite{Grigoriev:2003tn} which maximizes a function of the soft assignments.}.  This can be seen explicitly in Fig.~\ref{fig:densitymap} where the densities of all three clusters are everywhere nonzero, so $q_{ij}>0$ for all $j$.  The idea of probabilistic membership was recently studied in the context of the Q-jets algorithm~\cite{Ellis:2012sn} in which the same event is interpreted many times by injecting randomness into the clustering procedure.  Unlike Q-jets, fuzzy jets allocates the soft membership functions deterministically throughout the clustering procedure.  However, like Q-jets, there is an ambiguity in how to assign kinematic properties to the clustered jets.  Fuzzy jets are defined by their shape (and location), not their constituents.  This is in contrast to anti-$k_t$ jets, which are defined by their constituents without an explicit shape determined from the clustering procedure.  One simple assignment scheme is to define the momentum of a fuzzy jet $j$ as

%The kinematics of a fuzzy jet are {\it inherited} based on a procedure for mapping the set of particle momenta and their soft membership functions to jet properties.  

\begin{align}
p_\text{jet $j$}=\sum_{i=1}^m p_{i}\left\{\begin{matrix}1 & j=\text{argmax}_kq_{ik} \cr 0 & \text{else} \end{matrix}\right\} .
\end{align}

\noindent In other words, this procedure assigns every particle to its most probable associated jet.  This scheme will be known as the hard maximum likelihood (HML) scheme, but is not the only possible assignment algorithm.  The dual problem in sequential recombination is the jet area, which must be defined~\cite{areas}, whereas the jet kinematics are the `natural' coordinates.

\clearpage
\newpage

We now specialize the likelihood in Eq.~(\ref{eq:mm}) to the case of clustering particles into jets at a collider like the LHC.  Consider a mixture model in two dimensions\footnote{\label{second}One must take care in selecting a class of densities appropriate for the angular quantity $\phi$.  For more details on the wrapped Gaussian distribution and motivation for its use in this context, see Appendix~\ref{sec:wrapped}.} with $x_i=\rho_i$.  The resulting mixture model (MM) jets are inherently not IR safe: particle $p_T$ does not appear in the likelihood and therefore arbitrarily low energy particles can influence the clustering procedure.  Therefore, we add a modification to the log likelihood:

\begin{align}
\label{eq:mm2}
\log\mathcal{L}(\{p_{T,i},\rho_i\}|\theta)=\sum_{i=1}^m p_{T,i}^\alpha \log\left( \sum_{j=1}^k \pi_j f(\rho_i|\theta_j)\right),
\end{align}

\noindent where $\alpha$ is a weighting factor.  Equation~(\ref{eq:mm2}) is the log of Eq.~(\ref{eq:mm}) with the term $p_{T,i}^\alpha$  inserted in the outer sum.  For $\alpha > 0$, the resulting {\it modified} mixture model (mMM) jets are IR safe, and when $\alpha=1$, the jets are C safe.  Therefore, for $\alpha=1$, the jets are IRC safe.  Different choices of component densities $f$ in Eq.~(\ref{eq:mm2}) give rise to different IRC safe MM jet algorithms.  We have studied several possibilities for $f$, but for the remainder of this paper will specialize to (wrapped\cref{second}) Gaussian $f=\Phi$.  The resulting fuzzy jets are called modified Gaussian Mixture Model jets (mGMM) and are parameterized by the locations $\mu_j$, the covariance matrices $\Sigma_i$, and the cluster weights $\pi_j$.  We initialize $\pi_j=1/k$ and $\Sigma_j = I$.  %Deterministic values of $\pi_j=1/k$ and $\Sigma_j = I$ are used as seeds in order to maintain IRC safety of the prior initialization.

Since practical procedures for maximizing the modified likelihood in Eq.~(\ref{eq:mm2}) may converge to stationary points that are not globally optimal, the output of a fuzzy jet algorithm will depend on an initial setting of the cluster parameters $\theta$ and $\pi$.  One simple procedure, used exclusively for the rest of the paper, is to seed fuzzy jets based on the output of a sequential recombination jet algorithm. This guarantees an IRC safe initial condition and therefore the entire procedure is IRC safe. We now discuss practically how one can find the maximum of the fuzzy jets likelihood.

\clearpage
\newpage

\section{Clustering Fuzzy Jets: the EM Algorithm}
\label{sec:EMalgorithm}

One iterative procedure for maximizing the mixture model likelihood in Eq.~(\ref{eq:mm}) is the {\it Expectation-Maximization} (EM) algorithm~\cite{em1,em2,em3}.  After initializing the cluster locations and prior density $\pi$, the following two steps are repeated:

\begin{description}   
\item[Expectation] Given the current values of $\theta_j$, compute the fuzzy membership probabilities $q_{ij}=\pi_j\Phi(\vec{\rho}_i|\mu_j,\Sigma_j)/\sum_{j'}\pi_{j'}\Phi(\vec{\rho}_i|\mu_{j'},\Sigma_{j'})$.

\item[Maximization] Given $q_{ij}$, maximize the {\it expected modified complete log likelihood} over the parameters $\pi,\mu,\Sigma$.
\end{description}

\noindent The expected modified complete log likelihood has the form

\begin{align}
\label{eq:cll}
\sum_{i=1}^N\sum_{j=1}^kp_{Ti}^\alpha(q_{ij}\log\Phi(\vec{\rho}_i;\vec{\mu}_j,\Sigma_j)+q_{ij}\log\pi_j).
\end{align}

\begin{figure}[h!]

\begin{center}
\begin{overpic}[width=0.2\textwidth]{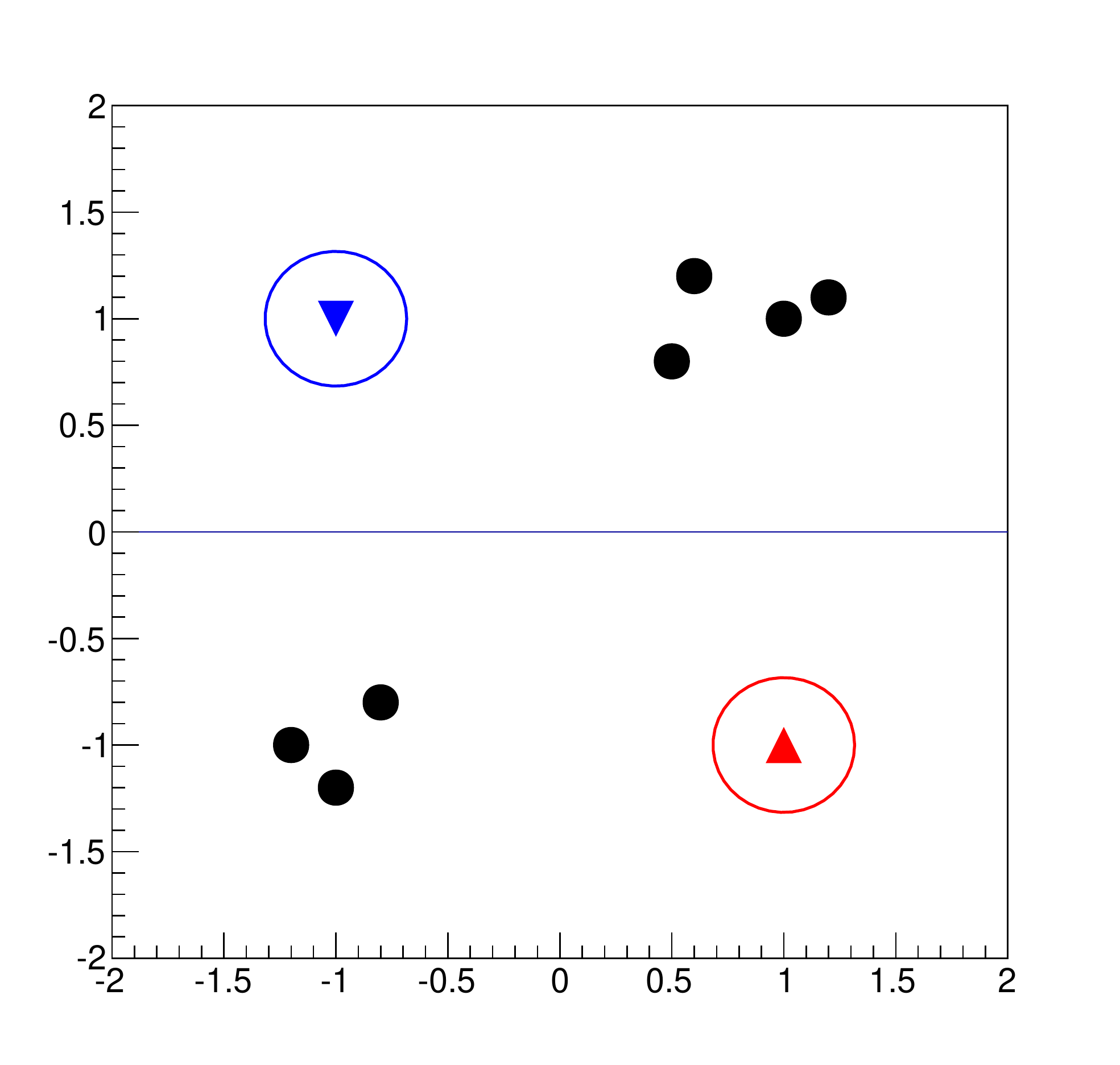}
\put(18,90){\scriptsize Initialization}
\end{overpic}\begin{overpic}[width=0.2\textwidth]{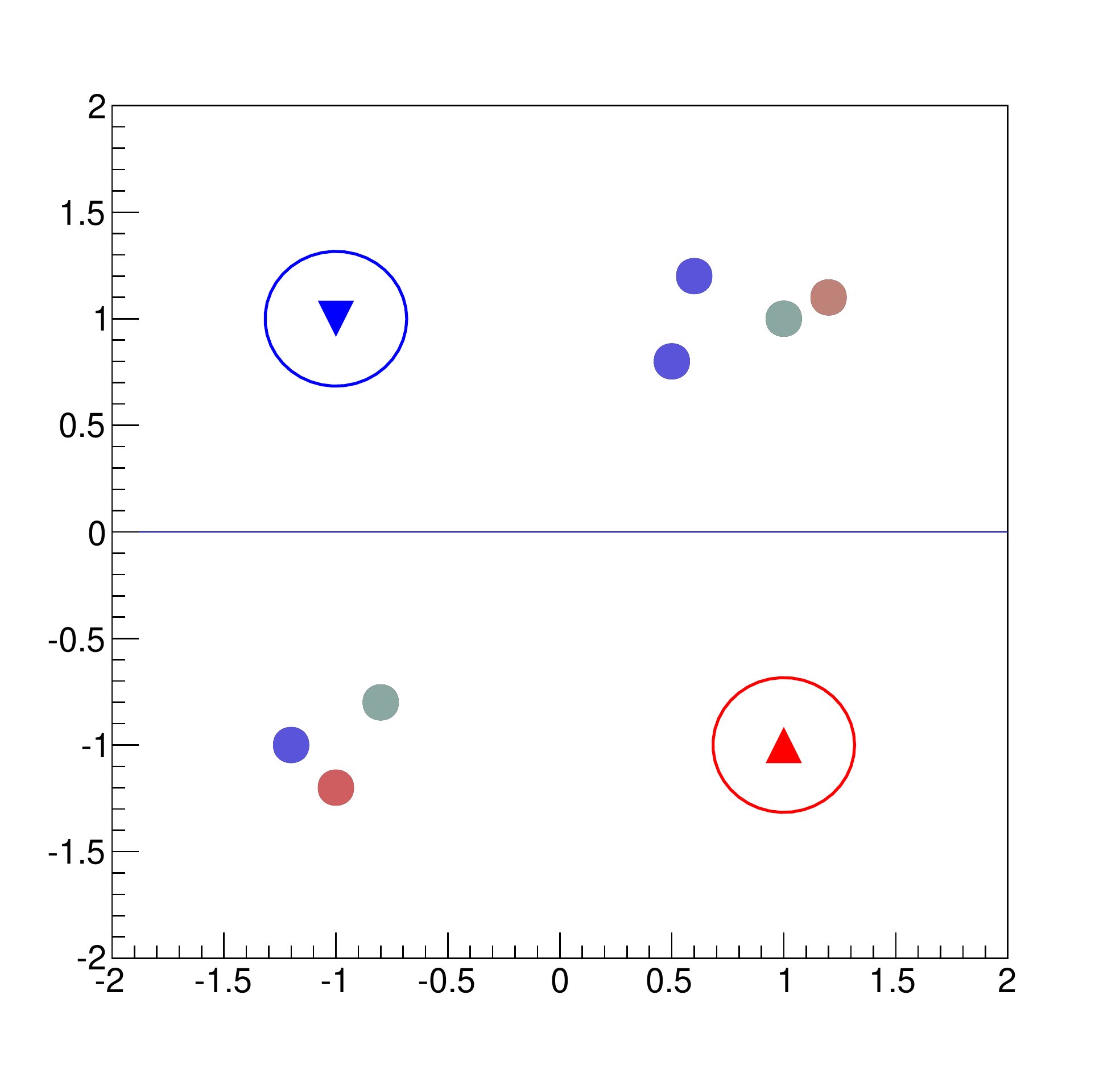}
\put(25,90){\scriptsize $1^\text{st}$ E step}
\end{overpic}\begin{overpic}[width=0.2\textwidth]{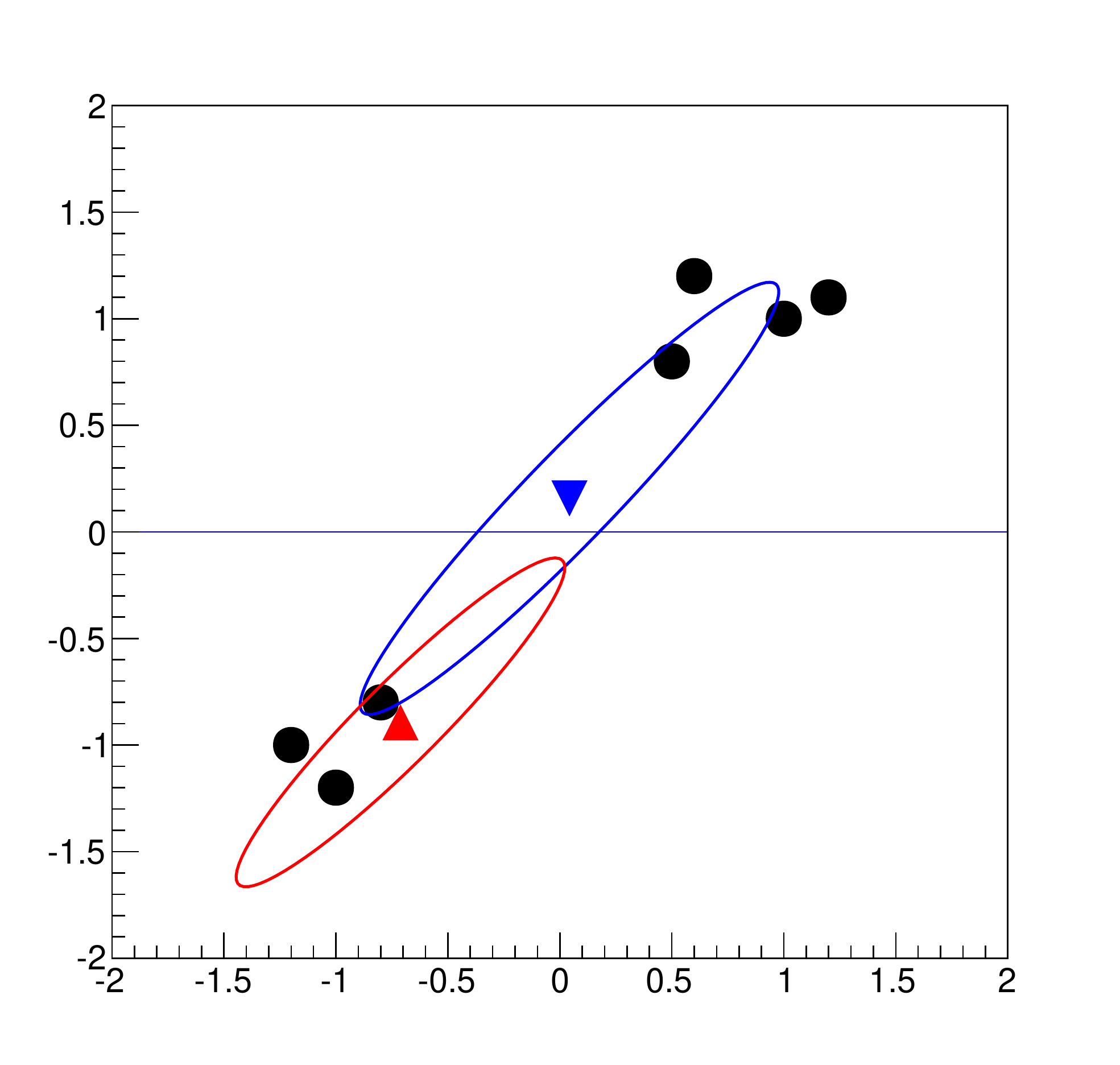}
\put(25,90){\scriptsize $1^\text{st}$ M step}
\end{overpic}\begin{overpic}[width=0.2\textwidth]{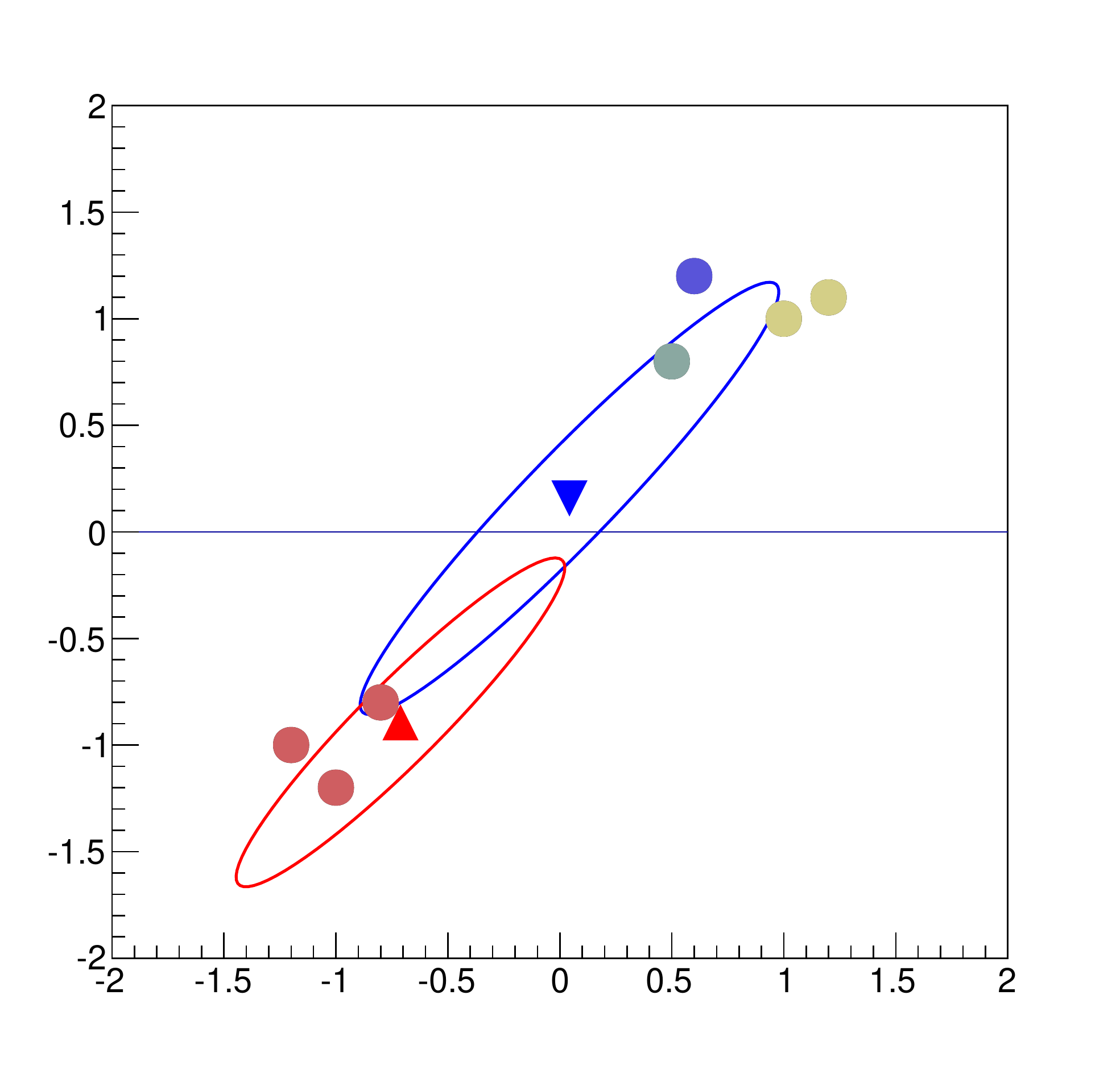}
\put(25,90){\scriptsize $2^\text{nd}$ E step}
\end{overpic}\begin{overpic}[width=0.2\textwidth]{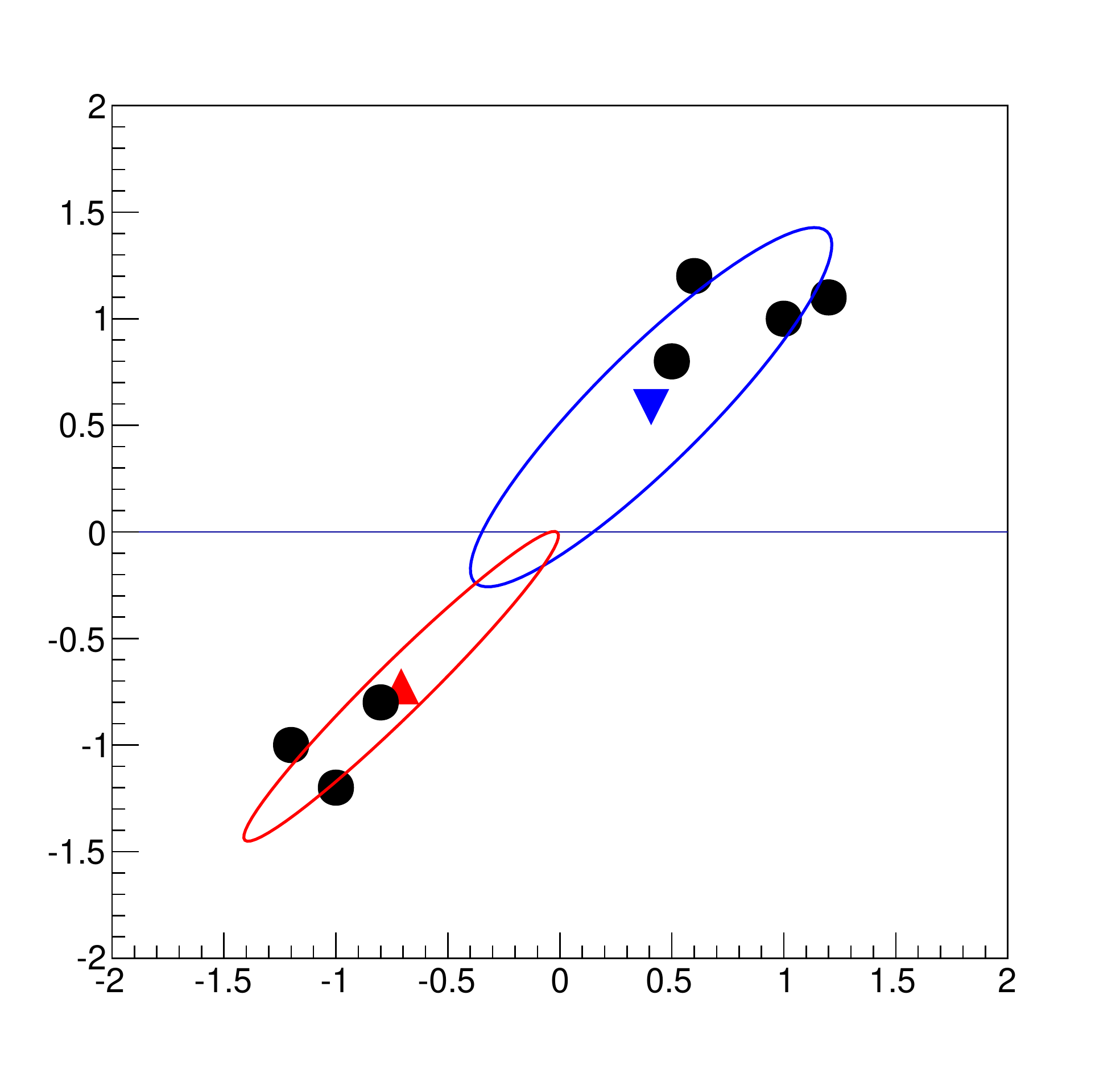}
\put(25,90){\scriptsize $2^\text{nd}$ M step}

\put(-305,0){

\begin{tikzpicture}
  \draw[thick,rounded corners=8pt] (0,1) -- (0,3) -- (6,3) 
   -- (6,0) -- (0,0) -- (0,1);
  \end{tikzpicture}

}

\put(-105,0){

\begin{tikzpicture}
  \draw[thick,rounded corners=8pt] (0,1) -- (0,3) -- (6,3) 
   -- (6,0) -- (0,0) -- (0,1);
  \end{tikzpicture}

}

\end{overpic}\\

\vspace{10mm}

\begin{overpic}[width=0.2\textwidth]{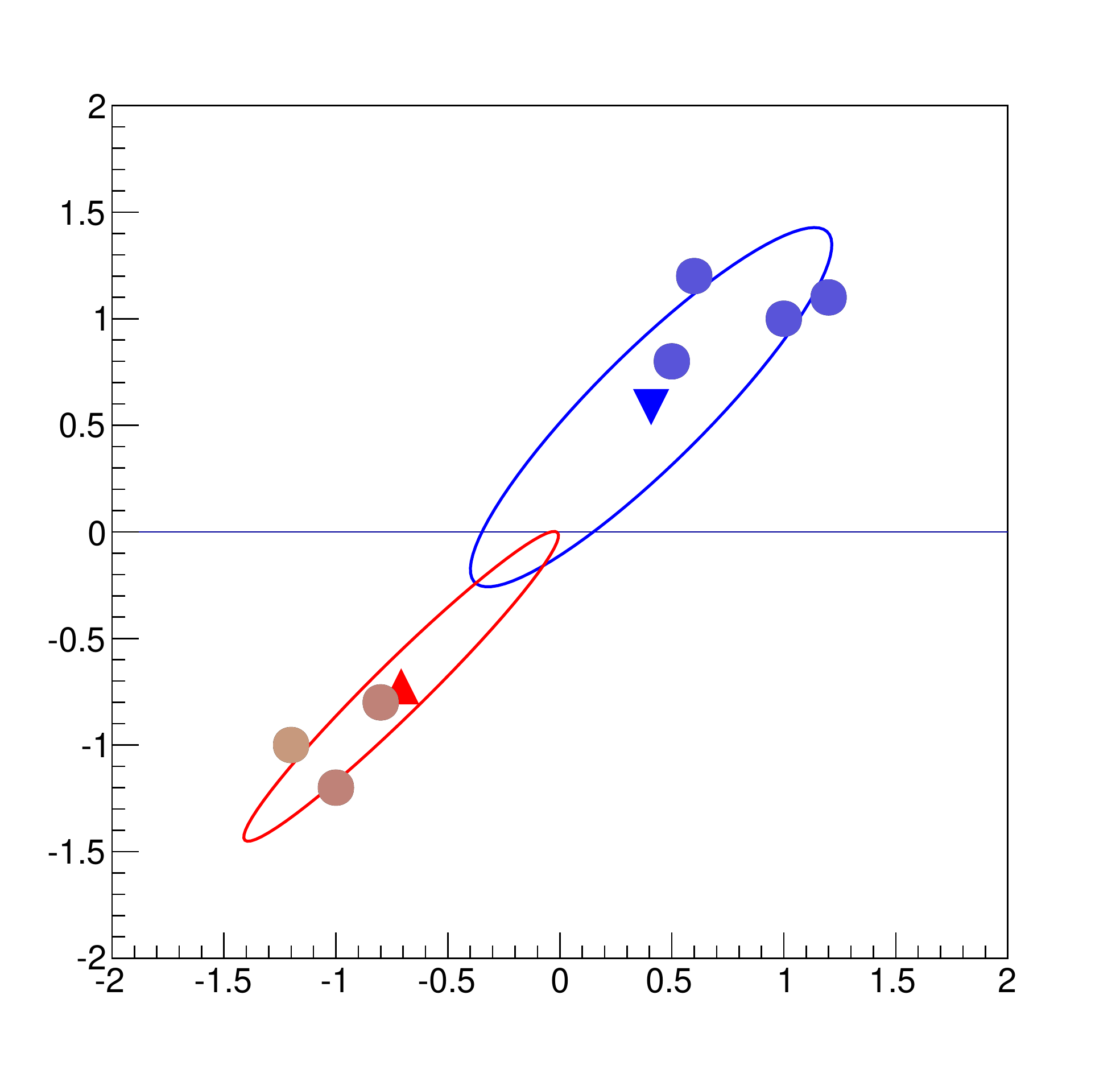}
\put(25,90){\scriptsize $3^\text{rd}$ E step}
\end{overpic}\begin{overpic}[width=0.2\textwidth]{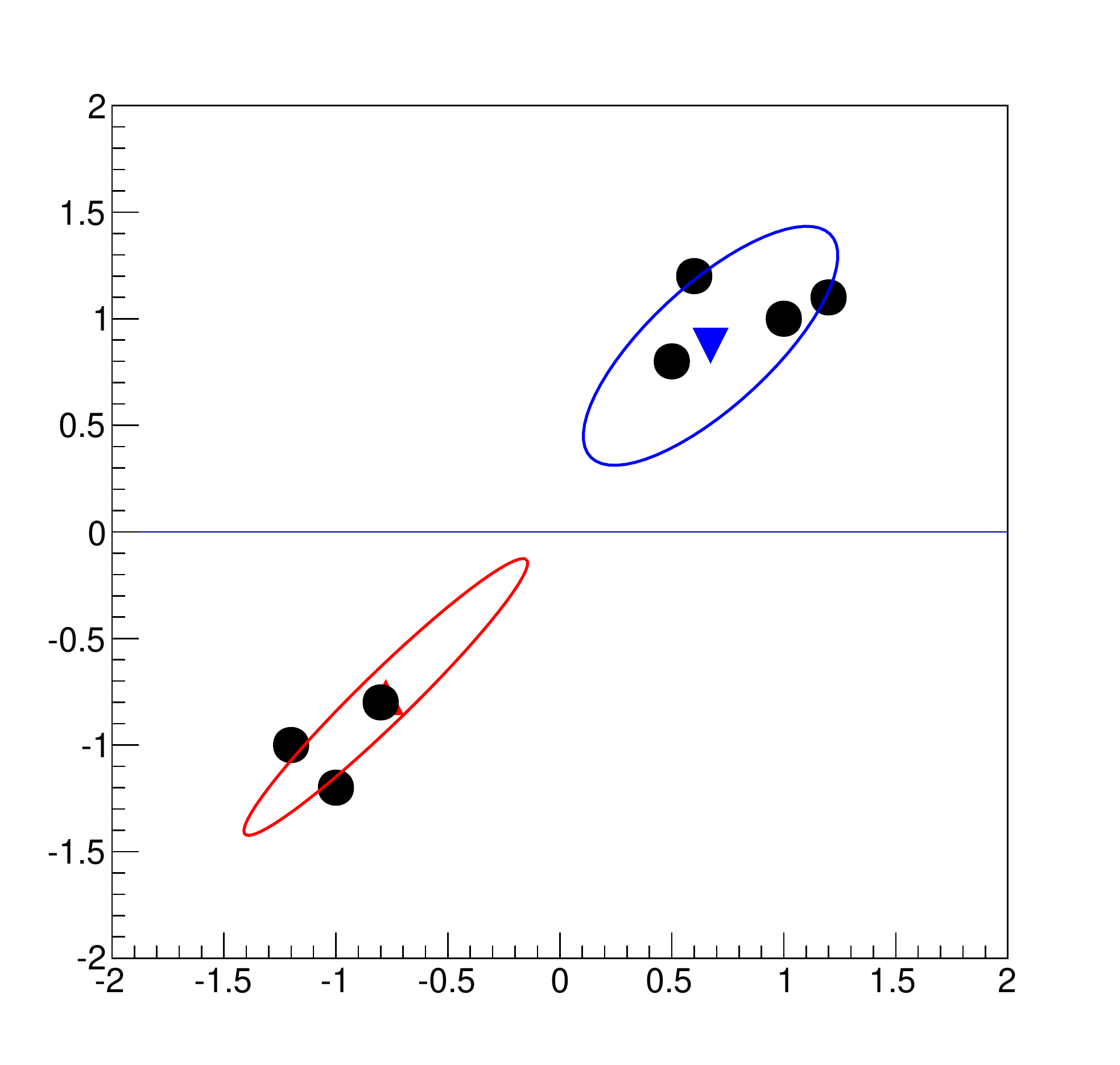}
\put(25,90){\scriptsize $3^\text{rd}$ M step}
\end{overpic}\begin{overpic}[width=0.2\textwidth]{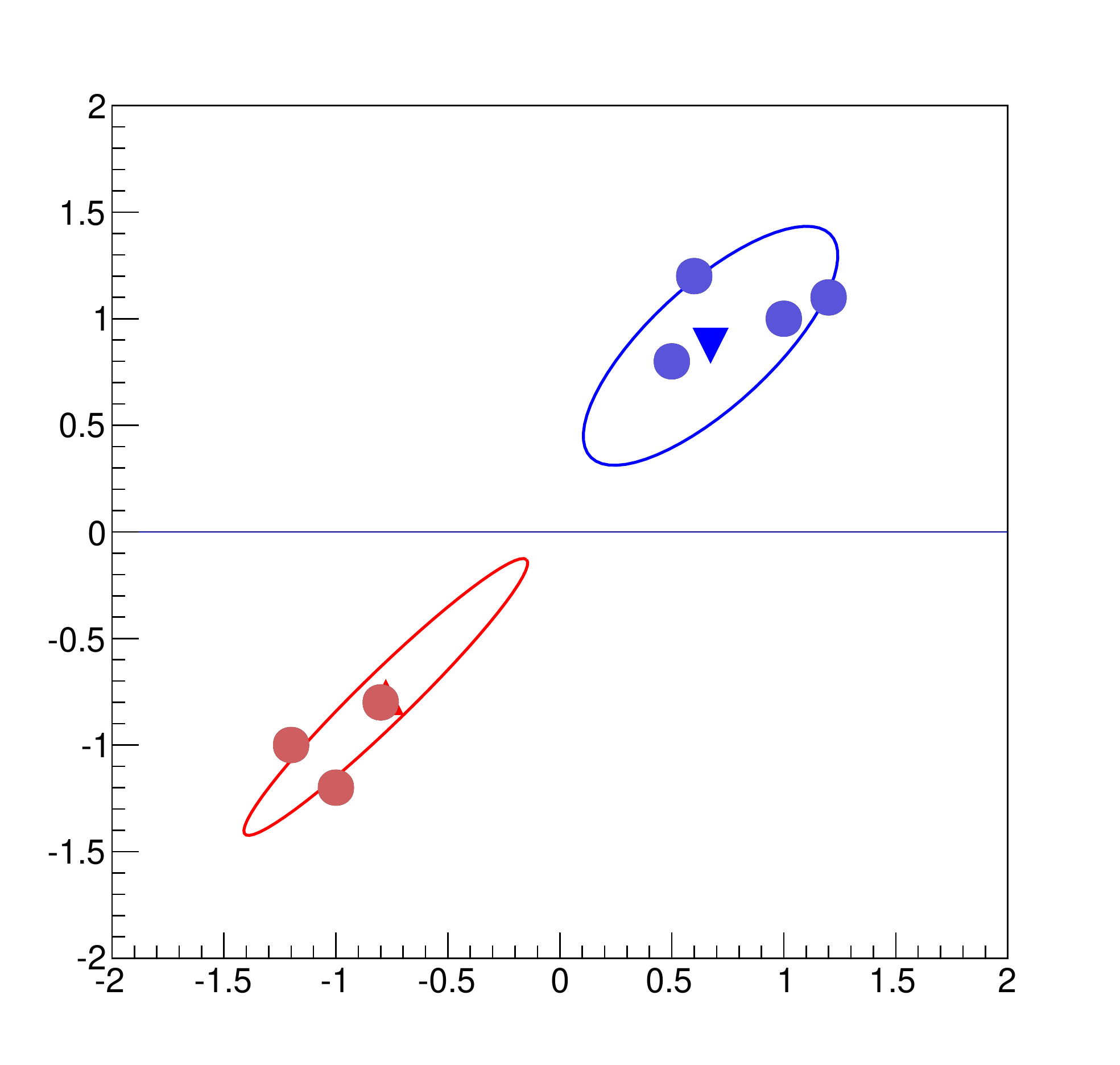}
\put(25,90){\scriptsize $4^\text{th}$ E step}
\end{overpic}\begin{overpic}[width=0.2\textwidth]{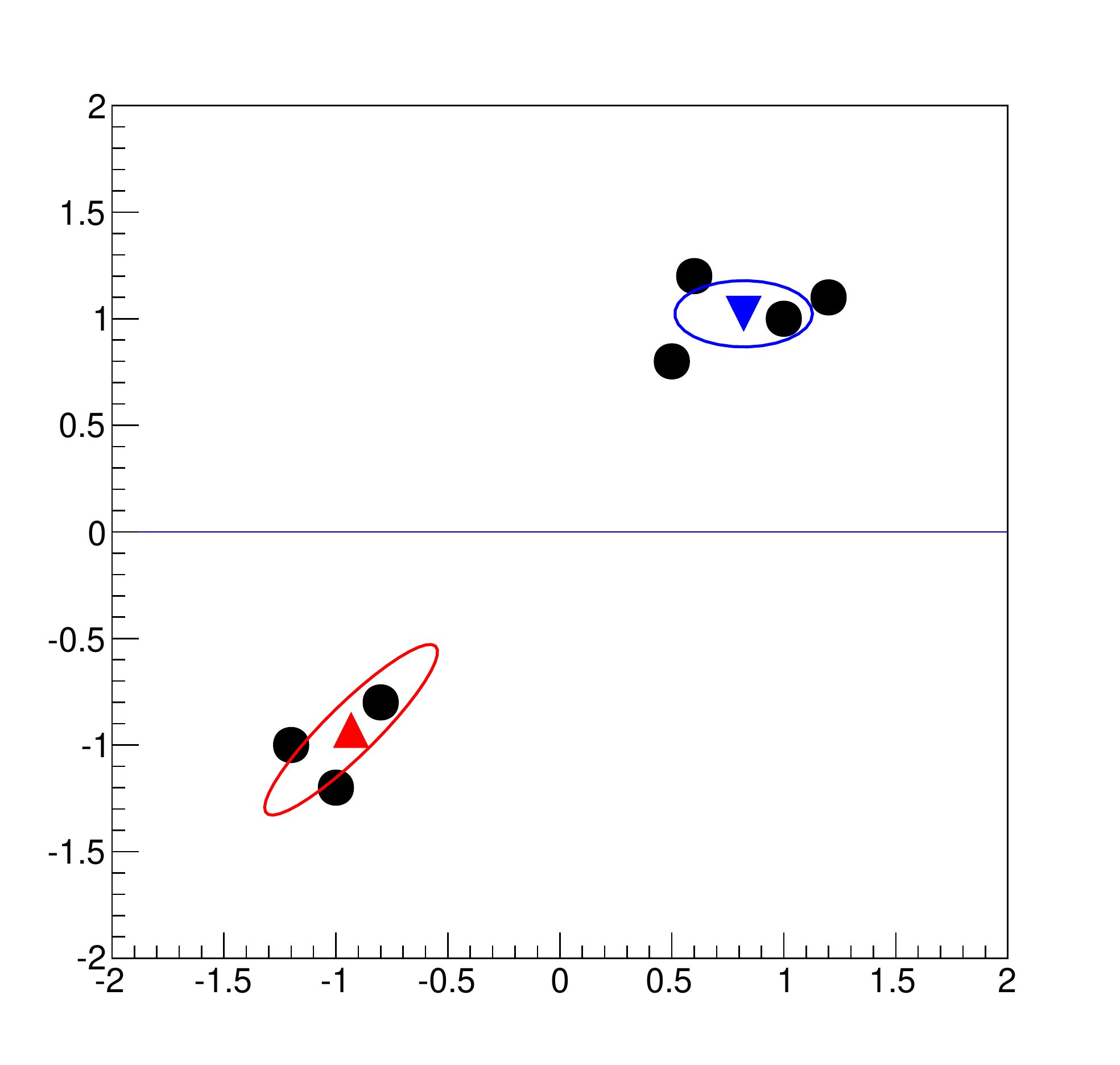}
\put(25,90){\scriptsize $4^\text{th}$ M step}
\end{overpic}\begin{overpic}[width=0.2\textwidth]{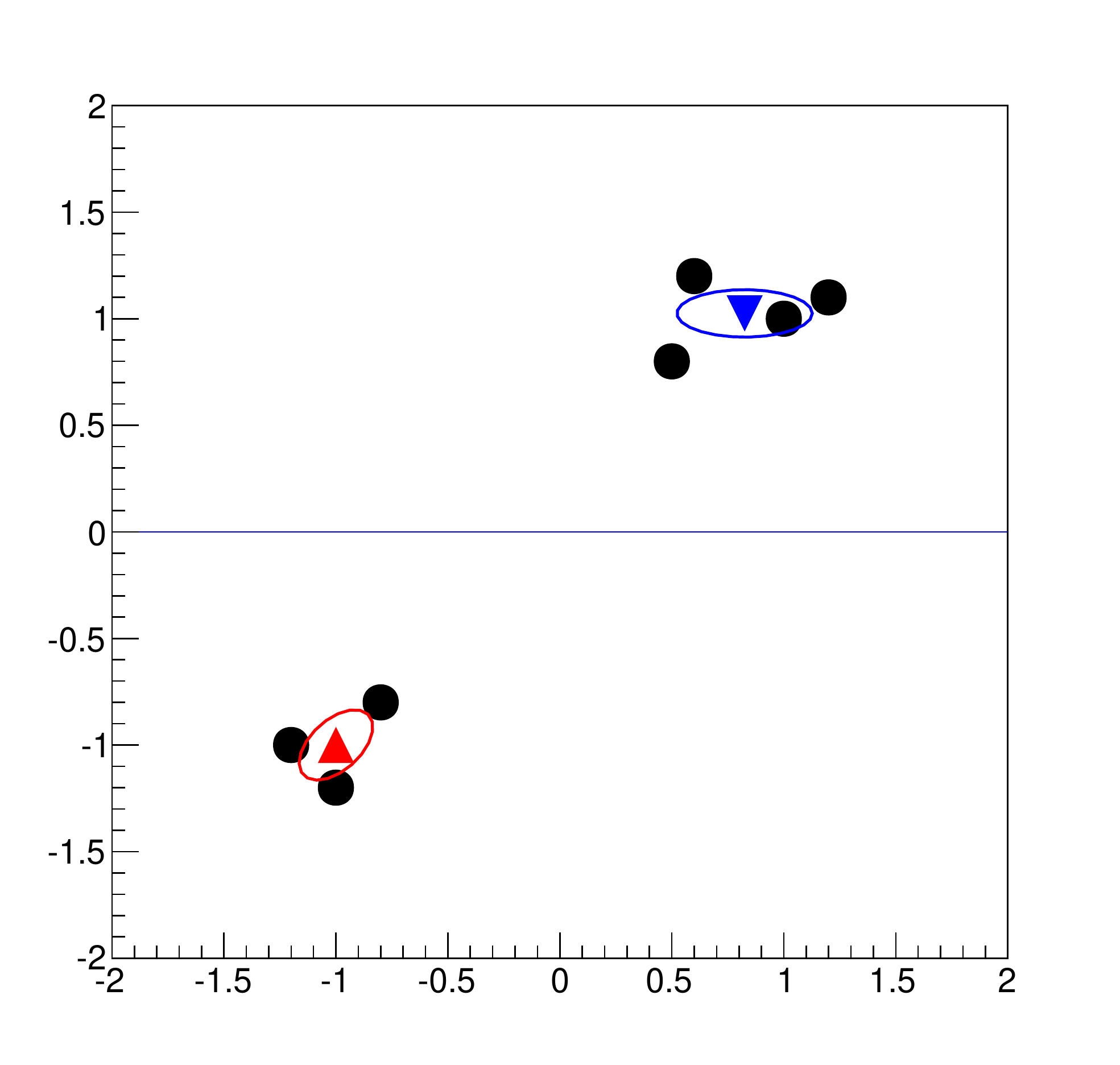}
\put(21,103){\scriptsize $5^\text{st}$ M step:}
\put(22,90){\scriptsize\color{red} Converged}

\put(-405,0){

\begin{tikzpicture}
  \draw[thick,rounded corners=8pt] (0,1) -- (0,3) -- (6,3) 
   -- (6,0) -- (0,0) -- (0,1);
  \end{tikzpicture}

}

\put(-205,0){

\begin{tikzpicture}
  \draw[thick,rounded corners=8pt] (0,1) -- (0,3) -- (6,3) 
   -- (6,0) -- (0,0) -- (0,1);
  \end{tikzpicture}

}

\end{overpic}

\caption{An illustration of of the EM algorithm for $k=2$.  The circles represent data points, the triangles represent the estimated cluster locations $\mu_j$, and the ellipsoids are equidensity contours describing the shapes $\Sigma_j$ of the learned cluster distributions.  In the E-step, bluer colors correspond to higher value of $p_{i,\text{\color{blue} blue jet}}$.}
\label{fig:em}
\end{center}
\end{figure}
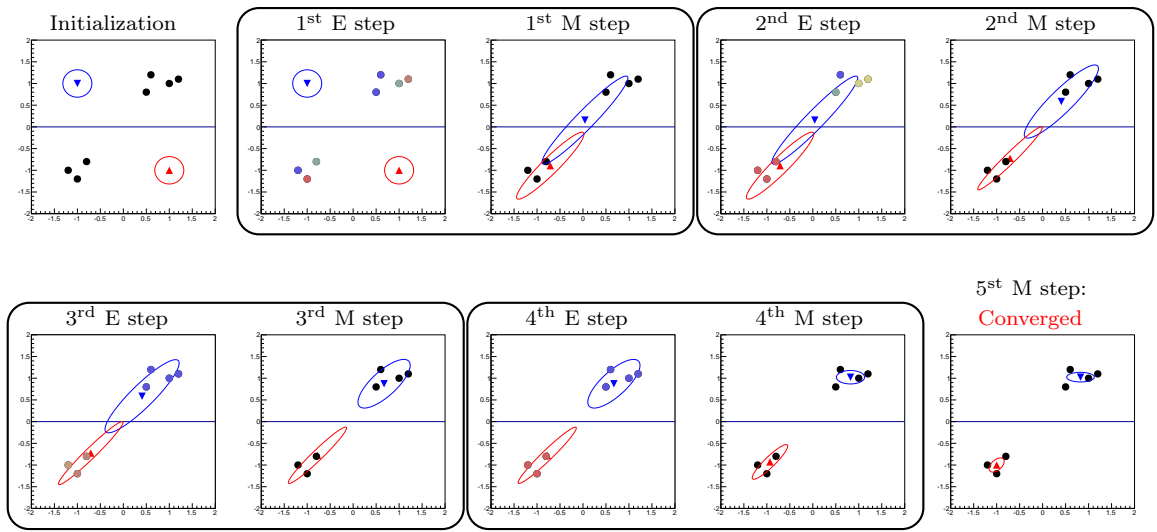

\noindent Note that the expected modified complete log likelihood is not the same as the expected modified log likelihood, shown in Eq.~(\ref{eq:mm2}).  They differ in that the complete log likelihood has the second sum outside the logarithm while Eq.~(\ref{eq:mm2}) has the sum inside the logarithm.  The power of the EM algorithm is that maximizing the complete log likelihood results in fixed point iteration to monotonically improve the original log likelihood.  This desirable property of the EM algortihm is still true when $\alpha>0$; for a proof, see Appendix~\ref{sec:emalgo}.  Many choices for $f$ have closed form maxima for the M step; in the Gaussian $f=\Phi$ case outlined above, the updates are given by

\begin{align}
\label{eq:emupdates}
\mu_j^* = \sum_{i=1}^n \tilde{q}_{ij}x_i \hspace{7mm}\Sigma_j^*=\sum_{i=1}^n  \tilde{q}_{ij}(x_i-\mu_j)(x_i-\mu_j)^\mathsf{T}\hspace{7mm}\pi_j^*=\frac{1}{\sum_{i=1}^np_{Ti}^\alpha}\sum_{i=1}^n p_{Ti}^\alpha  \tilde{q}_{ij},
\end{align}

\noindent where $ \tilde{q}_{ij}=q_{ij}p_{Ti}^\alpha/\sum_{l=1}^np_{lj}p_{Tl}^\alpha$.  The well-known $k$-means clustering algorithm~\cite{macqueen1967} can be recovered as the limit of expectation-maximization in a Gaussian mixture model with $\Sigma=\sigma^2 I, \sigma^2\rightarrow 0$.  Figure~\ref{fig:em} illustrates GMM clustering using the EM algorithm with $k = 2$ clusters.  The EM algorithm readily accommodates constraints on the model parameters.  One constraint we will consider throughout the rest of the paper is $\Sigma_j =\sigma_j ^2 I$ for all $j$, which requires the curves of constant likelihood in $(y,\phi)$ to be circular.  We will see in the next section that the learned value of $\sigma_j$ is useful for distinguishing jets originating from different physics processes.

\clearpage
\newpage

\section{Comparisons with Sequential Recombination and Jet Tagging}
\label{sec:tagging}

This section describes some numerical comparisons between sequential recombination and fuzzy jets.  Section~\ref{sec:details} summarizes the simulation details with some first event displays showing both fuzzy and sequential recombination jets.  These two approaches to jet clustering are studied over an ensemble of events in Sec.~\ref{sec:compare}.  A third subsection, Sec.~\ref{sec:newinfo},  illustrates that fuzzy jets captures new information about the hadronic final state, and in the fourth section, Sec~\ref{sec:tagging}, it is demonstrated that this new information can be used to classify the jet type.

\subsection{Details of the Simulation}
\label{sec:details}

In order to study fuzzy jets in a realistic scenario, we run Monte Carlo (MC) simulations.  Three physics processes are generated using \textsc{Pythia} 8.170~\cite{Pythia8,Pythia} at $\sqrt{s}=8$ TeV.  Hadronic $W$ boson and top quarks are used for studying hard 2- and 3-prong type jets, respectively.  To simulate high $p_T$ hadronic $W$ decays, $W'$ bosons are generated to decay exclusively into a $W$ and $Z$ boson which subsequently decay into quarks and leptons, respectively.  The $p_T$ scale of the hadronically decaying $W$ is set by the mass of the $W'$ which is tuned to 800~GeV for this study so that the $p_T^W \lesssim 400$ GeV. In this $p_T^W$ range, the $W$ decay products are expected to merge within a cone of $R~1.0$ where $\Delta R^2=\Delta\phi^2+\Delta\eta^2\sim 4m_W^2/p_{T,W}^2$.  A sample enriched in 3-prong type jets is generated with $Z'\rightarrow t\bar{t}$, where the $Z'$ mass sets the energy scale of the hadronically decaying top quarks.  In this analysis, we use $m_{Z'}=1.0$~TeV, which sets $p_T^t \gtrsim 500$ GeV.  To study the impact on signal versus background, QCD dijets are generated with a range of $\hat{p}_T$ that is approximately in the same range as the relevant signal process.  In all distributions, the QCD $p_T$ spectrum is weighted to exactly match that of the signal to control for differences between signal and background due only to the $p_T$ spectrum differences.  Pileup is simulated by overlaying additional independently generated minimum-bias interactions with each signal event.  For the rest of this section, the number of pileup interactions $n_\text{PU}=0$.  See Sec.~\ref{sec:pileup} for studies of $n_\text{PU}>0$.

For a comparison to fuzzy jets, anti-$k_t$ jets are clustered using \textsc{FastJet}~\cite{fastjet} 3.0.3.  The signal processes are chosen such that jets with radius parameter $R = 1$ are most appropriate in capturing the decay products of the heavy particles.  The anti-$k_t$ jets are trimmed~\cite{trimming} by re-clustering the constituents into $R=0.3$ $k_t$ subjets and dropping those which have $p_T^\text{subjet}<0.05\times p_T^\text{jet}$.  Anti-$k_t$ jets are also used to seed the fuzzy jet clustering; the $p_T$ threshold for this initialization is 5 GeV\footnote{This low threshold guarantees that there are enough seed jets around to capture the radiation from the underlying event.  Another strategy could be to use the Event Jet (see Sec.~\ref{sec:pileup}) even when there is no pileup.}, and the impact of this choice is studied in Appendix~\ref{sec:pt_multiplicity}.

\clearpage
\newpage

To model the discretization and finite acceptance of a real detector, a calorimeter of towers with size $0.1\times 0.1$ in $(y,\phi)$ extends out to $y=5.0$.  The total energy of the simulated particles incident upon a particular cell are added as scalars and the four-vector $p_j$ of any particular tower $j$ is given by

\begin{align}
\label{eq:calo}
p_j = \sum_{i\text{ incident on $j$}}E_i(\cos\phi_j/\cosh y_j,\sin\phi_j/\cosh y_j,\sinh y_j/\cosh y_j,1).
\end{align}

\noindent To simulate a particle flow reconstruction, the sum in
Eq.~(\ref{eq:calo}) contains only neutral particles for $|y|<2.5$ and
both charged and neutral particles for $2.5<|y|<5$.  Charged particles
within $|y|<2.5$ are individually added to the list of inputs for
clustering, unless they originate from a pileup collision.  Anti-$k_t$
jet momenta are corrected for pileup on average using area
subtraction~\cite{areas}.  The median pileup density, $\rho$, is
estimated by clustering hard scatter particles, neutral pileup particles, and charged pileup particles in the range
$|y|<2.5$ using $k_t$ $R = 0.4$ jets in
\textsc{FastJet} with ghosted areas. %The subset of these jets in
%the range $|y|<1.5$ is considered, and $\rho$ is defined to be the
%$p_T$ per unit ghosted area of the jet with median $p_T$ in this
%range. %This process is parameterized in
%\textsc{FastJet} 3.0.3 with the jet median background estimator
%implementation contained therein.

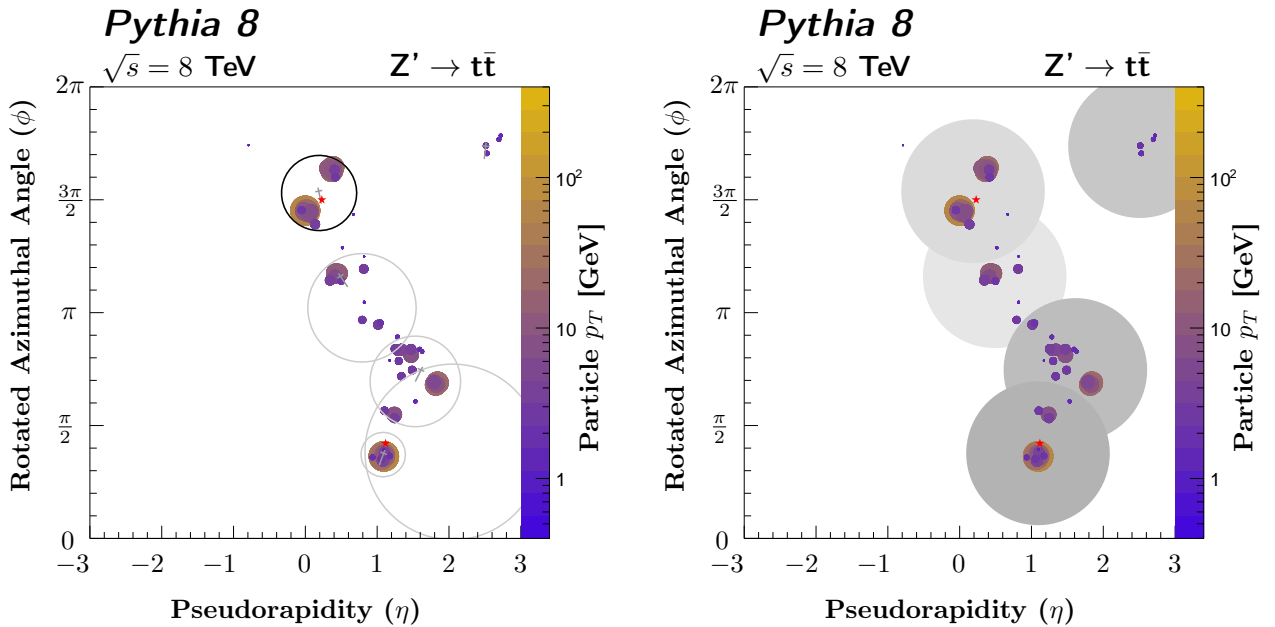
\begin{figure}[h!]
\vspace{1cm}
\begin{center}
\begin{tabular}{cc}
\begin{overpic}[width=0.5\textwidth]{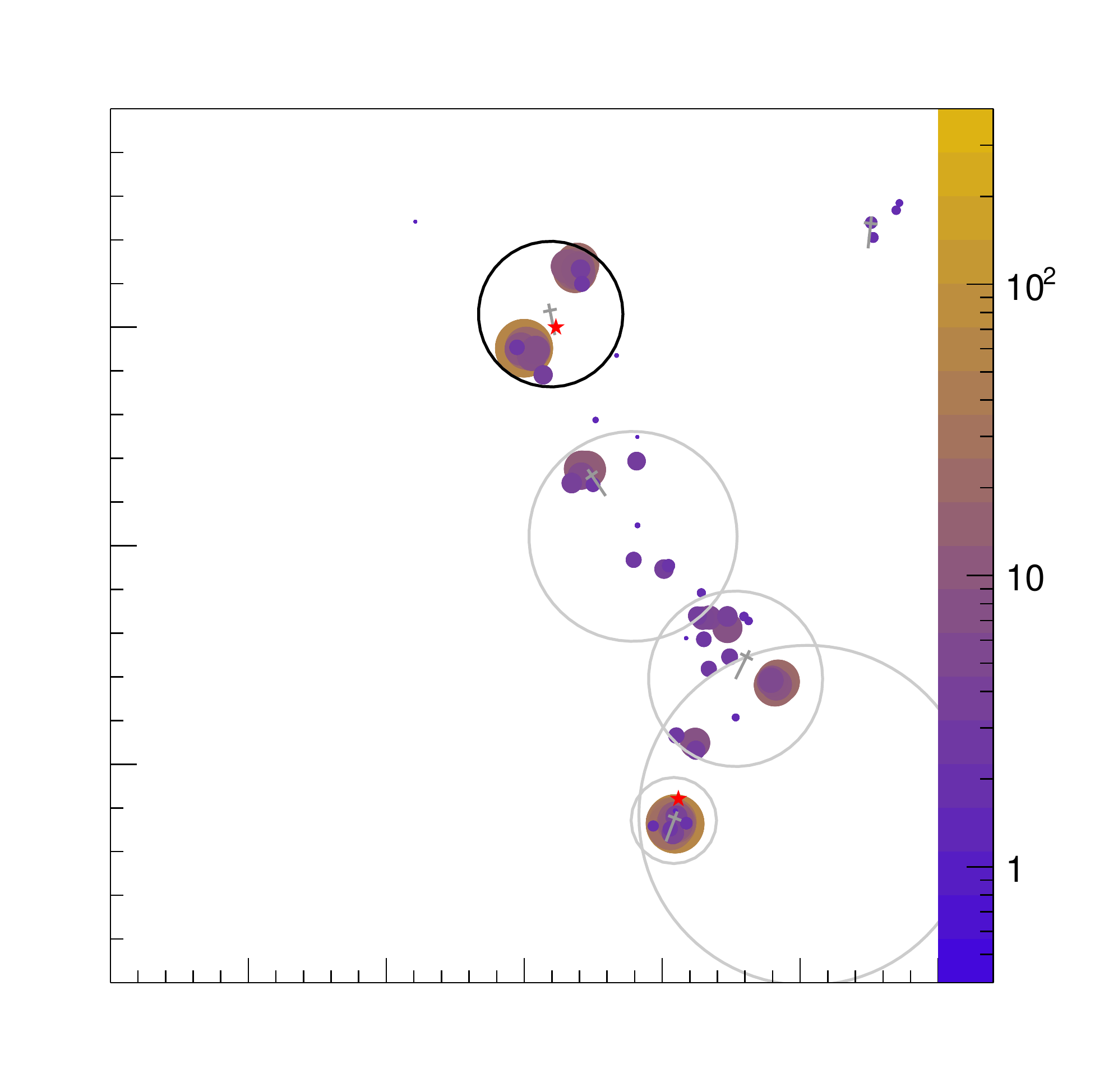}
% general labels
\put(12, 98){\bfseries \sffamily \Large \emph{Pythia 8}}
\put(12, 91){\bfseries \sffamily $\sqrt{s} = 8 \text{ TeV}$}
\put(62, 91){\bfseries \sffamily \large $\text{Z'}
    \rightarrow \text{t}\bar{\text{t}}$}

% axes labels
\put(24, -4){\bfseries \small Pseudorapidity ($\eta$)}
\put(-4, 18){\rotatebox{90}{\bfseries \small Rotated Azimuthal Angle
    ($\phi$)}}
\put(95, 25){\rotatebox{90}{\bfseries \small Particle $p_T \text{ [GeV]}$}}

% axes tick labels
\put(5, 8){\bfseries \small \sffamily $0$}
\put(4, 28.3){\bfseries \small \sffamily $\frac{\pi}{2}$}
\put(5, 48.3){\bfseries \small \sffamily $\pi$}
\put(4, 68){\bfseries \small \sffamily $\frac{3\pi}{2}$}
\put(4, 88){\bfseries \small \sffamily $2\pi$}

\put(4,  4){\bfseries \small \sffamily $-3$}
\put(17, 4){\bfseries \small \sffamily $-2$}
\put(29.6, 4){\bfseries \small \sffamily $-1$}
\put(46.2, 4){\bfseries \small \sffamily $0$}
\put(58.8, 4){\bfseries \small \sffamily $1$}
\put(71.3, 4){\bfseries \small \sffamily $2$}
\put(83.8, 4){\bfseries \small \sffamily $3$}

\end{overpic} &
\hspace{0.5cm}
\begin{overpic}[width=0.5\textwidth]{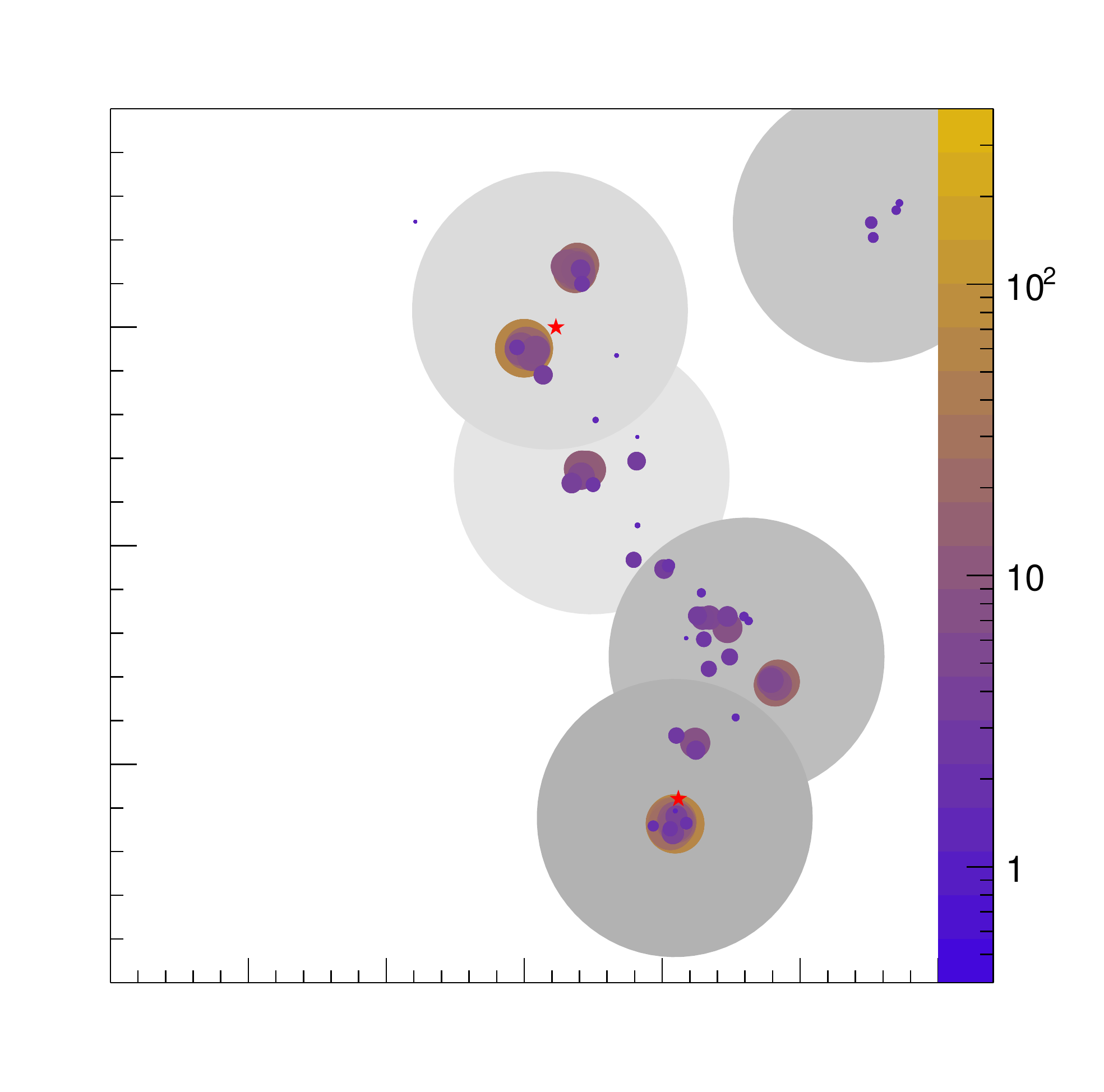}
% general labels
\put(12, 98){\bfseries \sffamily \Large \emph{Pythia 8}}
\put(12, 91){\bfseries \sffamily $\sqrt{s} = 8 \text{ TeV}$}
\put(62, 91){\bfseries \sffamily \large $\text{Z'}
    \rightarrow \text{t}\bar{\text{t}}$}

% axes labels
\put(24, -4){\bfseries \small Pseudorapidity ($\eta$)}
\put(-4, 18){\rotatebox{90}{\bfseries \small Rotated Azimuthal Angle
    ($\phi$)}}
\put(95, 25){\rotatebox{90}{\bfseries \small Particle $p_T \text{ [GeV]}$}}

% axes tick labels
\put(5, 8){\bfseries \small \sffamily $0$}
\put(4, 28.3){\bfseries \small \sffamily $\frac{\pi}{2}$}
\put(5, 48.3){\bfseries \small \sffamily $\pi$}
\put(4, 68){\bfseries \small \sffamily $\frac{3\pi}{2}$}
\put(4, 88){\bfseries \small \sffamily $2\pi$}

\put(4,  4){\bfseries \small \sffamily $-3$}
\put(17, 4){\bfseries \small \sffamily $-2$}
\put(29.6, 4){\bfseries \small \sffamily $-1$}
\put(46.2, 4){\bfseries \small \sffamily $0$}
\put(58.8, 4){\bfseries \small \sffamily $1$}
\put(71.3, 4){\bfseries \small \sffamily $2$}
\put(83.8, 4){\bfseries \small \sffamily $3$}
\end{overpic} \vspace{1.6cm}

\\
\end{tabular}

\end{center}
\vspace{-12mm}
\caption{A representative event display for a $Z'\rightarrow t\bar{t}$
  event. In the top left plot, gray circles show the location and
  size of mGMM fuzzy jets after clustering, with the size of the
  circle indicating 1-$\sigma$ contours in the detector; the black
  circle indicates the highest $p_T$ jet with HML particle
  assignment. The small filled colored circles are the particles, with the color and size indicating their energy.  In each case, the events have been rotated in $\phi$ to
  place the truth top quark at $\phi=3\pi/2$, which is indicated by a red
  star. Anti-$k_t$ jet locations are shown with gray crosses in the
  left hand plot, the long tail of which points towards the mGMM jet
  for which it was a seed. In the top right plot, anti-$k_t$ R = 1.0 jets passing
  a $5$ GeV $p_T$ cut are shown as discs under the particles
  indicating their active area, with centers the same as the crosses
  in the left hand side. Shades of gray in the anti-$k_t$ discs have
  no scale and are meant to aid the eye, but go from low $p_T$
  (lighter) to high $p_T$ (darker).}  %The bottom plot gives a sense of the fuzzy jet catchment area through a Voronoi diagram, colored to indicate the (HML) jet assignment.}
\label{fig:eventdisplay}
\end{figure}

A representative event display for a $Z'\rightarrow t\bar{t}$ event is
shown in Figure~\ref{fig:eventdisplay}.  The top right plot in Figure~\ref{fig:eventdisplay} shows the anti-$k_t$ jets with $p_T>5$ GeV as filled in (partial) circles.  The filled area is determined by the jet area and there are deviations from circles only one a low $p_T$ jet is close to a higher $p_T$ jet.  The two top quarks are depicted as red stars, each of which sits at the center of two high $p_T$ jets.  The top left plot in Figure~\ref{fig:eventdisplay} shows mGMM fuzzy jets.  The fuzzy jets are depicted by their 1-$\sigma$ contours.  In contrast to the anti-$k_t$ jets, fuzzy jets vary widely in radial size.  Gray crosses in the top left plot indicate the locations of the anti-$k_t$ jets shown in the top right plot.  The long tail of the crosses point toward the fuzzy jet for which they were the seed.  The two jets closest to the top quarks did not move a long distance from the seed location, though the size did change significantly from $R=1$.  The lowest $p_T$ fuzzy jet moved a long distance from the seed to the final location.   %To give a sense for the catchment are of fuzzy jets, a Voronoi diagram built from all particles in the event is shown in the bottom plot of Figure~\ref{fig:eventdisplay}.  The different colors in the diagram indicate the assignment, under the HML procedure, of particles to fuzzy jets.  The blue jet in the bottom plot corresponds to the black circle in the top left plot - it is the highest $p_T$ fuzzy jet under HML.   The large fuzzy jet at $(\eta,\phi)\sim(2,-\pi/4)$ in the top left plot provides a catchment for all the particles in the event which are far away from the other fuzzy jets, as seen by the light gray patches on the four corners of the Voronoi diagram.  

Another new feature of fuzzy jets compared to anti-$k_t$ jets is that they can overlap with each other.  This is seen by the four jets with overlapping 1-$\sigma$ contours in the top left plot of Figure~\ref{fig:eventdisplay}.  Overlapping mGMM jets are an expression of structure inadequately captured with a single Gaussian shape. The ability to learn features at different
scales in the same event without relying on a size parameter like the anti-$k_t$ radius parameter can give mGMM fuzzy jets additional
descriptive power over anti-$k_t$ and other traditional jet algorithms.  This particular event will be used
again for reference in Section~\ref{sec:pileup} during a discussion on the
performance of the technique in the presence of pileup interactions.

\subsection{Kinematic Properties of Fuzzy Jets}
\label{sec:compare}
\begin{figure}[h!]
\begin{center}
\begin{tabular}{cc}
\begin{tikzpicture}
  \node[anchor=south west,inner sep=0] (image) at (0,0)
  {\includegraphics[width=0.43\textwidth]{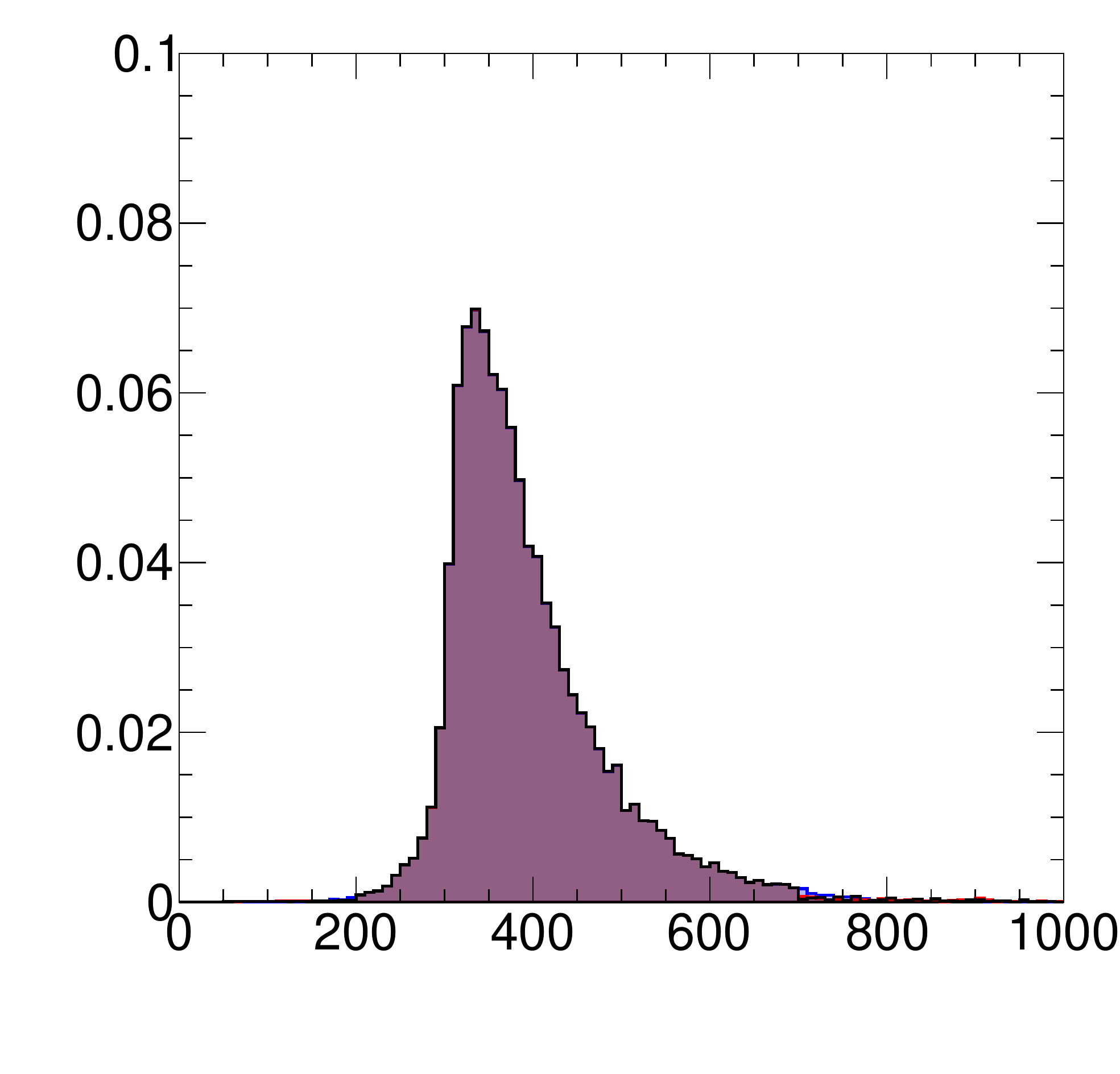}};
  \begin{scope}[x={(image.south east)},y={(image.north west)}]
    % legend
    \draw[blue,fill=white!72!blue,thick] (0.58,0.85) rectangle
    (0.65,0.9);
    \draw[red,fill=white!72!red,thick] (0.58,0.78) rectangle
    (0.65,0.83);
    \draw[black,fill=white!72!black,thick] (0.58,0.71) rectangle
    (0.65,0.76);
    \node[draw=none, anchor=west] at (0.65, 0.875) {\bfseries \sffamily
      \normalsize $\text{QCD}$};
    \node[draw=none, anchor=west] at (0.65, 0.805) {\bfseries \sffamily
      \normalsize $\text{Z'} \rightarrow \text{t}\bar{\text{t}}$};
    \node[draw=none, anchor=west] at (0.65, 0.73) {\bfseries \sffamily
      \normalsize $\text{W} \rightarrow \text{qq'}$};

    % labels
    \node[draw=none] at (0.56,0.065) {\bfseries \small Leading
      anti-$k_t$ Jet $p_T$ [GeV]};
    \node[draw=none, rotate=90] at (0.02, 0.55){\bfseries \small Arbitrary
    Units};
    \node[draw=none, anchor=west] at (0.17,0.88) {\bfseries \sffamily
      \Large \emph{Pythia 8}};
    \node[draw=none, anchor=west] at (0.17,0.80) {\bfseries \sffamily
      $\sqrt{s} = 8 \text{ TeV}$};
  \end{scope}
\end{tikzpicture} &
\begin{tikzpicture}
  \node[anchor=south west,inner sep=0] (image) at (0,0)
  {\includegraphics[width=0.43\textwidth]{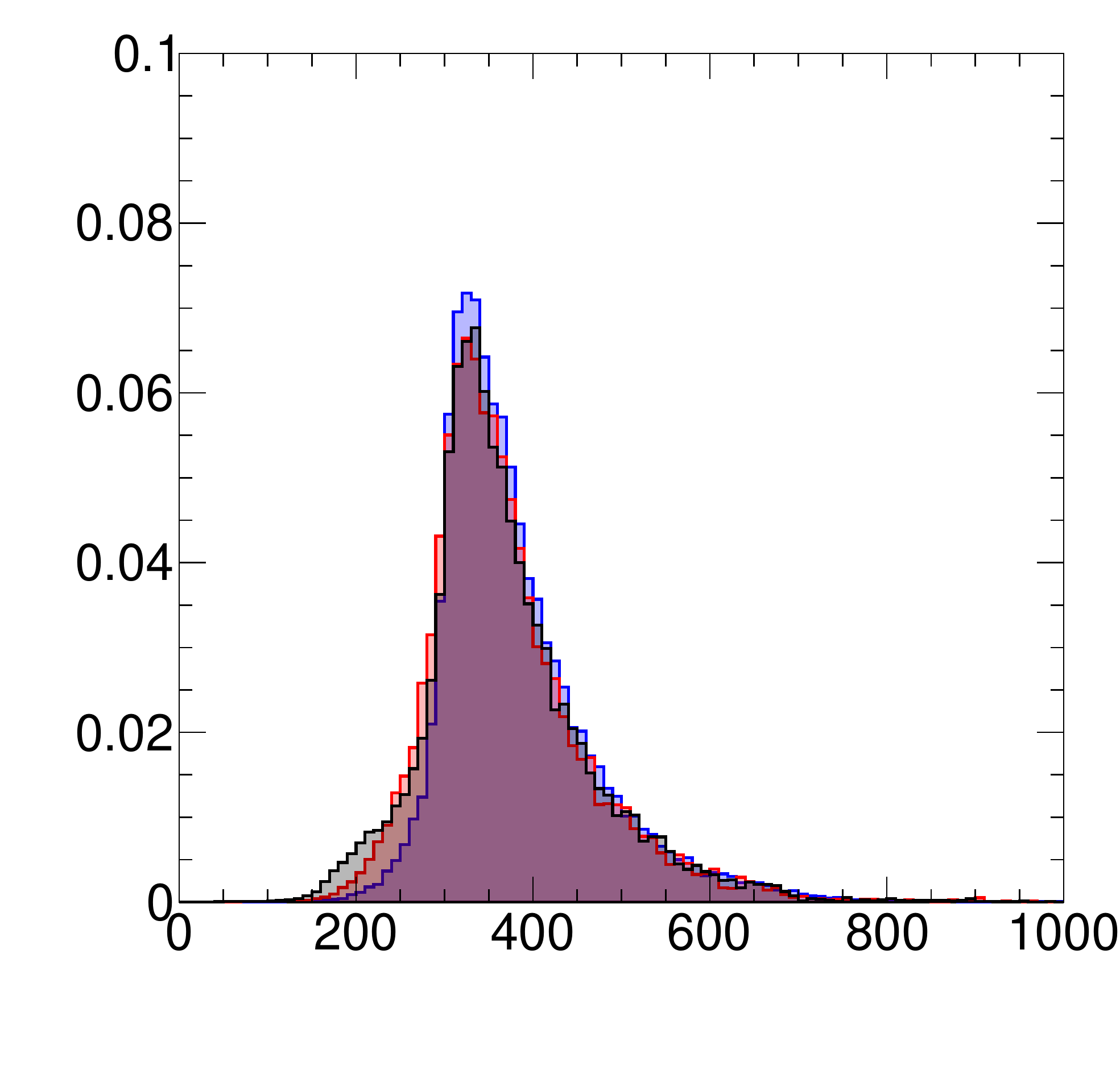}};
  \begin{scope}[x={(image.south east)},y={(image.north west)}]
    % legend
    \draw[blue,fill=white!72!blue,thick] (0.58,0.85) rectangle
    (0.65,0.9);
    \draw[red,fill=white!72!red,thick] (0.58,0.78) rectangle
    (0.65,0.83);
    \draw[black,fill=white!72!black,thick] (0.58,0.71) rectangle
    (0.65,0.76);
    \node[draw=none, anchor=west] at (0.65, 0.875) {\bfseries \sffamily
      \normalsize $\text{QCD}$};
    \node[draw=none, anchor=west] at (0.65, 0.805) {\bfseries \sffamily
      \normalsize $\text{Z'} \rightarrow \text{t}\bar{\text{t}}$};
    \node[draw=none, anchor=west] at (0.65, 0.73) {\bfseries \sffamily
      \normalsize $\text{W} \rightarrow \text{qq'}$};

    % labels
    \node[draw=none] at (0.56,0.065) {\bfseries \small Leading
      mGMM Jet $p_T$ [GeV]};
    \node[draw=none, rotate=90] at (0.02, 0.55){\bfseries \small Arbitrary
    Units};
    \node[draw=none, anchor=west] at (0.17,0.88) {\bfseries \sffamily
      \Large \emph{Pythia 8}};
    \node[draw=none, anchor=west] at (0.17,0.80) {\bfseries \sffamily
      $\sqrt{s} = 8 \text{ TeV}$};
  \end{scope}
\end{tikzpicture} \\
\end{tabular}
\caption{The jet $p_T$ for the leading anti-$k_t$ jet (left) and leading fuzzy jet under the HML particle assignment scheme (right).  All the processes are re-weighted so that the anti-$k_t$ $p_T$ spectra are the same.}
\label{fig:kinematics_pt}
\end{center}
\end{figure}
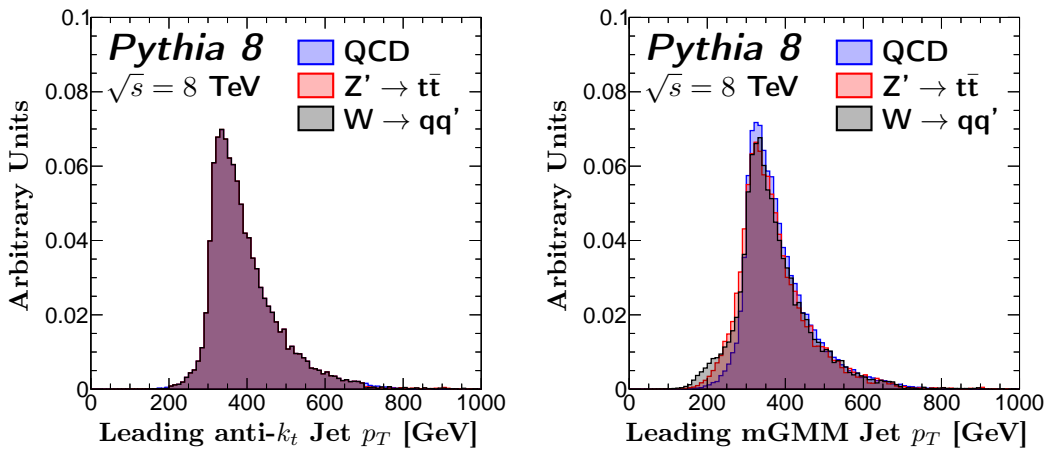

Jets clustered according to the mGMM algorithm capture similar hard jet locations and jet energy (under HML) as those clustered by anti-$k_t$
$R = 1$. In Figure~\ref{fig:kinematics_pt}, the $p_T$ distribution for the
highest $p_T$ jets for three different physics processes
are plotted as given by anti-$k_t$ $R = 1.0$ and
mGMM jets. The anti-$k_t$
$p_T$ distributions are re-weighted so that all three processes have identical distributions in the left plot. On the
right, the distributions are in good correspondence with those in the
left plot, though there is a slight shift of the peak.  Additionally, the $(y, \phi)$ locations of the highest $p_T$
mGMM jets are in excellent correspondence with the locations of the
anti-$k_t$ jets as was already discussed in reference to Figure~\ref{fig:eventdisplay}.

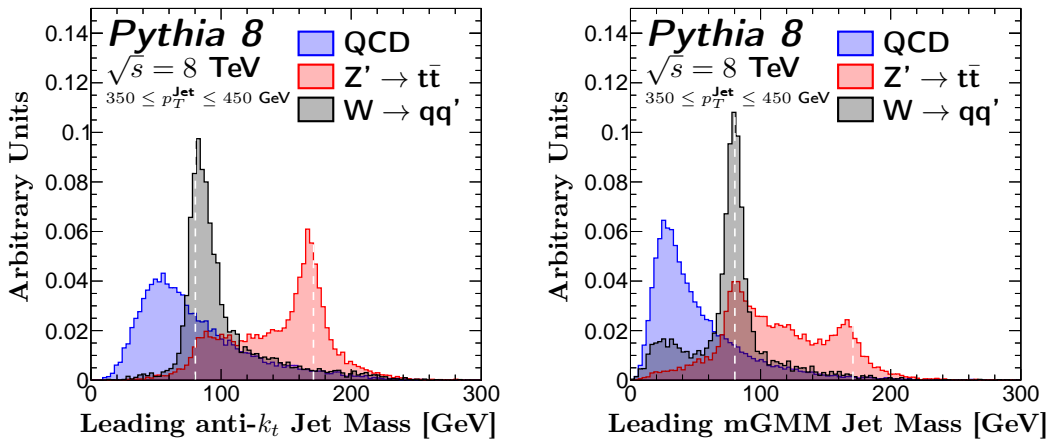
\begin{figure}[h!]
\begin{center}
\begin{tabular}{cc}
\begin{tikzpicture}
  \node[anchor=south west,inner sep=0] (image) at (0,0)
  {\includegraphics[width=0.43\textwidth]{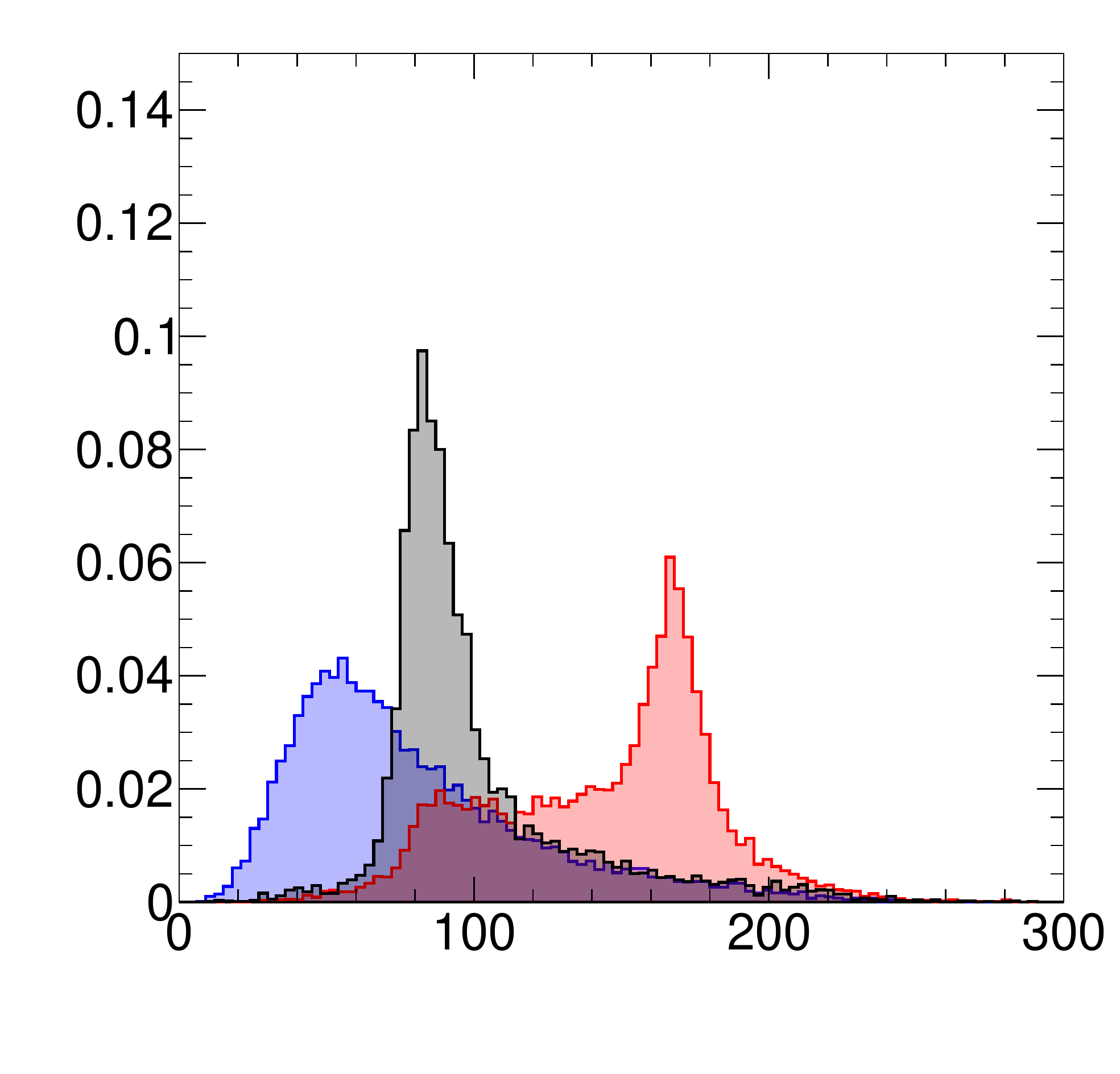}};
  \begin{scope}[x={(image.south east)},y={(image.north west)}]
    % legend
    \draw[blue,fill=white!72!blue,thick] (0.58,0.85) rectangle
    (0.65,0.9);
    \draw[red,fill=white!72!red,thick] (0.58,0.78) rectangle
    (0.65,0.83);
    \draw[black,fill=white!72!black,thick] (0.58,0.71) rectangle
    (0.65,0.76);
    \node[draw=none, anchor=west] at (0.65, 0.875) {\bfseries \sffamily
      \normalsize $\text{QCD}$};
    \node[draw=none, anchor=west] at (0.65, 0.805) {\bfseries \sffamily
      \normalsize $\text{Z'} \rightarrow \text{t}\bar{\text{t}}$};
    \node[draw=none, anchor=west] at (0.65, 0.73) {\bfseries \sffamily
      \normalsize $\text{W} \rightarrow \text{qq'}$};

    % helpers
    \draw [dashed, color=white, semithick] (0.3715, 0.1622) -- (0.3715, 0.76);
    \draw [dashed, color=white, semithick] (0.61, 0.1622) -- (0.61, 0.66);

    % labels
    \node[draw=none] at (0.56,0.065) {\bfseries \small Leading
      anti-$k_t$ Jet Mass [GeV]};
    \node[draw=none, rotate=90] at (0.02, 0.55){\bfseries \small Arbitrary
    Units};
    \node[draw=none, anchor=west] at (0.17,0.89) {\bfseries \sffamily
      \Large \emph{Pythia 8}};
    \node[draw=none, anchor=west] at (0.17,0.82) {\bfseries \sffamily
      $\sqrt{s} = 8 \text{ TeV}$};
    \node[draw=none, anchor=west] at (0.17, 0.76) {\sffamily \tiny $350 \leq
      p_T^{\text{Jet}} \leq 450 \text{ GeV}$};
  \end{scope}
\end{tikzpicture} &
\begin{tikzpicture}
  \node[anchor=south west,inner sep=0] (image) at (0,0)
  {\includegraphics[width=0.43\textwidth]{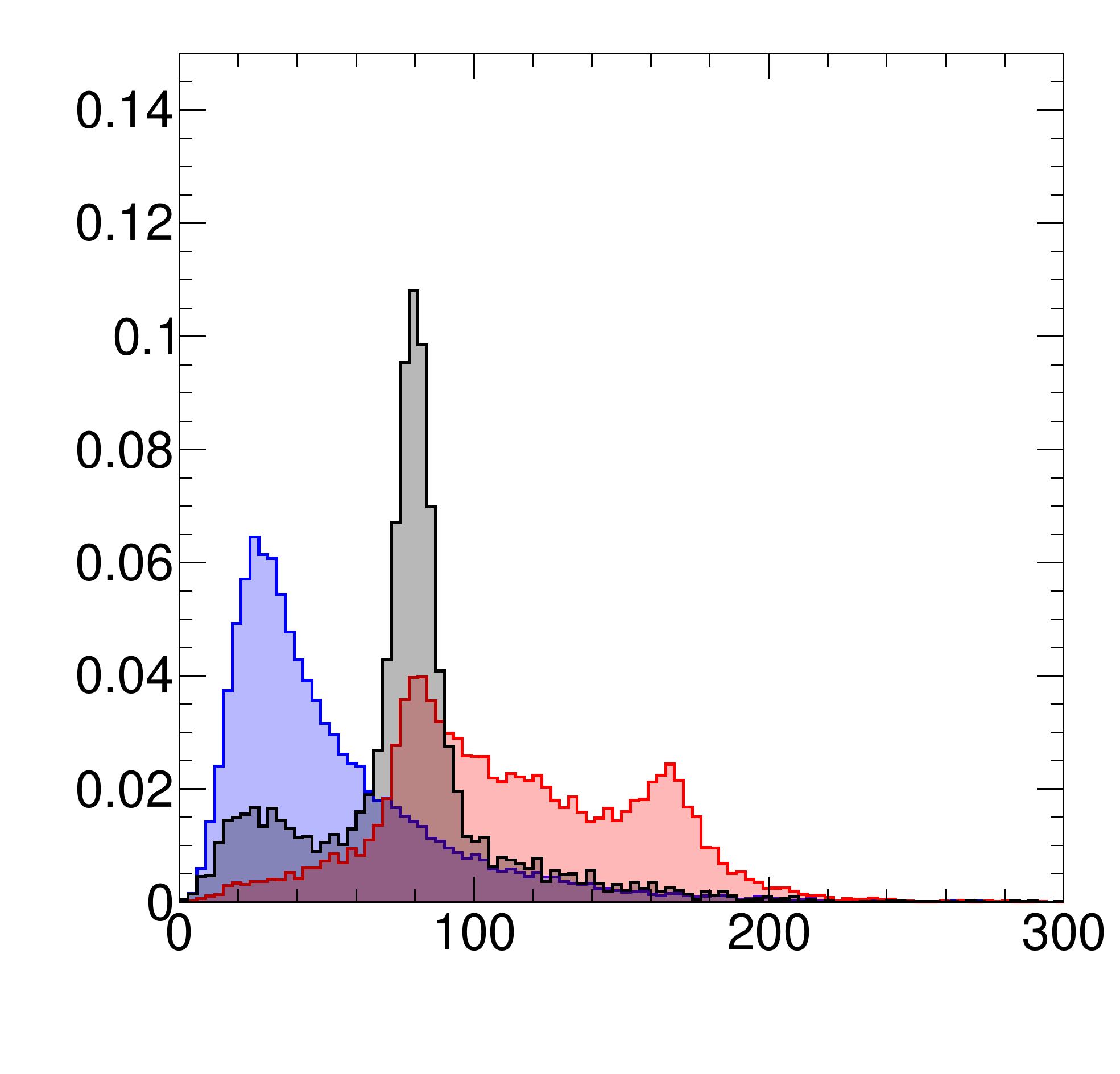}};
  \begin{scope}[x={(image.south east)},y={(image.north west)}]
    % legend
    \draw[blue,fill=white!72!blue,thick] (0.58,0.85) rectangle
    (0.65,0.9);
    \draw[red,fill=white!72!red,thick] (0.58,0.78) rectangle
    (0.65,0.83);
    \draw[black,fill=white!72!black,thick] (0.58,0.71) rectangle
    (0.65,0.76);
    \node[draw=none, anchor=west] at (0.65, 0.875) {\bfseries \sffamily
      \normalsize $\text{QCD}$};
    \node[draw=none, anchor=west] at (0.65, 0.805) {\bfseries \sffamily
      \normalsize $\text{Z'} \rightarrow \text{t}\bar{\text{t}}$};
    \node[draw=none, anchor=west] at (0.65, 0.73) {\bfseries \sffamily
      \normalsize $\text{W} \rightarrow \text{qq'}$};

    % helpers
    \draw [dashed, color=white, semithick] (0.3715, 0.1622) -- (0.3715, 0.76);
    \draw [dashed, color=white, semithick] (0.61, 0.1622) -- (0.61, 0.66);

    % labels
    \node[draw=none] at (0.56,0.065) {\bfseries \small Leading
      mGMM Jet Mass [GeV]};
    \node[draw=none, rotate=90] at (0.02, 0.55){\bfseries \small Arbitrary
    Units};
    \node[draw=none, anchor=west] at (0.17,0.89) {\bfseries \sffamily
      \Large \emph{Pythia 8}};
    \node[draw=none, anchor=west] at (0.17,0.82) {\bfseries \sffamily
      $\sqrt{s} = 8 \text{ TeV}$};
    \node[draw=none, anchor=west] at (0.17, 0.76) {\sffamily \tiny $350 \leq
      p_T^{\text{Jet}} \leq 450 \text{ GeV}$};
  \end{scope}
\end{tikzpicture} \\
\end{tabular}
\caption{The jet mass for the leading anti-$k_t$ (left) and
  leading fuzzy jet under the HML particle assignment scheme (right),
  in an anti-$k_t$ leading jet $p_T$ window of 350 to 450 GeV.
  All the processes are re-weighted so that the anti-$k_t$ $p_T$
  distributions are the same.  The dashed white lines mark $m_W = 80.4 \text{
    GeV}$ and $m_t = 173.3 \text{ GeV}$. }
\label{fig:kinematics_mass}
\end{center}
\end{figure}

The mGMM algorithm differs from the anti-$k_t$ algorithm in how the size and structure of clustered jets.  This was already shown qualitatively in
Figure~\ref{fig:eventdisplay}: fuzzy jets come in a variety
of sizes, and can overlap in complex ways. The matter is further
complicated by the choice of particle assignment scheme for defining
kinematic properties in the mGMM family of algorithms. The catchment area's
volume and shape of a fuzzy jet depends in general on the full set of
learned jet locations and model parameters, $\Sigma$.  In contrast, for anti-$k_t$ jets,
the catchment area is bounded from above by $R$ and is only smaller when another high $p_T$ jet is
nearby.  The nonlocality of the
mGMM clustering model can be observed quantitatively by examining jet mass, given in Eq.~(\ref{eq:mass}), which
is sensitive to the distribution of energy within a jet.
\noindent The jet mass distributions for both mGMM (HML assignment) and anti-$k_t$ jets are shown in Figure~\ref{fig:kinematics_mass}, with the same $p_T$ weighting as in Figure~\ref{fig:kinematics_pt}.  Even though fuzzy jets learn the same core (i.e. $p_T$) for jets as
anti-$k_t$, they do not learn the
same mass.  The white dashed lines in Figure~\ref{fig:kinematics_mass} mark the locations of the $W$ boson and top quark masses at about 80 GeV and 175 GeV, respectively~\cite{Agashe:2014kda}.  For both anti-$k_t$ and fuzzy jets, there are clear peaks at the $W$ mass for the boosted $W\rightarrow qq'$ from $W'$ simulated events and at the top quark mass for $Z'\rightarrow t\bar{t}$ simulated events.  However, there are clear differences in the shape of these distributions.  The $W$ mass peak for $W'$ events is more peaked for fuzzy jets, though there is also a low-mass contribution to the distribution.  For $Z'$ events, the top quark mass peak is less populated for fuzzy jets, which instead has shifted events to the $W$ mass peak.  This often happens when the tree-prong structure is learned by two (overlapping) fuzzy jets.  The QCD multi-jet jet mass distribution is also qualitatively different between fuzzy jets and anti-$k_t$ jets, with the former shifted to lower values of the mass.

\begin{align}
\label{eq:mass}
m_\text{jet}^2 = \left(\sum_{i\in\text{jet}}
E_i\right)^2-\left(\sum_{i\in\text{jet}} \vec{p}_i\right)^2
\end{align}

\subsection{New Information from Fuzzy Jets}
\label{sec:newinfo}

The properties $\Sigma$ of a fuzzy jet can be useful in distinguishing jets resulting from different physics processes.  In the simplest realization of mGMM jets already described above, $\Sigma = \sigma^2 I $, where $\sigma$ is a measure of the size of the core of a jet.  Although $\sigma$ is a simple variable to construct from the
wealth of data available after clustering with the mGMM algorithm, it captures at least some of the
schematic differences in the likelihood for $Z'\rightarrow
t\bar{t}$ and $W'\rightarrow WZ$ relative to a QCD multijet background (shown below).

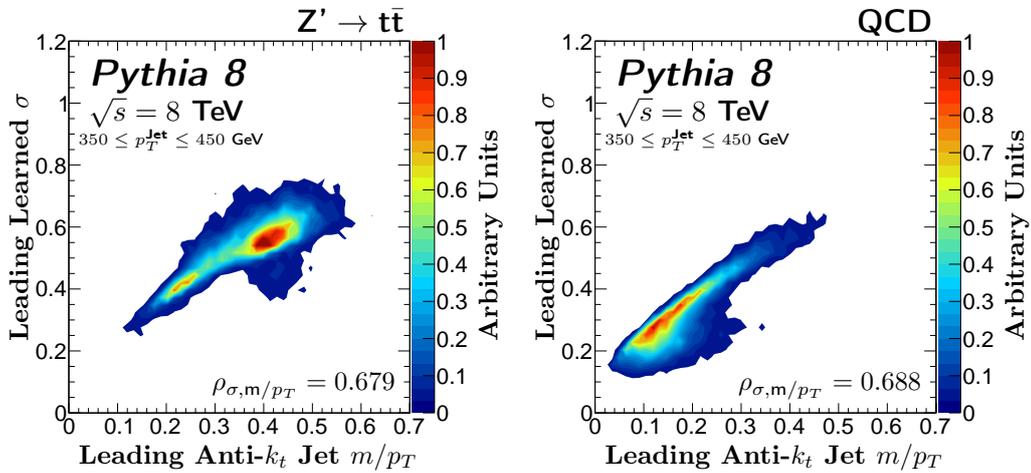
\begin{figure}[h!]
\begin{center}
\begin{tabular}{cc}
\begin{overpic}[width=0.43\textwidth]{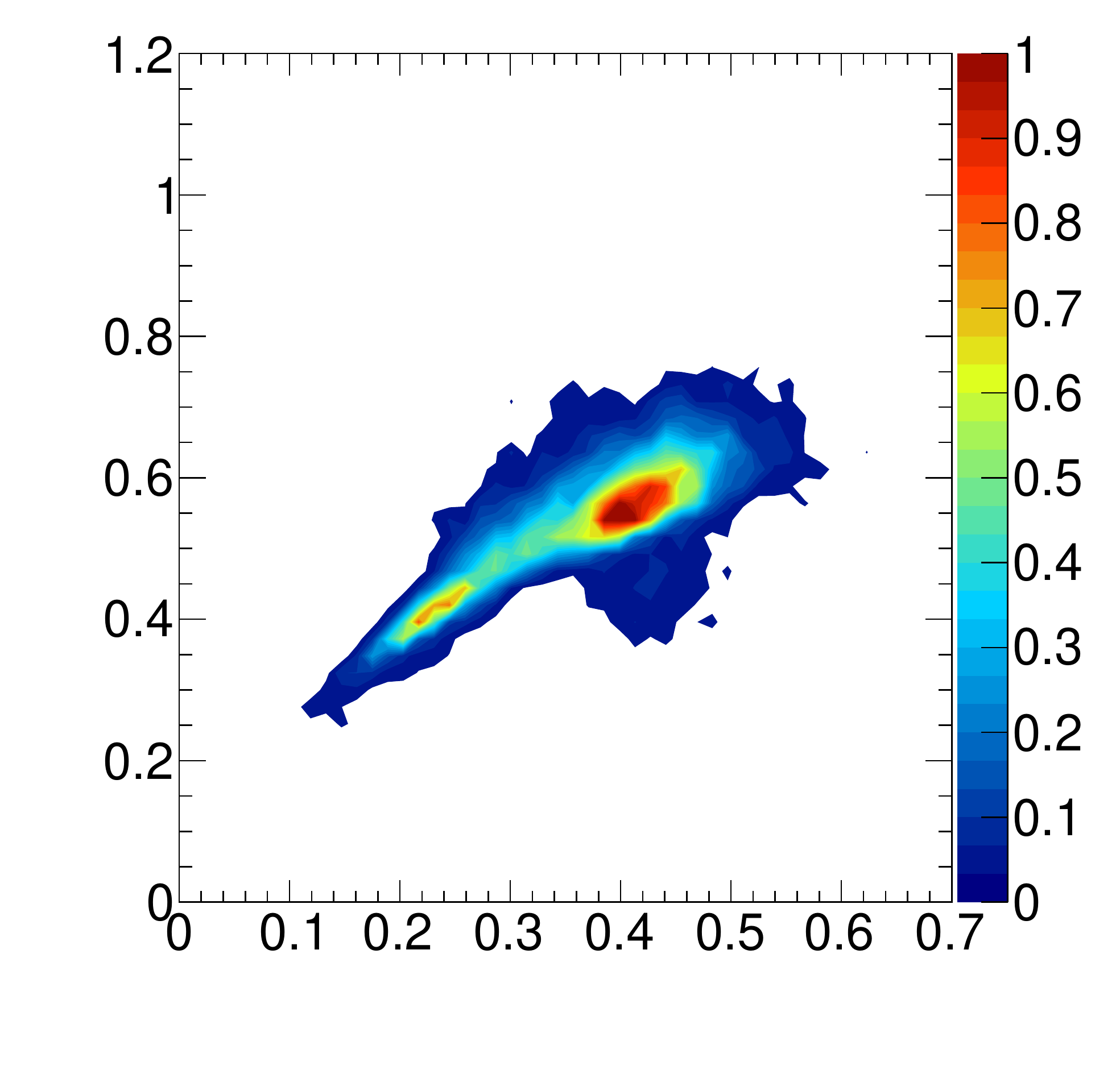}
\put(18, 5){\bfseries \small Leading Anti-$k_t$ Jet $m/p_T$}
\put(4, 30){\rotatebox{90}{\bfseries \small Leading Learned $\sigma$}}
\put(99, 30){\rotatebox{90}{\bfseries \small Arbitrary Units}}
\put(62, 93){\bfseries \sffamily \large $\text{Z'}
    \rightarrow \text{t}\bar{\text{t}}$}

\put(20, 82){\bfseries \sffamily \Large \emph{Pythia 8}}
\put(20, 75){\bfseries \sffamily $\sqrt{s} = 8 \text{ TeV}$}
\put(18, 70){\sffamily \tiny $350 \leq
  p_T^{\text{Jet}} \leq 450 \text{ GeV}$}

\put(45, 20){\sffamily \small $\rho_{\sigma,\text{m}/p_T} = 0.679$}
\end{overpic} &
\begin{overpic}[width=0.43\textwidth]{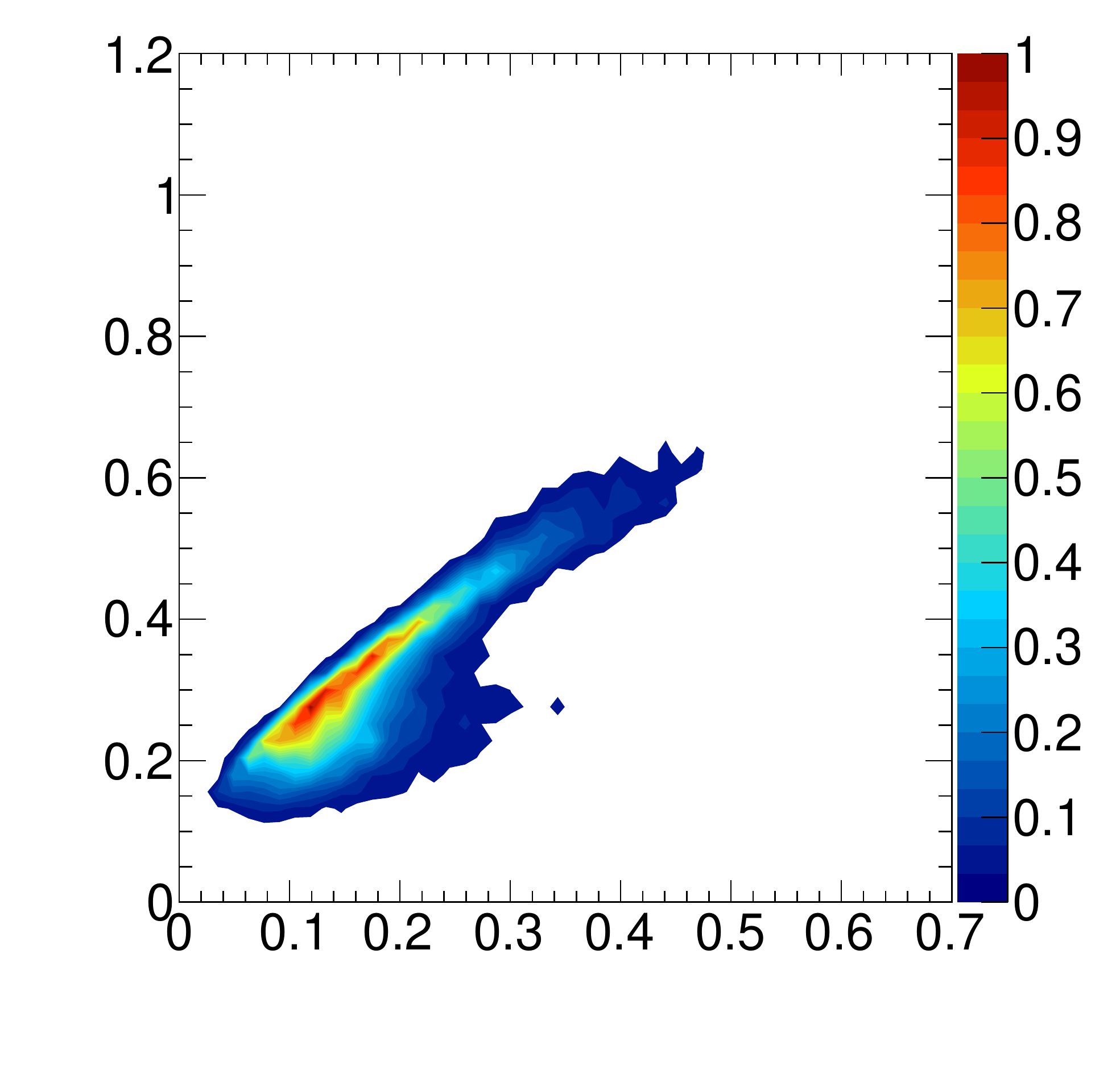}
\put(18, 5){\bfseries \small Leading Anti-$k_t$ Jet $m/p_T$}
\put(4, 30){\rotatebox{90}{\bfseries \small Leading Learned $\sigma$}}
\put(99, 30){\rotatebox{90}{\bfseries \small Arbitrary Units}}
\put(69, 93){\bfseries \sffamily \large $\text{QCD}$}

\put(20, 82){\bfseries \sffamily \Large \emph{Pythia 8}}
\put(20, 75){\bfseries \sffamily $\sqrt{s} = 8 \text{ TeV}$}
\put(18, 70){\sffamily \tiny $350 \leq
  p_T^{\text{Jet}} \leq 450 \text{ GeV}$}

\put(45, 20){\sffamily \small $\rho_{\sigma,\text{m}/p_T} = 0.688$}
\end{overpic} \\
\end{tabular}
\caption{The left and right plots show the correlation between
  $\sigma$ and the leading jet anti-$k_t$ mass divided by $p_T$ in
  an anti-$k_t$ $p_T$ window of $350$ to $450$ GeV for $Z'\rightarrow
  t\bar{t}$ and QCD events, respectively. Indicated in the
  lower right of each figure is the linear correlation between the variables. }
\label{fig:corr_mpt}
\end{center}
\end{figure}

The left plot of Figure~\ref{fig:sigma_hist} also shows the average $\sigma$ over all fuzzy jets in an event.  The generic fuzzy jet is rather independent of the physics process and tends to be quite large.  This is because fuzzy jets capturing hard radiation tend to be small, but most of the fuzzy jets needed to capture the sparse radiation pattern from the underlying event need to be large.  In contrast, the $\sigma$ for the leading mGMM jets are shown the right plot of Figure~\ref{fig:sigma_hist} for each of the three physics
processes.  As expected, the decay relative size of the highest $p_T$ jets depends on the physics process.  For the decay of a boosted heavy particle with mass $m$ and $p_T$, the radial size of the decay products scales as $2m/p_T$ and thus since the $p_T$ distribution in Figure~\ref{fig:sigma_hist} is fixed, one would expect that the top quark jets have a larger $\sigma$ than the $W$ boson jets, which are in turn larger than the quark and gluon jets.  This is reflected\footnote{At leading order, there is an exact relationship between $\sigma$ and the jet mass - See Appendix~\ref{sec:theory}.} in the three peaks in the left plot of Figure~\ref{fig:sigma_hist}.  The separation between the three physics processes it not 100\% correlated with the naive scaling $m/p_T$ of the corresponding leading anti-$k_t$ jets.  Figure~\ref{fig:corr_mpt} shows that there is a strong positive correlation between $\sigma$ and the corresponding anti-$k_t$ mass over $p_T$ as expected.   There are two peaks in the
  correlation for the $Z'\rightarrow t\bar{t}$ events because the
  anti-$k_t$ mass spectrum has peaks at both the top mass, and the $W$ boson mass.  While the correlations between the fuzzy jet $\sigma$ and the anti-$k_t$ $m/p_T$ are non-negligible, they are far from unity and thus there may be additional information contained in the fuzzy jet $\sigma$ that is useful for tagging the flavour of a jet.

%The histogram for $Z'\rightarrow t\bar{t}$ is
%particularly well separated from that for QCD, suggesting that
%$\sigma$ may offer additional discriminating information useful for
%jet tagging.

 %Unless otherwise stated, the event $\sigma$ is simply the width parameter corresponding to the highest $p_T$ mGMM jet under the ML assignment scheme.

\begin{figure}[h!]
\begin{center}
\begin{tabular}{cc}
\begin{tikzpicture}
  \node[anchor=south west,inner sep=0] (image) at (0,0)
  {\includegraphics[width=0.43\textwidth]{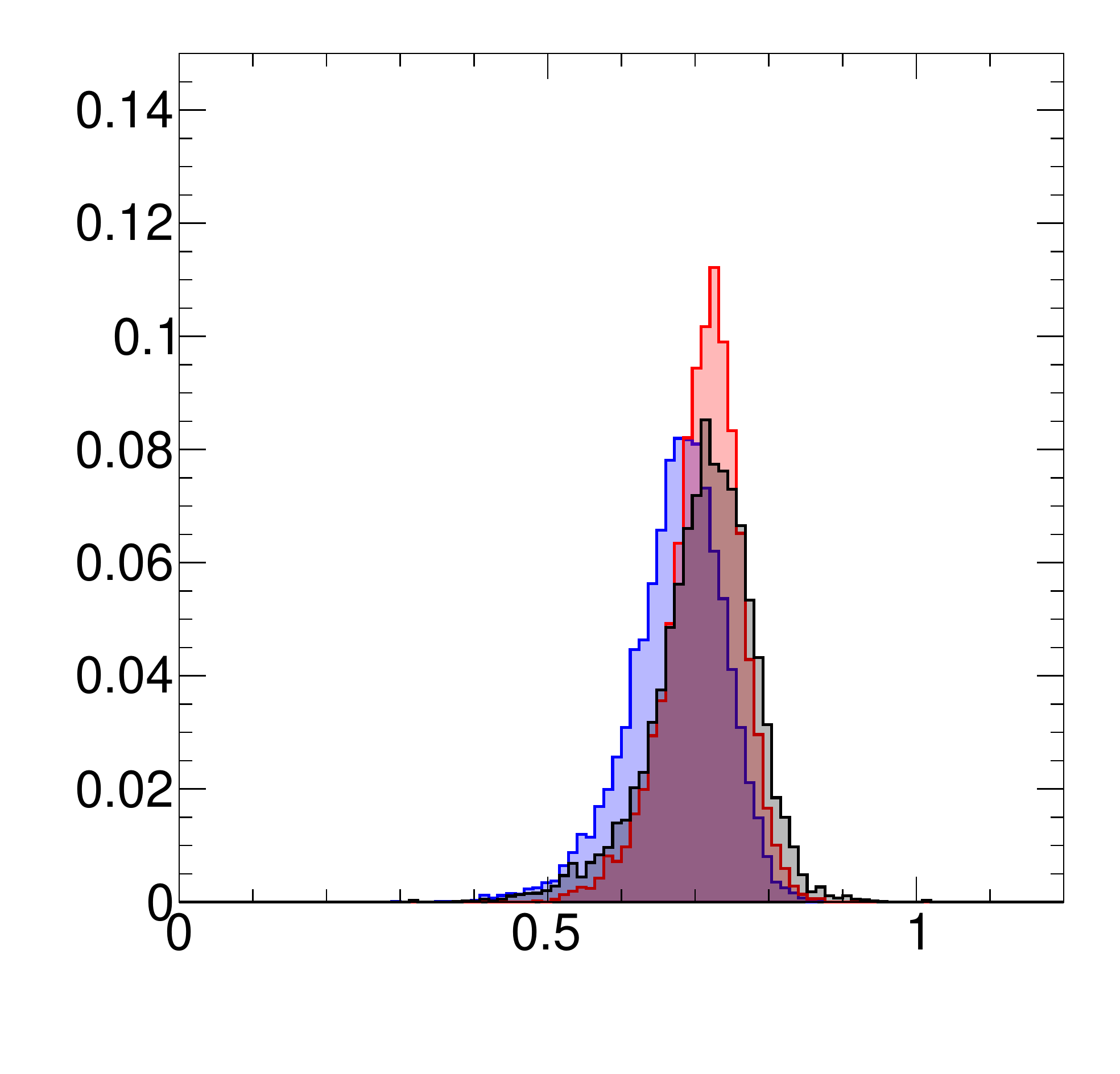}};
  \begin{scope}[x={(image.south east)},y={(image.north west)}]
    % legend
    \draw[blue,fill=white!72!blue,thick] (0.58,0.85) rectangle
    (0.65,0.9);
    \draw[red,fill=white!72!red,thick] (0.58,0.78) rectangle
    (0.65,0.83);
    \draw[black,fill=white!72!black,thick] (0.58,0.71) rectangle
    (0.65,0.76);
    \node[draw=none, anchor=west] at (0.65, 0.875) {\bfseries \sffamily
      \normalsize $\text{QCD}$};
    \node[draw=none, anchor=west] at (0.65, 0.805) {\bfseries \sffamily
      \normalsize $\text{Z'} \rightarrow \text{t}\bar{\text{t}}$};
    \node[draw=none, anchor=west] at (0.65, 0.73) {\bfseries \sffamily
      \normalsize $\text{W} \rightarrow \text{qq'}$};

    % labels
    \node[draw=none] at (0.56,0.065) {\bfseries \small Average
      Learned $\sigma$};
    \node[draw=none, rotate=90] at (0.02, 0.55){\bfseries \small Arbitrary
    Units};
    \node[draw=none, anchor=west] at (0.17,0.88) {\bfseries \sffamily
      \Large \emph{Pythia 8}};
    \node[draw=none, anchor=west] at (0.17,0.80) {\bfseries \sffamily
      $\sqrt{s} = 8 \text{ TeV}$};
    \node[draw=none, anchor=west] at (0.17, 0.74) {\sffamily \tiny $350 \leq
      p_T^{\text{Jet}} \leq 450 \text{ GeV}$};
  \end{scope}
\end{tikzpicture} &

\begin{tikzpicture}
  \node[anchor=south west,inner sep=0] (image) at (0,0)
  {\includegraphics[width=0.43\textwidth]{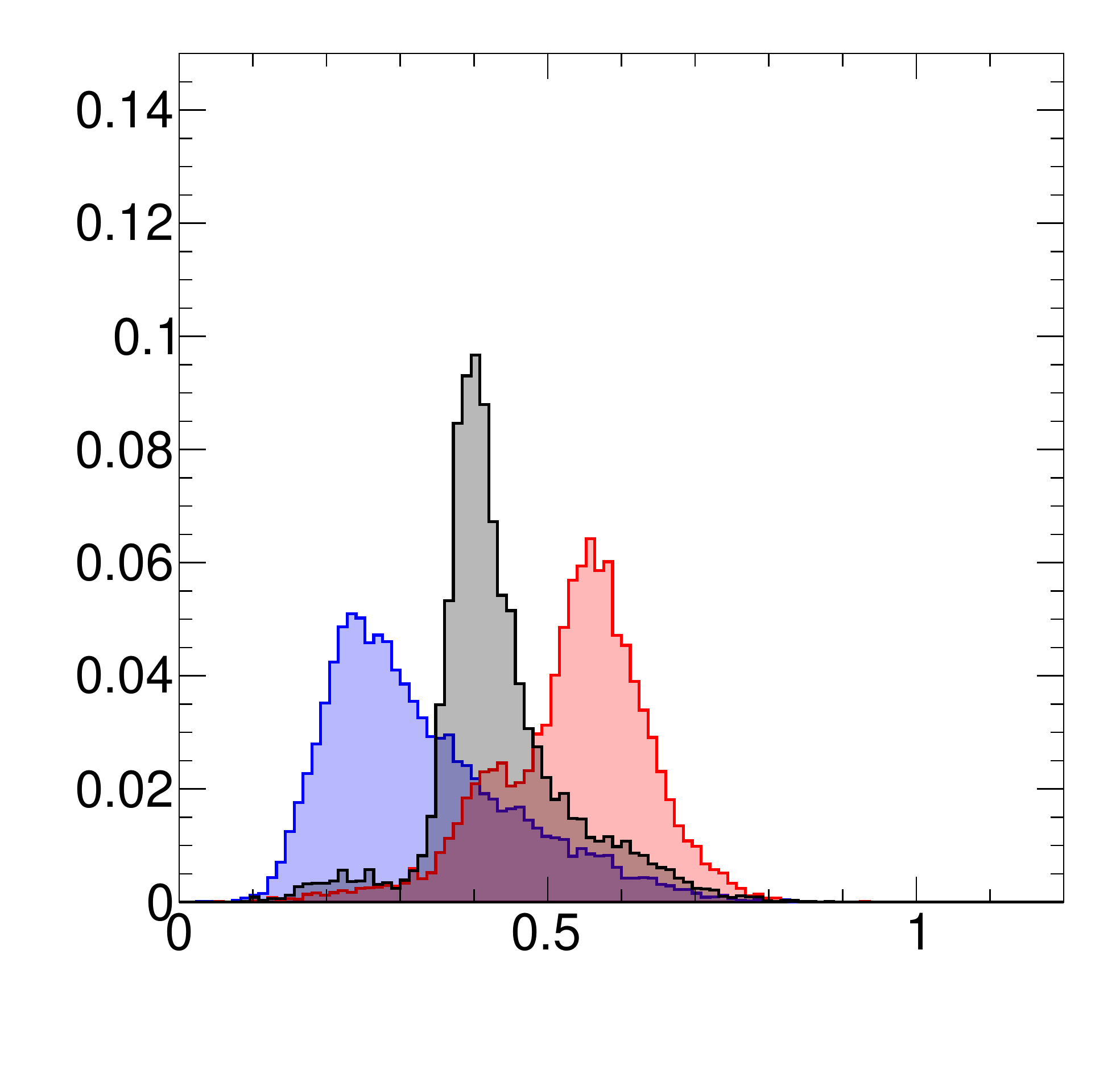}};
  \begin{scope}[x={(image.south east)},y={(image.north west)}]
    % legend
    \draw[blue,fill=white!72!blue,thick] (0.58,0.85) rectangle
    (0.65,0.9);
    \draw[red,fill=white!72!red,thick] (0.58,0.78) rectangle
    (0.65,0.83);
    \draw[black,fill=white!72!black,thick] (0.58,0.71) rectangle
    (0.65,0.76);
    \node[draw=none, anchor=west] at (0.65, 0.875) {\bfseries \sffamily
      \normalsize $\text{QCD}$};
    \node[draw=none, anchor=west] at (0.65, 0.805) {\bfseries \sffamily
      \normalsize $\text{Z'} \rightarrow \text{t}\bar{\text{t}}$};
    \node[draw=none, anchor=west] at (0.65, 0.73) {\bfseries \sffamily
      \normalsize $\text{W} \rightarrow \text{qq'}$};

    % labels
    \node[draw=none] at (0.56,0.065) {\bfseries \small Leading
      Learned $\sigma$};
    \node[draw=none, rotate=90] at (0.02, 0.55){\bfseries \small Arbitrary
    Units};
    \node[draw=none, anchor=west] at (0.17,0.88) {\bfseries \sffamily
      \Large \emph{Pythia 8}};
    \node[draw=none, anchor=west] at (0.17,0.80) {\bfseries \sffamily
      $\sqrt{s} = 8 \text{ TeV}$};
    \node[draw=none, anchor=west] at (0.17, 0.74) {\sffamily \tiny $350 \leq
      p_T^{\text{Jet}} \leq 450 \text{ GeV}$};
  \end{scope}
\end{tikzpicture}
 \\
\end{tabular}
\caption{The learned value of $\sigma$ for the highest $p_T$ jet under the HML scheme (left) and for all jets (right) for various physics processes.}
\label{fig:sigma_hist}
\end{center}
\end{figure}
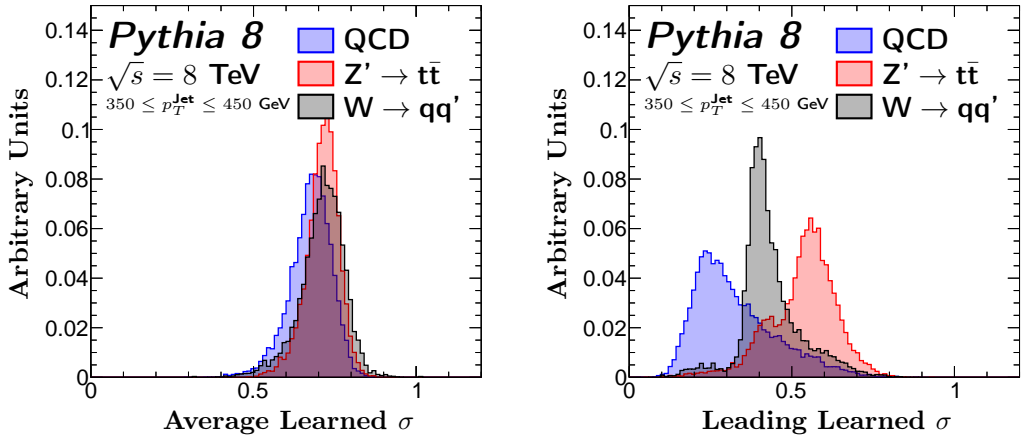

\begin{comment}
On the right hand side of Figure~\ref{fig:sigma_hist}, we note that
the average jet in an event is more or less uniform in size across the
different physics processes, suggesting that the effect we are seeing
in the case of the leading jet $\sigma$ does not appear as a result of
the number of jets changing substantially between the
samples. Instead, there is an actual difference in the extent of the
hard core of the different types of jets. Physically, this should not
surprise us because of the different decay structures expressed in the
samples: we should expect on average that at moderate jet $p_T$ the
full hard core of jets from a $Z'\rightarrow t\bar{t}$ decay is wider
because this decay has a 3-pronged structure.
\end{comment}

\clearpage

\subsection{Fuzzy Jets for Tagging}
\label{sec:tagging}

In this section, $\sigma$ is compared with another class of jet substructure variables known to be useful for tagging: the
$N$-subjettiness ratios~\cite{nsub}. $N$-subjettiness moments are defined over a
set of $N$ axes\footnote{We use the ``one-pass'' $k_{t}$ axes
  optimization technique, which uses an exclusive $k_t$ algorithm to
  find $N$ axes and then refines them by minimizing the $N$-subjettiness
  value.}, and calculated as:
\begin{equation}
\tau_N = \frac{1}{d_0} \sum_k p_{T,k} \min \{ \Delta R_{1,k}, \Delta R_{2,k}, \ldots \Delta R_{N,k} \},
\end{equation}
where $d_0$ is the normalization
\begin{equation}
d_0 = \sum_k p_{T,k} R_0,
\end{equation}
and $R_0$ is the radius of the jet. In practice, the useful variables for determining how much more $i$-pronged a jet is compared to $j$-pronged are the $N$-subjettiness ratios:
\begin{equation}
\tau_{ij} = \frac{\tau_i}{\tau_j}.
\end{equation}
The variable $\tau_{21}$ is often used for the separation of $W$ from QCD
jets~\cite{wbosonCMS,wbosonATLAS} and is a measure of the
compatibility of a jet with a 2-prong hypothesis compared to a 1-prong
hypothesis.  Low value of $\tau_{21}$ indicates that the jet likely has a
2-prong structure. Similarly, $\tau_{32}$ is useful for top tagging in that it measures whether a
3-prong structure is a better description of a jet relative to a 2-prong structure.

%\cite{boost2011,boost2012}

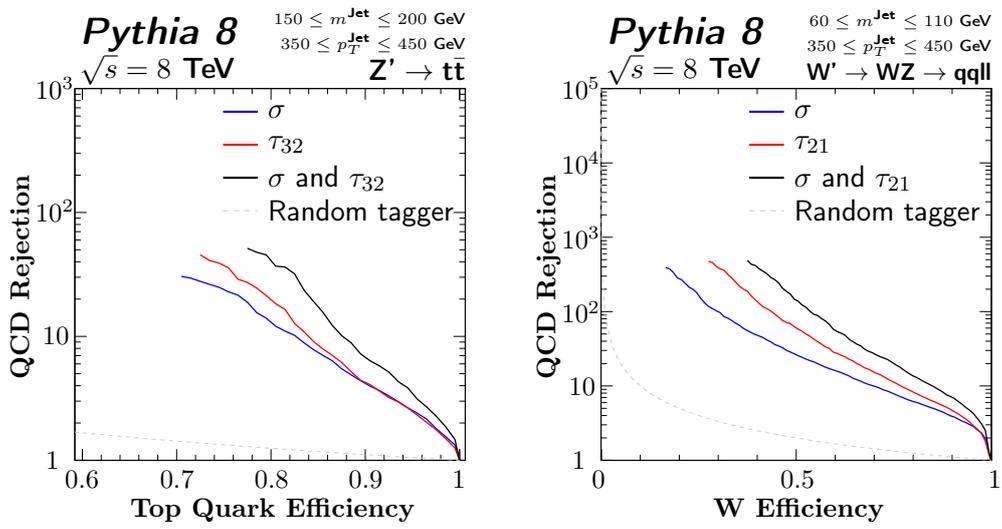
\begin{figure}[h!]
\begin{center}
\begin{tabular}{cc}
\begin{tikzpicture}
  \node[anchor=south west,inner sep=0] (image) at (0,0)
  {\includegraphics[width=0.43\textwidth]{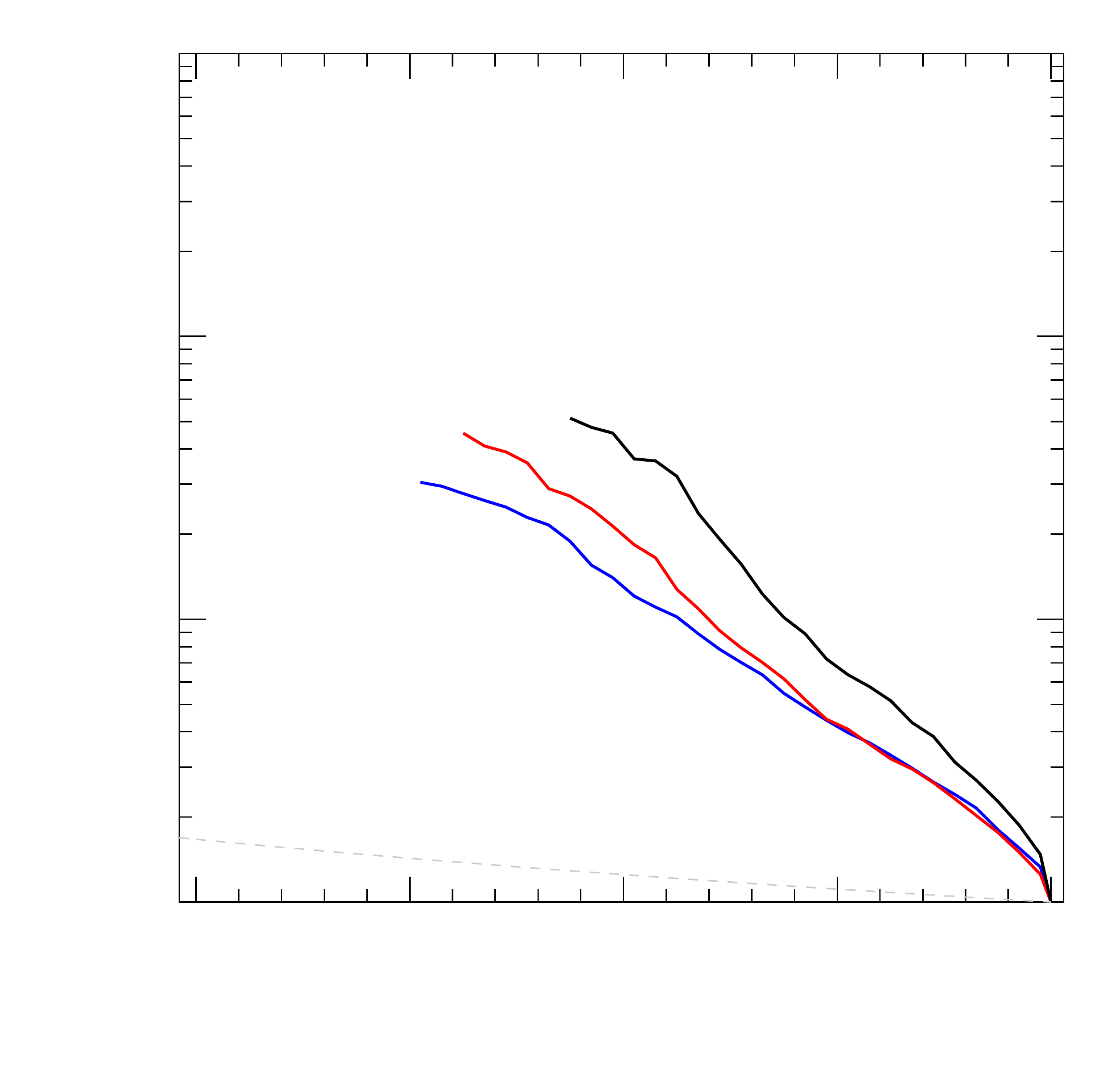}};
  \begin{scope}[x={(image.south east)},y={(image.north west)}]
    % legend
    %\draw[blue,thick] (0.46,0.92) -- (0.53,0.88);
    %\draw[red,thick] (0.46,0.85) -- (0.53,0.81);
    %\draw[black,thick] (0.46,0.78) -- (0.53,0.74);
    %\draw[white!80!black,dash pattern=on 2pt off 2pt] (0.46,0.71) -- (0.53,0.67);

    \draw[blue,thick] (0.46,0.9) -- (0.53,0.9);
    \draw[red,thick] (0.46,0.83) -- (0.53,0.83);
    \draw[black,thick] (0.46,0.76) -- (0.53,0.76);
    \draw[white!80!black,dash pattern=on 2pt off 2pt] (0.46,0.69) -- (0.53,0.69);

    \node[draw=none, anchor=west] at (0.53, 0.9) {\sffamily
      \normalsize $\sigma$};
    \node[draw=none, anchor=west] at (0.53, 0.83) {\sffamily
      \normalsize $\tau_{32}$};
    \node[draw=none, anchor=west] at (0.53, 0.76) {\sffamily
      \normalsize $\sigma \text{ and } \tau_{32}$};
    \node[draw=none, anchor=west] at (0.53, 0.685) {\sffamily
      \normalsize Random tagger};

    % labels
    \node[draw=none] at (0.56,0.055) {\bfseries \small Top Quark Efficiency};
    \node[draw=none, rotate=90] at (0.055, 0.55){\bfseries \small QCD Rejection};
    \node[draw=none, anchor=west] at (0.15,1.065) {\bfseries \sffamily
      \Large \emph{Pythia 8}};
    \node[draw=none, anchor=west] at (0.15,0.99) {\bfseries \sffamily
      $\sqrt{s} = 8 \text{ TeV}$};
    \node[draw=none, anchor=east] at (0.97,0.99) {\bfseries \sffamily
      $\text{Z'} \rightarrow \text{t}\bar{\text{t}}$};

    \node[draw=none, anchor=east] at (0.97, 1.045) {\sffamily \tiny $350 \leq
      p_T^{\text{Jet}} \leq 450 \text{ GeV}$};
    \node[draw=none, anchor=east] at (0.97, 1.1) {\sffamily \tiny $150 \leq
      m^{\text{Jet}} \leq 200 \text{ GeV}$};

    \node[draw=none, anchor=east] at (0.15, 0.16) {\bfseries \small
      \sffamily $1$};
    \node[draw=none, anchor=east] at (0.17, 0.425) {\bfseries \small
      \sffamily $10$};
    \node[draw=none, anchor=east] at (0.18, 0.69) {\bfseries \small
      \sffamily $10^2$};
    \node[draw=none, anchor=east] at (0.18, 0.95) {\bfseries \small
      \sffamily $10^3$};

    \node[draw=none, anchor=east] at (0.23, 0.12) {\bfseries \small
      \sffamily $0.6$};
    \node[draw=none, anchor=east] at (0.422, 0.12) {\bfseries \small
      \sffamily $0.7$};
    \node[draw=none, anchor=east] at (0.613, 0.12) {\bfseries \small
      \sffamily $0.8$};
    \node[draw=none, anchor=east] at (0.803, 0.12) {\bfseries \small
      \sffamily $0.9$};
    \node[draw=none, anchor=east] at (0.97, 0.12) {\bfseries \small
      \sffamily $1$};
  \end{scope}
\end{tikzpicture} &
\begin{tikzpicture}
  \node[anchor=south west,inner sep=0] (image) at (0,0)
  {\includegraphics[width=0.43\textwidth]{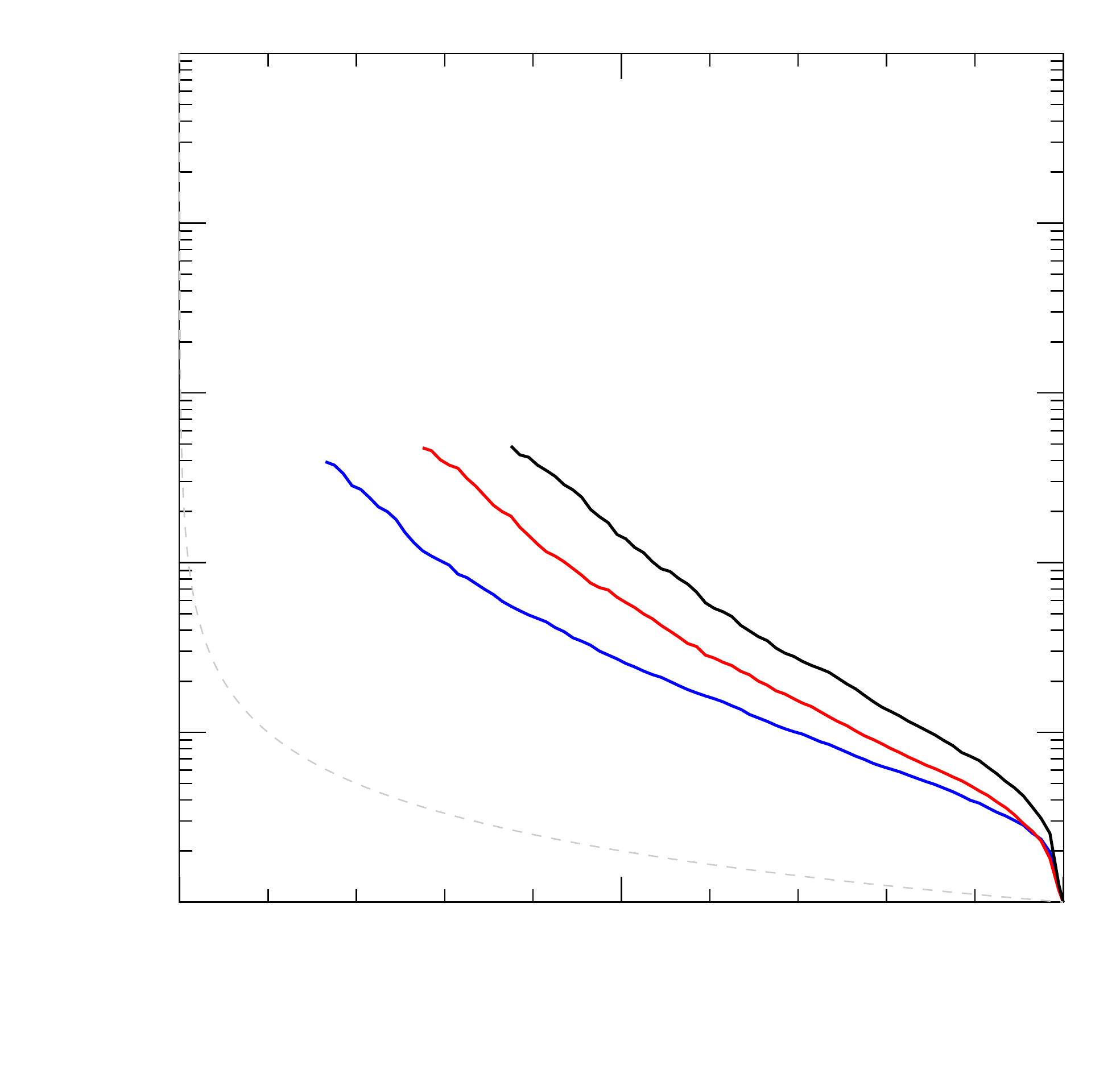}};
  \begin{scope}[x={(image.south east)},y={(image.north west)}]
    % legend
    %\draw[blue,thick] (0.46,0.92) -- (0.53,0.88);
    %\draw[red,thick] (0.46,0.85) -- (0.53,0.81);
    %\draw[black,thick] (0.46,0.78) -- (0.53,0.74);
    %\draw[white!80!black,dash pattern=on 2pt off 2pt] (0.46,0.71) -- (0.53,0.67);

    \draw[blue,thick] (0.46,0.9) -- (0.53,0.9);
    \draw[red,thick] (0.46,0.83) -- (0.53,0.83);
    \draw[black,thick] (0.46,0.76) -- (0.53,0.76);
    \draw[white!80!black,dash pattern=on 2pt off 2pt] (0.46,0.69) -- (0.53,0.69);

    \node[draw=none, anchor=west] at (0.53, 0.9) {\sffamily
      \normalsize $\sigma$};
    \node[draw=none, anchor=west] at (0.53, 0.83) {\sffamily
      \normalsize $\tau_{21}$};
    \node[draw=none, anchor=west] at (0.53, 0.76) {\sffamily
      \normalsize $\sigma \text{ and } \tau_{21}$};
    \node[draw=none, anchor=west] at (0.53, 0.685) {\sffamily
      \normalsize Random tagger};

    % labels
    \node[draw=none] at (0.56,0.055) {\bfseries \small W Efficiency};
    \node[draw=none, rotate=90] at (0.055, 0.55){\bfseries \small QCD Rejection};
    \node[draw=none, anchor=west] at (0.15,1.065) {\bfseries \sffamily
      \Large \emph{Pythia 8}};
    \node[draw=none, anchor=west] at (0.15,0.99) {\bfseries \sffamily
      $\sqrt{s} = 8 \text{ TeV}$};
    \node[draw=none, anchor=east] at (0.97,0.98) {\bfseries \footnotesize \sffamily
      $\text{W'} \rightarrow \text{WZ} \rightarrow \text{qqll}$};

    \node[draw=none, anchor=east] at (0.97, 1.04) {\sffamily \tiny $350 \leq
      p_T^{\text{Jet}} \leq 450 \text{ GeV}$};
    \node[draw=none, anchor=east] at (0.97, 1.095) {\sffamily \tiny $60 \leq
      m^{\text{Jet}} \leq 110 \text{ GeV}$};

    \node[draw=none, anchor=east] at (0.15, 0.16) {\bfseries \small
      \sffamily $1$};
    \node[draw=none, anchor=east] at (0.17, 0.318) {\bfseries \small
      \sffamily $10$};
    \node[draw=none, anchor=east] at (0.18, 0.476) {\bfseries \small
      \sffamily $10^2$};
    \node[draw=none, anchor=east] at (0.18, 0.634) {\bfseries \small
      \sffamily $10^3$};
    \node[draw=none, anchor=east] at (0.18, 0.792) {\bfseries \small
      \sffamily $10^4$};
    \node[draw=none, anchor=east] at (0.18, 0.95) {\bfseries \small
      \sffamily $10^5$};

    \node[draw=none, anchor=east] at (0.19, 0.12) {\bfseries \small
      \sffamily $0$};
    \node[draw=none, anchor=east] at (0.613, 0.12) {\bfseries \small
      \sffamily $0.5$};
    \node[draw=none, anchor=east] at (0.99, 0.12) {\bfseries \small
      \sffamily $1$};
  \end{scope}
\end{tikzpicture} \\
\end{tabular}
\caption{The tagging performance of $\sigma$ relative to $\tau_{32}$ ($\tau_{21}$) for distinguishing top quarks ($W$ bosons) from a QCD background is shown on the left (right).  The {\it random tagger} keeps a fixed fraction of all events, regardless of their origin and is a lower bound on the performance of any tagger.}
\label{fig:tagging_tmva}
\end{center}
\end{figure}

The rest of his section contains comparisons of the performance of $\sigma$ relative to $\tau_{21}$
for separating $W$ from QCD jets, as well as $\sigma$ relative to
$\tau_{32}$ for tagging $Z'\rightarrow t\bar{t}$ amongst a QCD jet
background. In Figure~\ref{fig:tagging_tmva}, a $k$-nearest
neighbors classifier was trained with 2-fold cross validation in TMVA~\cite{Hocker:2007ht}. The left plot in Figure~\ref{fig:tagging_tmva} demonstrates an increase in performance for discriminating
$Z' \rightarrow t\bar{t}$ from QCD relative to using $\tau_{32}$
alone. The fuzzy jet $\sigma$ is roughly equally useful to the $N$-subjettiness
ratio at a sigma efficiency of $0.85$, and using both variables
greatly improves background rejection. Similar results can be seen in
the right plot of Figure~\ref{fig:tagging_tmva}, where $\sigma$ boosts background
rejection relative to $\tau_{21}$ alone. In each case, the training
and classification was performed in a mass window around the
particles of interest, the top quark mass in the $Z'\rightarrow
t\bar{t}$ sample and the $W$ boson mass to discriminate $W\rightarrow
qq'$ from QCD.

%In each case, $\sigma$ offers
%  additional performance, but does not outperform $N$-subjettiness,
 % except at a signal efficiency of roughly 0.85 in the $Z'$ sample,
 % where equal performance to $\tau_{32}$ is observed.

The comparisons of the fuzzy jet $\sigma$ and $N$-subjettiness are intended to be an illustrative example.   As discussed in the opening of this section, $\sigma$ is just \emph{one} variable that can be
constructed by using mGMM clustered jets.  Expanded studies of the various learned parameters could come up with additional
variables, or the full learned parameter set could be thrown into an
off the shelf classifier or machine learning model.

\clearpage
\newpage

\section{Underlying Event and Pileup}
\label{sec:pileup}

As with any new jet algorithm or jet variable, understanding the
effect of pileup vertices from additional proton-proton collisions is
essential to make meaningful statements about how the method will
be applicable to real data analyses at the LHC.   Studying pileup in the context of mGMM jets is complicated by the effective catchment area of the jets. For
hierarchical-agglomerative algorithms like anti-$k_t$, the catchment
area scales with the radius parameter.  However fuzzy
jets can have infinite catchment area because the likelihood for
particle membership is nonzero for any finite distance and arrangement
of Gaussian jets and particles. Furthermore, the catchment area can
change depending on the other jets in an event.  Although this effect also occurs in
the hierarchical-agglomerative case, the effect is much more
pronounced in the mGMM clustering algorithm, with some jets having
finite catchment areas while others cluster infinite area.

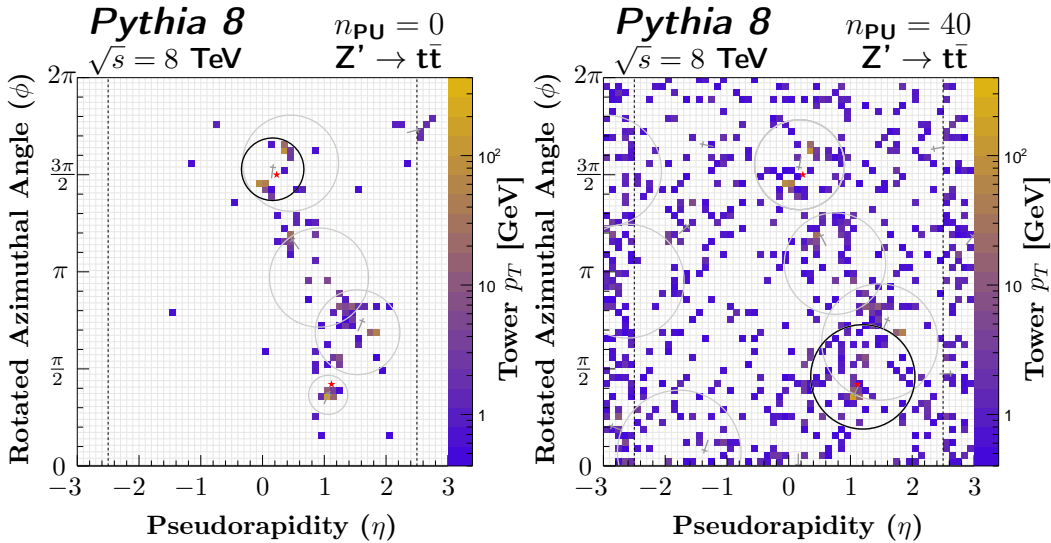
\begin{figure}[h!]
\vspace{1cm}
\begin{center}
\begin{tabular}{cc}
\begin{overpic}[width=0.43\textwidth]{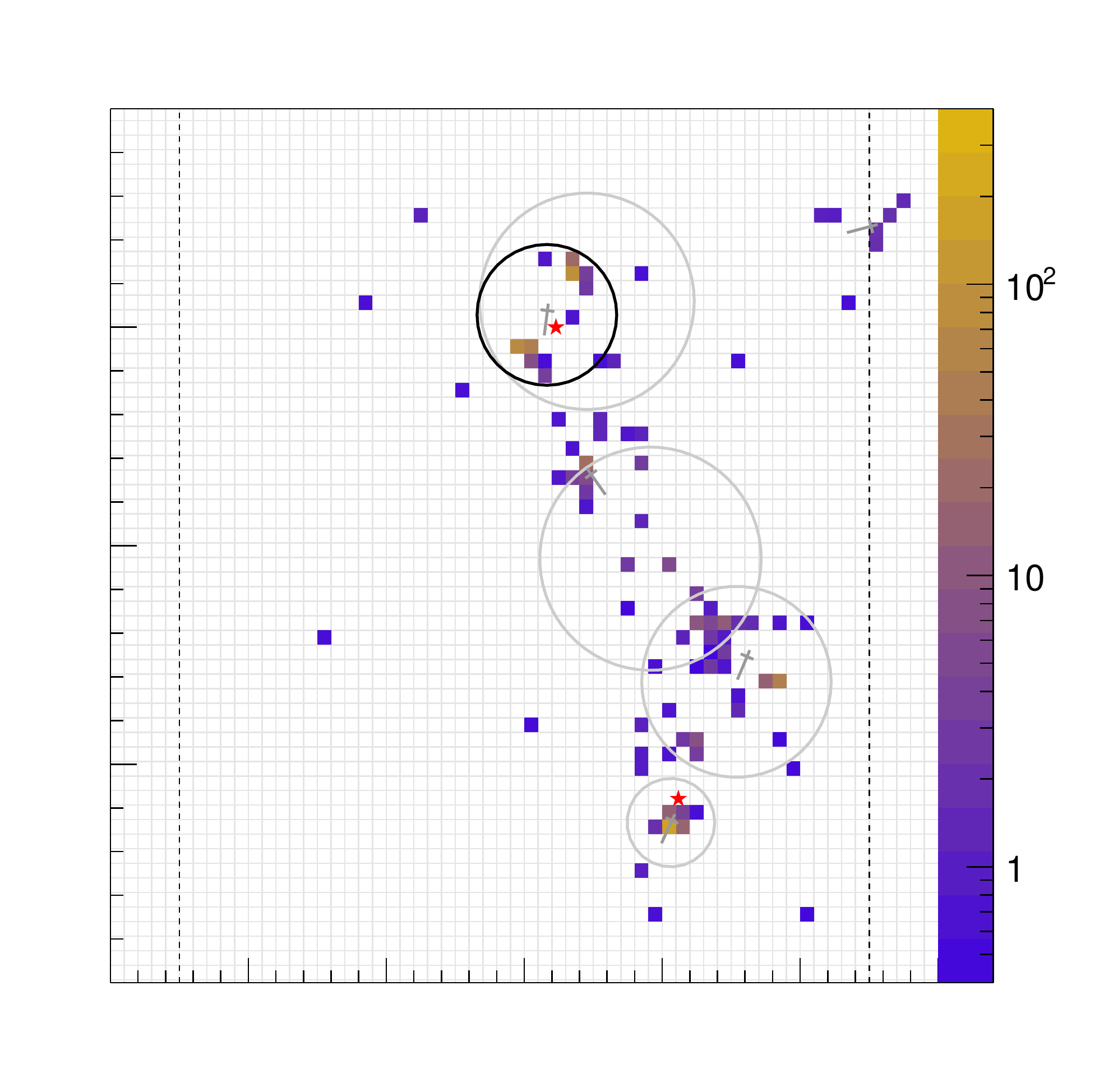}
% general labels
\put(12, 98){\bfseries \sffamily \Large \emph{Pythia 8}}
\put(12, 91){\bfseries \sffamily $\sqrt{s} = 8 \text{ TeV}$}
\put(62, 91){\bfseries \sffamily \large $\text{Z'}
    \rightarrow \text{t}\bar{\text{t}}$}
\put(62, 98){\bfseries \sffamily \large $n_{\text{PU}} = 0$}

% axes labels
\put(24, -4){\bfseries \small Pseudorapidity ($\eta$)}
\put(-4, 10){\rotatebox{90}{\bfseries \small Rotated Azimuthal Angle
    ($\phi$)}}
\put(95, 27){\rotatebox{90}{\bfseries \small Tower $p_T \text{ [GeV]}$}}

% axes tick labels
\put(5, 8){\bfseries \small \sffamily $0$}
\put(4, 28.3){\bfseries \small \sffamily $\frac{\pi}{2}$}
\put(5, 48.3){\bfseries \small \sffamily $\pi$}
\put(4, 68){\bfseries \small \sffamily $\frac{3\pi}{2}$}
\put(4, 88){\bfseries \small \sffamily $2\pi$}

\put(4,  4){\bfseries \small \sffamily $-3$}
\put(17, 4){\bfseries \small \sffamily $-2$}
\put(29.6, 4){\bfseries \small \sffamily $-1$}
\put(46.2, 4){\bfseries \small \sffamily $0$}
\put(58.8, 4){\bfseries \small \sffamily $1$}
\put(71.3, 4){\bfseries \small \sffamily $2$}
\put(83.8, 4){\bfseries \small \sffamily $3$}
\end{overpic} &
\begin{overpic}[width=0.43\textwidth]{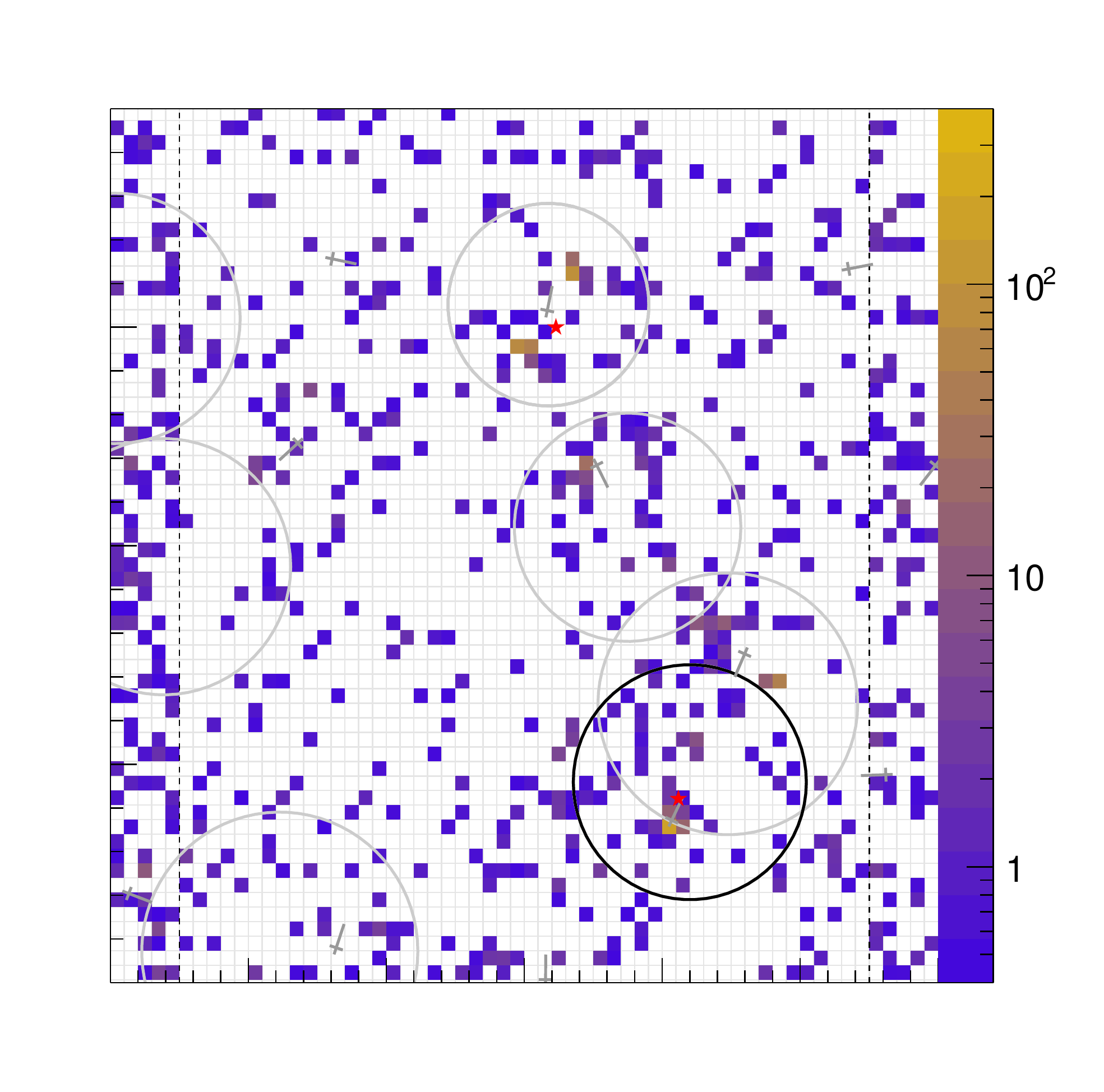}
% general labels
\put(12, 98){\bfseries \sffamily \Large \emph{Pythia 8}}
\put(12, 91){\bfseries \sffamily $\sqrt{s} = 8 \text{ TeV}$}
\put(62, 91){\bfseries \sffamily \large $\text{Z'}
    \rightarrow \text{t}\bar{\text{t}}$}
\put(59, 98){\bfseries \sffamily \large $n_{\text{PU}} = 40$}

% axes labels
\put(24, -4){\bfseries \small Pseudorapidity ($\eta$)}
\put(-4, 10){\rotatebox{90}{\bfseries \small Rotated Azimuthal Angle
    ($\phi$)}}
\put(95, 27){\rotatebox{90}{\bfseries \small Tower $p_T \text{ [GeV]}$}}

% axes tick labels
\put(5, 8){\bfseries \small \sffamily $0$}
\put(4, 28.3){\bfseries \small \sffamily $\frac{\pi}{2}$}
\put(5, 48.3){\bfseries \small \sffamily $\pi$}
\put(4, 68){\bfseries \small \sffamily $\frac{3\pi}{2}$}
\put(4, 88){\bfseries \small \sffamily $2\pi$}

\put(4,  4){\bfseries \small \sffamily $-3$}
\put(17, 4){\bfseries \small \sffamily $-2$}
\put(29.6, 4){\bfseries \small \sffamily $-1$}
\put(46.2, 4){\bfseries \small \sffamily $0$}
\put(58.8, 4){\bfseries \small \sffamily $1$}
\put(71.3, 4){\bfseries \small \sffamily $2$}
\put(83.8, 4){\bfseries \small \sffamily $3$}
\end{overpic} \\
\end{tabular}
\end{center}
\caption{mGMM jets defined according to Section~\ref{sec:stats} with
  an isotropic kernel are broadened as a result
  of the introduction of additional $pp$ pileup vertices. The same
  hard scatter is clustered twice, on the left with $n_{\text{PU}} =
  0$ and on the right with $n_{\text{PU}} = 40$. Vertical dashed lines
  at $\eta = \pm 2.5$ show the extent of a simulated tracker with the
  same $\eta$ extent as that used at ATLAS and CMS. Charged pileup
  falling within the extent of the simulated tracker is discarded
  before clustering and the aggregation of particles into towers.}
\label{fig:pileup_ed}
\end{figure}

The challenge of pileup for fuzzy jets is illustrated in
Figure~\ref{fig:pileup_ed}, where the same event is shown
with $n_{\text{PU}} = 0$, and with $n_{\text{PU}} = 40$. The event
displays show the central region of the detector, where most of the decay
products of the hard scatter lie. Qualitatively, it can be seen that
the introduction of additional interaction vertices broadens all of
the mGMM jets. This
broadening clearly impacts the power of $\sigma$ for differentiating QCD
background from signal processes.

The next sections explore two methods for mitigating the impact of pileup in relation to fuzzy jets, illustrated with the variable $\sigma$.  %Section~\ref{sec:alpha_tuning} describes how changing fuzzy jets parameter $\alpha$
%can be used to deal with pileup, at the cost of sacrificing IRC
%safety. Finally, in section~\ref{sec:event_jet}, we propose a small change to
%the algorithm by introducing a background jet in the event and study a
%choice of mGMM parameters and corrections to the tower $p_T$s before
%clustering which is effective in dealing with pileup consistent with LHC
%Run I conditions.

\subsection{Changing $\alpha$ for Pileup Suppression}
\label{sec:alpha_tuning}
\begin{figure}[h!]
\vspace{1cm}
\begin{center}
\begin{tabular}{cc}
\begin{overpic}[width=0.43\textwidth]{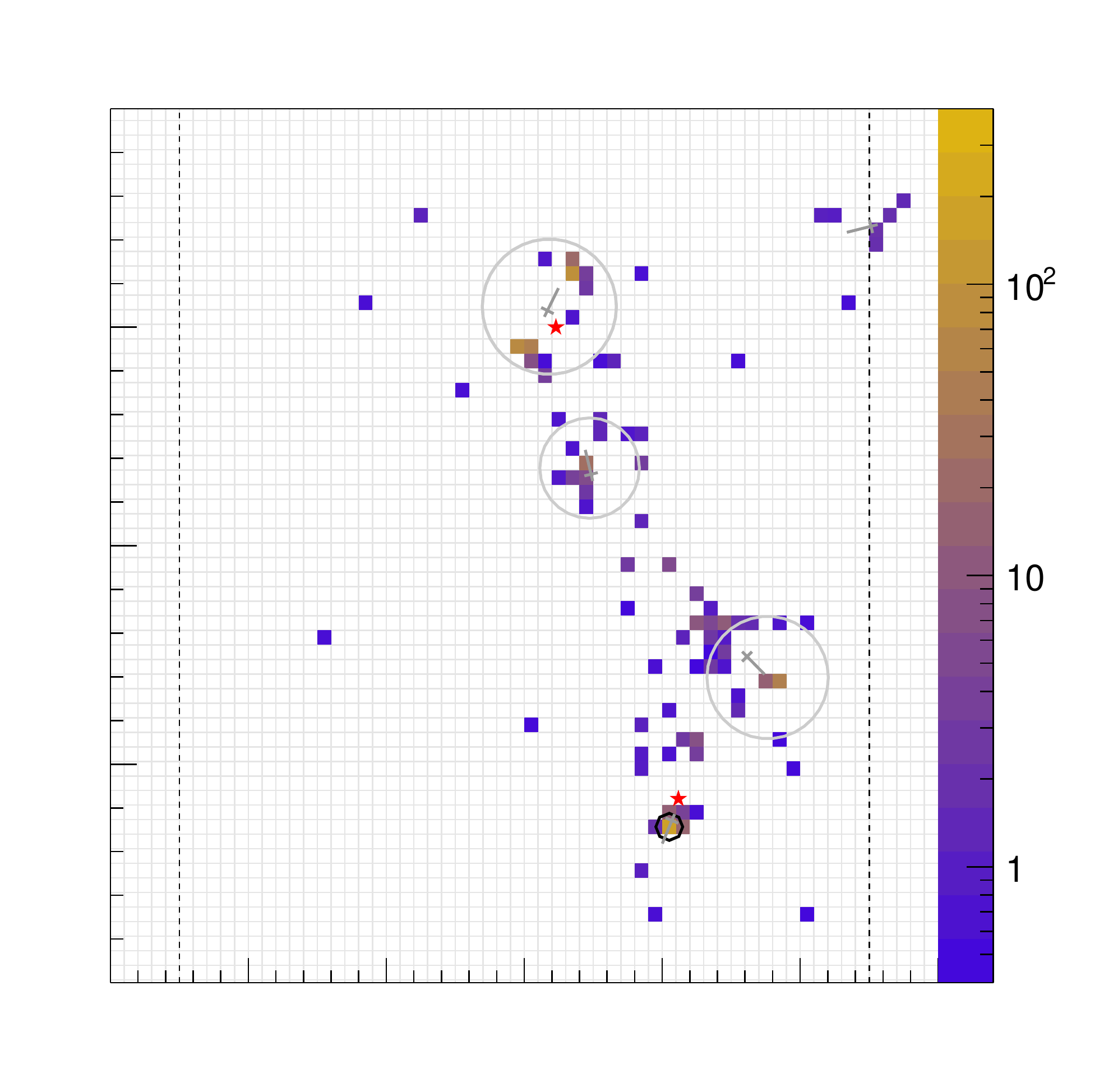}
% general labels
\put(12, 98){\bfseries \sffamily \Large \emph{Pythia 8}}
\put(12, 91){\bfseries \sffamily $\sqrt{s} = 8 \text{ TeV}$}
\put(62, 91){\bfseries \sffamily \large $\text{Z'}
    \rightarrow \text{t}\bar{\text{t}}$}
\put(62, 98){\bfseries \sffamily \large $n_{\text{PU}} = 0$}
\put(68.8, 105){\bfseries \sffamily \large $\alpha = 2$}

% axes labels
\put(24, -4){\bfseries \small Pseudorapidity ($\eta$)}
\put(-4, 10){\rotatebox{90}{\bfseries \small Rotated Azimuthal Angle
    ($\phi$)}}
\put(95, 27){\rotatebox{90}{\bfseries \small Tower $p_T \text{ [GeV]}$}}

% axes tick labels
\put(5, 8){\bfseries \small \sffamily $0$}
\put(4, 28.3){\bfseries \small \sffamily $\frac{\pi}{2}$}
\put(5, 48.3){\bfseries \small \sffamily $\pi$}
\put(4, 68){\bfseries \small \sffamily $\frac{3\pi}{2}$}
\put(4, 88){\bfseries \small \sffamily $2\pi$}

\put(4,  4){\bfseries \small \sffamily $-3$}
\put(17, 4){\bfseries \small \sffamily $-2$}
\put(29.6, 4){\bfseries \small \sffamily $-1$}
\put(46.2, 4){\bfseries \small \sffamily $0$}
\put(58.8, 4){\bfseries \small \sffamily $1$}
\put(71.3, 4){\bfseries \small \sffamily $2$}
\put(83.8, 4){\bfseries \small \sffamily $3$}
\end{overpic} &
\begin{overpic}[width=0.43\textwidth]{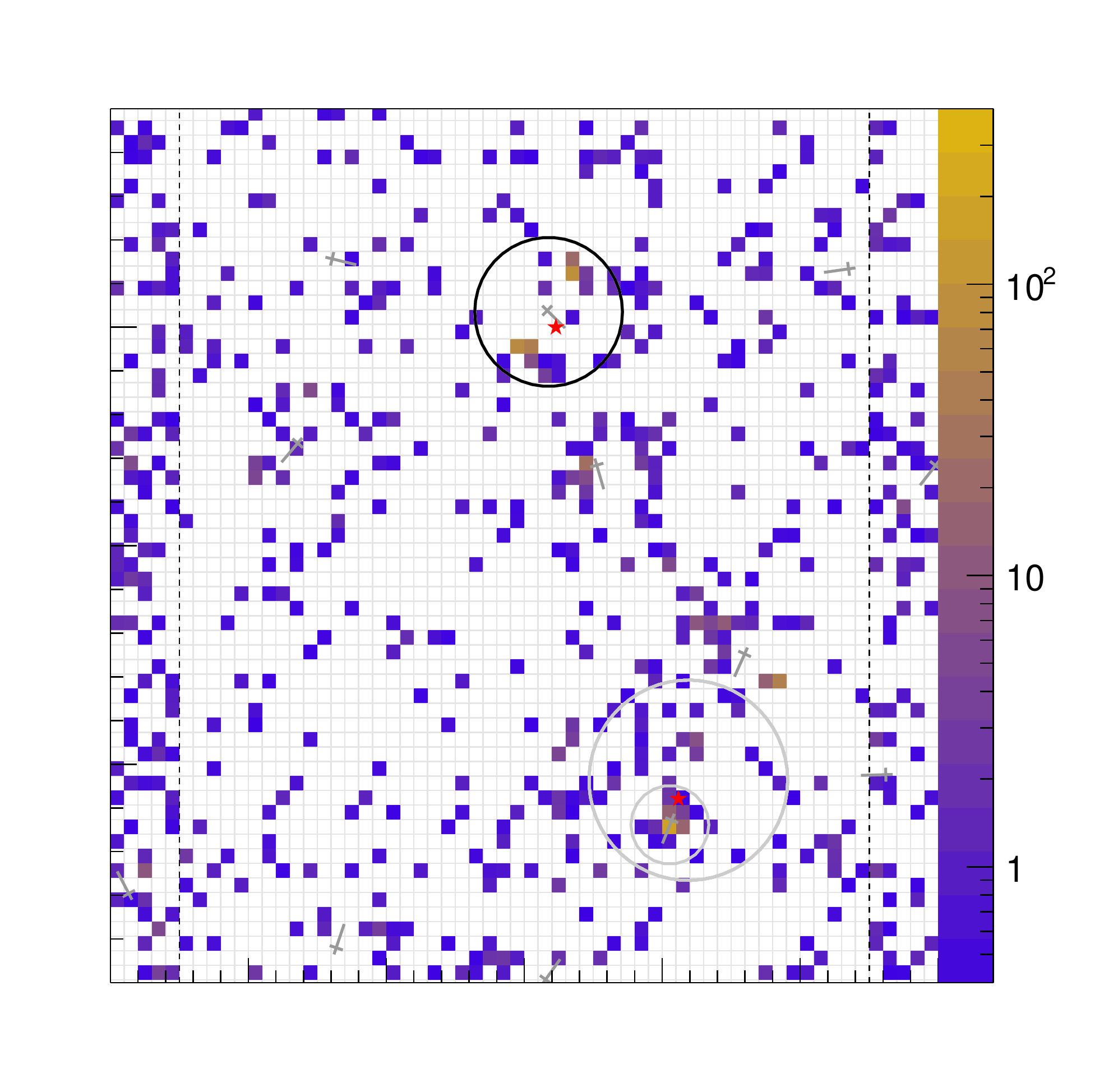}
% general labels
\put(12, 98){\bfseries \sffamily \Large \emph{Pythia 8}}
\put(12, 91){\bfseries \sffamily $\sqrt{s} = 8 \text{ TeV}$}
\put(62, 91){\bfseries \sffamily \large $\text{Z'}
    \rightarrow \text{t}\bar{\text{t}}$}
\put(59, 98){\bfseries \sffamily \large $n_{\text{PU}} = 40$}
\put(68.8, 105){\bfseries \sffamily \large $\alpha = 2$}

% axes labels
\put(24, -4){\bfseries \small Pseudorapidity ($\eta$)}
\put(-4, 10){\rotatebox{90}{\bfseries \small Rotated Azimuthal Angle
    ($\phi$)}}
\put(95, 27){\rotatebox{90}{\bfseries \small Tower $p_T \text{ [GeV]}$}}

% axes tick labels
\put(5, 8){\bfseries \small \sffamily $0$}
\put(4, 28.3){\bfseries \small \sffamily $\frac{\pi}{2}$}
\put(5, 48.3){\bfseries \small \sffamily $\pi$}
\put(4, 68){\bfseries \small \sffamily $\frac{3\pi}{2}$}
\put(4, 88){\bfseries \small \sffamily $2\pi$}

\put(4,  4){\bfseries \small \sffamily $-3$}
\put(17, 4){\bfseries \small \sffamily $-2$}
\put(29.6, 4){\bfseries \small \sffamily $-1$}
\put(46.2, 4){\bfseries \small \sffamily $0$}
\put(58.8, 4){\bfseries \small \sffamily $1$}
\put(71.3, 4){\bfseries \small \sffamily $2$}
\put(83.8, 4){\bfseries \small \sffamily $3$}
\end{overpic} \\
\end{tabular}
\end{center}
\caption{Clustering in the mGMM model with $\alpha = 2$. There is
  little broadening between the $n_{\text{PU}} = 0$ (left) and
  $n_{\text{PU}} = 40$ (right) cases, but jets at the locations of the
  tops in the event are substantially narrower than in the case where
  $\alpha = 1$, even with $n_{\text{PU}} = 0$ (compare to
  Figure~\ref{fig:pileup_ed}). Under the ML particle assignment, the
  $\alpha = 2$ algorithm identifies the other top as the highest $p_T$
  jet in the event, demonstrating the difficulty in dealing with fuzzy
  jet kinematics.}
\label{fig:pileup_alpha2_ed}
\end{figure}
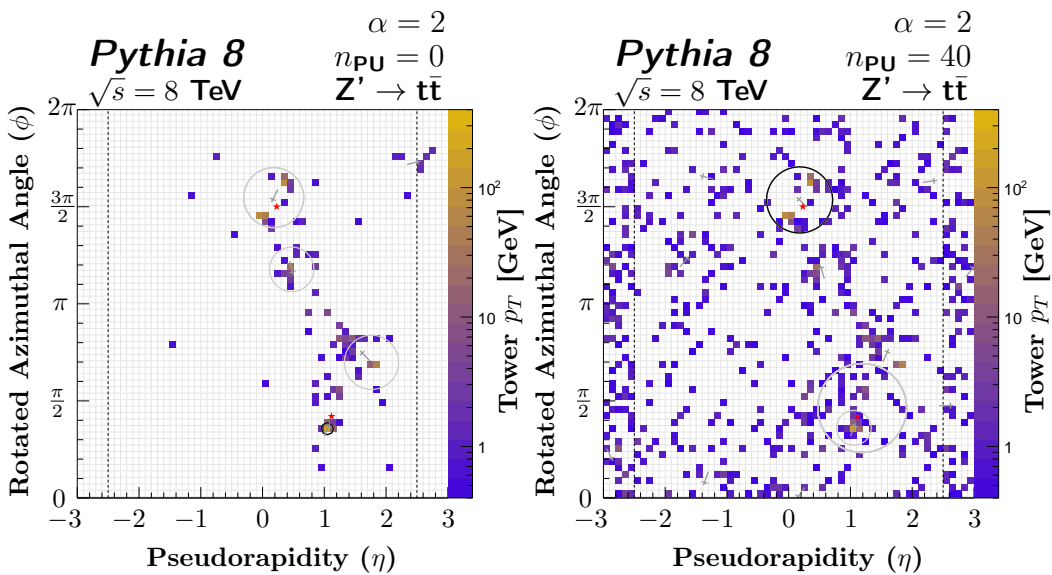

In Section~\ref{sec:stats}, it was discussed that choosing $\alpha = 1$ in
the likelihood (Eq.~(\ref{eq:mm2})) guarantees IRC safety.  With $\alpha = 1$, the mGMM algorithm treats hard structure and
soft structure linearly in
the particle or tower $p_T$. However, one can exploit the fact that
$\sigma$ is disproportionately a measure of the shape and extent of
the leading jet hard structure to make the variable more resilient to
the effects of pileup. In particular, choosing $\alpha > 1$ stabilizes
$\sigma$ at high $n_{\text{PU}}$ because so long as the average input
particle $p_T$ due to pileup is significantly smaller than the $p_T$
of the particles constituting the leading jet hard structure, the
change in likelihood will be suppressed roughly according to $\left(
  p_{T,\text{hs}}/p_{T,\text{PU}} \right)^\alpha$. An example of this
effect is illustrated in Figure~\ref{fig:pileup_alpha2_ed}, which shows
the same event as in Figure~\ref{fig:pileup_ed}.  The price for adjusting $\alpha$ is the loss of collinear safety.  Varying $\alpha$ is not explored further, as Section~\ref{sec:event_jet} demonstrates a method for dealing with pileup
effectively that does not rely on moving $\alpha$ away from the IRC safe value of one.

\subsection{Tower Subtraction and the Event Jet: Effective Pileup Correction}
\label{sec:event_jet}

Recent developments in pileup mitigation have led to several algorithms for correcting jet inputs before jet clustering beings.  Such techniques include Pileup Per Particle Identification (PUPPI), Constituent
Subtraction, and SoftKiller~\cite{puppi,constsub,softkill}.  One simple input-correction scheme is to subtract from each calorimeter tower the estimated pileup $p_T$ density per unit area multiplied by the size of the tower in the detector.  As a first step, $\rho$ is calculated in the same way as described in Sec.~\ref{sec:details}.  Tower momenta
are then corrected according to Eq.~(\ref{eq:tower_sub}), where
$p_{T,\text{s}}$ is the corrected momentum, $p_{T,\text{o}}$ is the
original momentum, and $A$ is the area of the tower. In this study,
all towers have area $0.1\times0.1$ in $y$-$\phi$ space.

\begin{align}
\label{eq:tower_sub}
p_{T,\text{s}} = \text{max}\left( p_{T,\text{o}} - \rho
  A, 0 \right).
\end{align}

\noindent While subtracting the average $p_T$ background from towers before
clustering is a relatively safe way of reducing the effect of pileup,
at least when the $p_T$ scales of the tower to tower fluctuations are
small compared to the hard scatter $p_T$ scale, it would still be
helpful to systematically address the question of catchment areas. The
mGMM clustering algorithm provides a natural framework in which to
think about pileup, however, because the algorithm deals fundamentally
with likelihoods, and the pileup likelihood is to leading order uniform over the detector (this is the motivation for the area-subtraction technique).  This is the motivation for modifying the mGMM likelihood using a technique we call the {\it event jet}.

In addition to learning $k$ mGMM jets throughout clustering, the event
jet includes another background contribution to the likelihood which
attempts to capture the intuition of a uniform contribution of
particle likelihood due to pileup. Constraints are further imposed on the likelihood
on the event level jet so that it has constant likelihood during the
clustering process, making the necessary modifications to the
algorithm procedures simpler.

Practically, the effect of the event jet can be parameterized through
the introduction of an algorithmic parameter $\gamma$.  Particle membership probabilities change according to
Eq.~(\ref{eq:event_jet_probability}) with corresponding changes to the
analytical M step for the Gaussian kernel type.  The choice of $\gamma$ is important, and it should reflect the fact
that not all events are created equal in the sense that not all events
have the same contributions due to pileup. Although there is no strict
way of dealing with this issue, it is reasonable to replace $\gamma$
by a meaningful combination of parameters which is sensitive to our
estimates of the amount of pileup in a particular event. We have
chosen to take $\gamma = \rho A \gamma_w$ where $\rho$ is our estimate
of the $p_T$ density due to pileup, $A$ is the calorimeter area, and
$\gamma_w$ is a parameter of the algorithm controlling the
strength of the event jet.  Initial studies with the event jet indicate that introducing a $\rho$ dependent $\gamma$ is much more effective than a $\rho$ independent one.

\begin{align}
\label{eq:event_jet_probability}
q_{ij} \rightarrow \frac{q_{ij}}{\gamma + \sum_k p_{ik}}
\end{align}

\begin{figure}[h!]
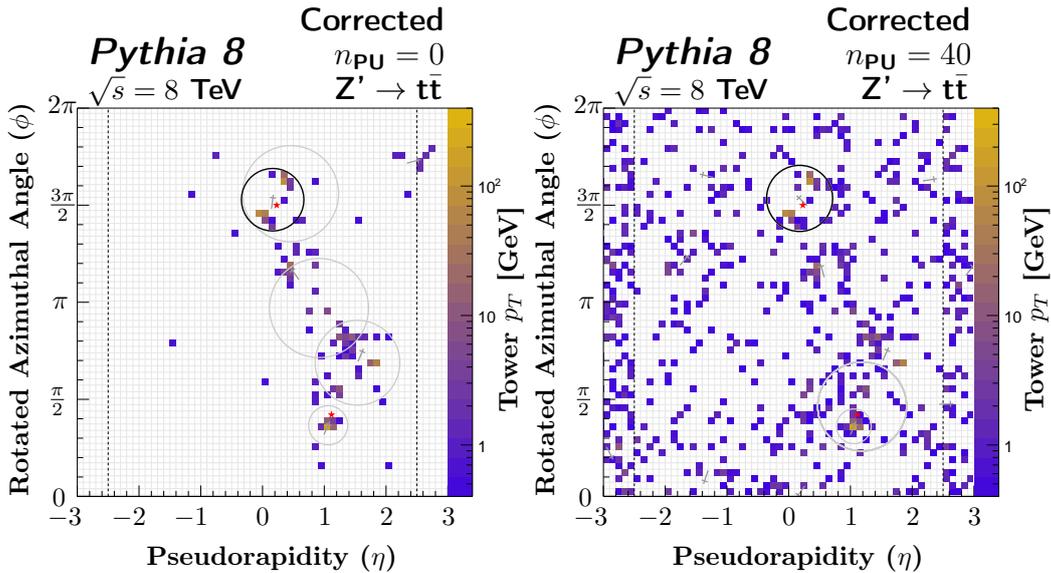

\vspace{1cm}
\begin{center}
\begin{tabular}{cc}
\begin{overpic}[width=0.43\textwidth]{figs/draft_figs/PaperED_mGMMc_1_0_corr.pdf}
% general labels
\put(12, 98){\bfseries \sffamily \Large \emph{Pythia 8}}
\put(12, 91){\bfseries \sffamily $\sqrt{s} = 8 \text{ TeV}$}
\put(62, 91){\bfseries \sffamily \large $\text{Z'}
    \rightarrow \text{t}\bar{\text{t}}$}
\put(62, 98){\bfseries \sffamily \large $n_{\text{PU}} = 0$}
\put(55, 105){\bfseries \sffamily \large Corrected}

% axes labels
\put(24, -4){\bfseries \small Pseudorapidity ($\eta$)}
\put(-4, 10){\rotatebox{90}{\bfseries \small Rotated Azimuthal Angle
    ($\phi$)}}
\put(95, 27){\rotatebox{90}{\bfseries \small Tower $p_T \text{ [GeV]}$}}

% axes tick labels
\put(5, 8){\bfseries \small \sffamily $0$}
\put(4, 28.3){\bfseries \small \sffamily $\frac{\pi}{2}$}
\put(5, 48.3){\bfseries \small \sffamily $\pi$}
\put(4, 68){\bfseries \small \sffamily $\frac{3\pi}{2}$}
\put(4, 88){\bfseries \small \sffamily $2\pi$}

\put(4,  4){\bfseries \small \sffamily $-3$}
\put(17, 4){\bfseries \small \sffamily $-2$}
\put(29.6, 4){\bfseries \small \sffamily $-1$}
\put(46.2, 4){\bfseries \small \sffamily $0$}
\put(58.8, 4){\bfseries \small \sffamily $1$}
\put(71.3, 4){\bfseries \small \sffamily $2$}
\put(83.8, 4){\bfseries \small \sffamily $3$}
\end{overpic} &
\begin{overpic}[width=0.43\textwidth]{figs/draft_figs/PaperED_mGMMc_1_40_corr.pdf}
% general labels
\put(12, 98){\bfseries \sffamily \Large \emph{Pythia 8}}
\put(12, 91){\bfseries \sffamily $\sqrt{s} = 8 \text{ TeV}$}
\put(62, 91){\bfseries \sffamily \large $\text{Z'}
    \rightarrow \text{t}\bar{\text{t}}$}
\put(59, 98){\bfseries \sffamily \large $n_{\text{PU}} = 40$}
\put(55, 105){\bfseries \sffamily \large Corrected}

% axes labels
\put(24, -4){\bfseries \small Pseudorapidity ($\eta$)}
\put(-4, 10){\rotatebox{90}{\bfseries \small Rotated Azimuthal Angle
    ($\phi$)}}
\put(95, 27){\rotatebox{90}{\bfseries \small Tower $p_T \text{ [GeV]}$}}

% axes tick labels
\put(5, 8){\bfseries \small \sffamily $0$}
\put(4, 28.3){\bfseries \small \sffamily $\frac{\pi}{2}$}
\put(5, 48.3){\bfseries \small \sffamily $\pi$}
\put(4, 68){\bfseries \small \sffamily $\frac{3\pi}{2}$}
\put(4, 88){\bfseries \small \sffamily $2\pi$}

\put(4,  4){\bfseries \small \sffamily $-3$}
\put(17, 4){\bfseries \small \sffamily $-2$}
\put(29.6, 4){\bfseries \small \sffamily $-1$}
\put(46.2, 4){\bfseries \small \sffamily $0$}
\put(58.8, 4){\bfseries \small \sffamily $1$}
\put(71.3, 4){\bfseries \small \sffamily $2$}
\put(83.8, 4){\bfseries \small \sffamily $3$}
\end{overpic} \\
\end{tabular}
\end{center}
\caption{Jet correction using tower subtraction and the event jet with
  parameter $\gamma_w = 0.01$. The two leading $p_T$ jets are almost
  identical in size in the left and right insets, which show the
  $n_{\text{PU}} = 0$ and $n_{\text{PU}} = 40$ cases
  respectively. Although many of the other jets change (including the
  migration of jets to higher $\left|\eta\right|$ as a result of the
  simulated tracker), those that give the $\sigma$ and sub-leading
  $\sigma$ variables are insensitive to the effect of pileup.}
\label{fig:pileup_corr}
\end{figure}

Studies of the pileup conditions similar to LHC Run I, with $\sim 20$ pileup interactions, indicate that with a 5 GeV $p_T$ cut, $\gamma_w = 0.01$ provides reasonable stability of the learned $\sigma$.  This is demonstrated qualitatively in Figure~\ref{fig:pileup_corr}, in which the tower and event jet corrections are applied to the same event shown in Figure~\ref{fig:eventdisplay} at both $n_{\text{PU}} = 0$ and $n_{\text{PU}} = 40$. Unlike any of
the methods discussed previously, this method for correction maintains
IRC safety, demonstrates very little jet broadening at $n_{\text{PU}}
= 40$, and is not drastically different in its qualitative features by
comparison to the standard mGMM algorithm.  Note that the assignment of towers to jets under the HML scheme is impacted with the event jet because many towers belong to the event jet with higher probability than any of the other fuzzy jets.  To preserve tower-to-jet assignments under pileup, a smaller value of $\gamma_w$ should be chosen.  The event jet is useful instead because it changes the dynamics
of clustering, making jets less sensitive to soft radiation far away
from the jet axis during the EM update steps, and therefore increasing
the stability of the hard core that is eventually clustered.

\begin{figure}[h!]
\begin{center}
\begin{tabular}{cc}
\begin{tikzpicture}
  \node[anchor=south west,inner sep=0] (image) at (0,0)
  {\includegraphics[width=0.43\textwidth]{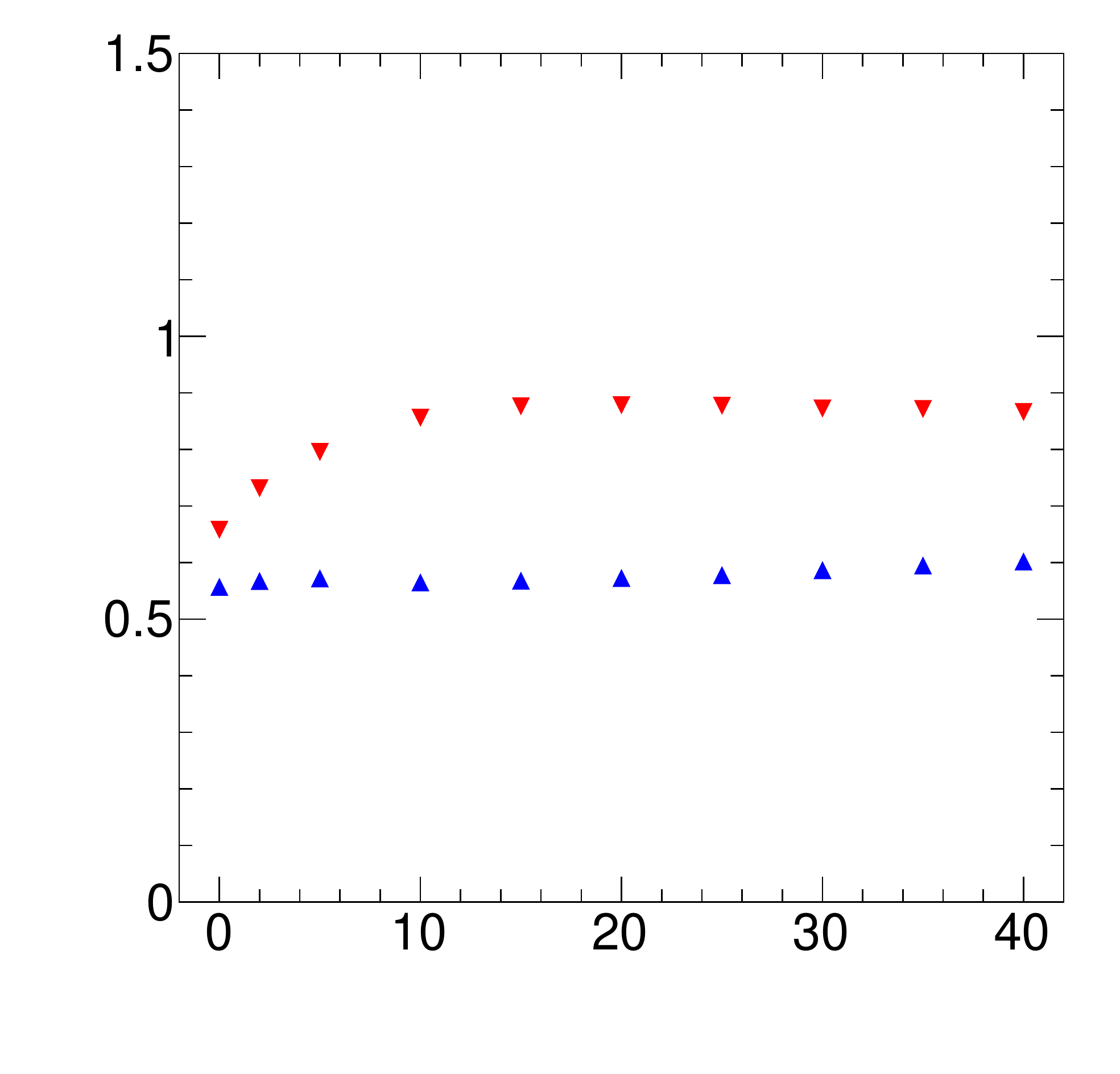}};
  \begin{scope}[x={(image.south east)},y={(image.north west)}]
    % legend
    \draw[fill=red,draw=none] (0.625,0.883)--(0.645,0.883)--(0.635,0.863)--cycle;
    \draw[fill=blue,draw=none] (0.625,0.793)--(0.645,0.793)--(0.635,0.813)--cycle;
    \node[draw=none, anchor=west] at (0.71, 1) {\bfseries \sffamily \large $\text{Z'}
    \rightarrow \text{t}\bar{\text{t}}$};
    \node[draw=none, anchor=west] at (0.65, 0.875) {\bfseries \sffamily
      \footnotesize Uncorrected};
    \node[draw=none, anchor=west] at (0.65, 0.805) {\bfseries \sffamily
      \footnotesize Corrected};

    % labels
    \node[draw=none] at (0.56,0.065) {\bfseries \small $n_{\text{PU}}$};
    \node[draw=none, rotate=90] at (0.05, 0.55){\bfseries \small Mean
      of $\sigma$};
    \node[draw=none, anchor=west] at (0.17,0.88) {\bfseries \sffamily
      \Large \emph{Pythia 8}};
    \node[draw=none, anchor=west] at (0.17,0.80) {\bfseries \sffamily
      $\sqrt{s} = 8 \text{ TeV}$};
  \end{scope}
\end{tikzpicture} &
\begin{tikzpicture}
  \node[anchor=south west,inner sep=0] (image) at (0,0)
  {\includegraphics[width=0.43\textwidth]{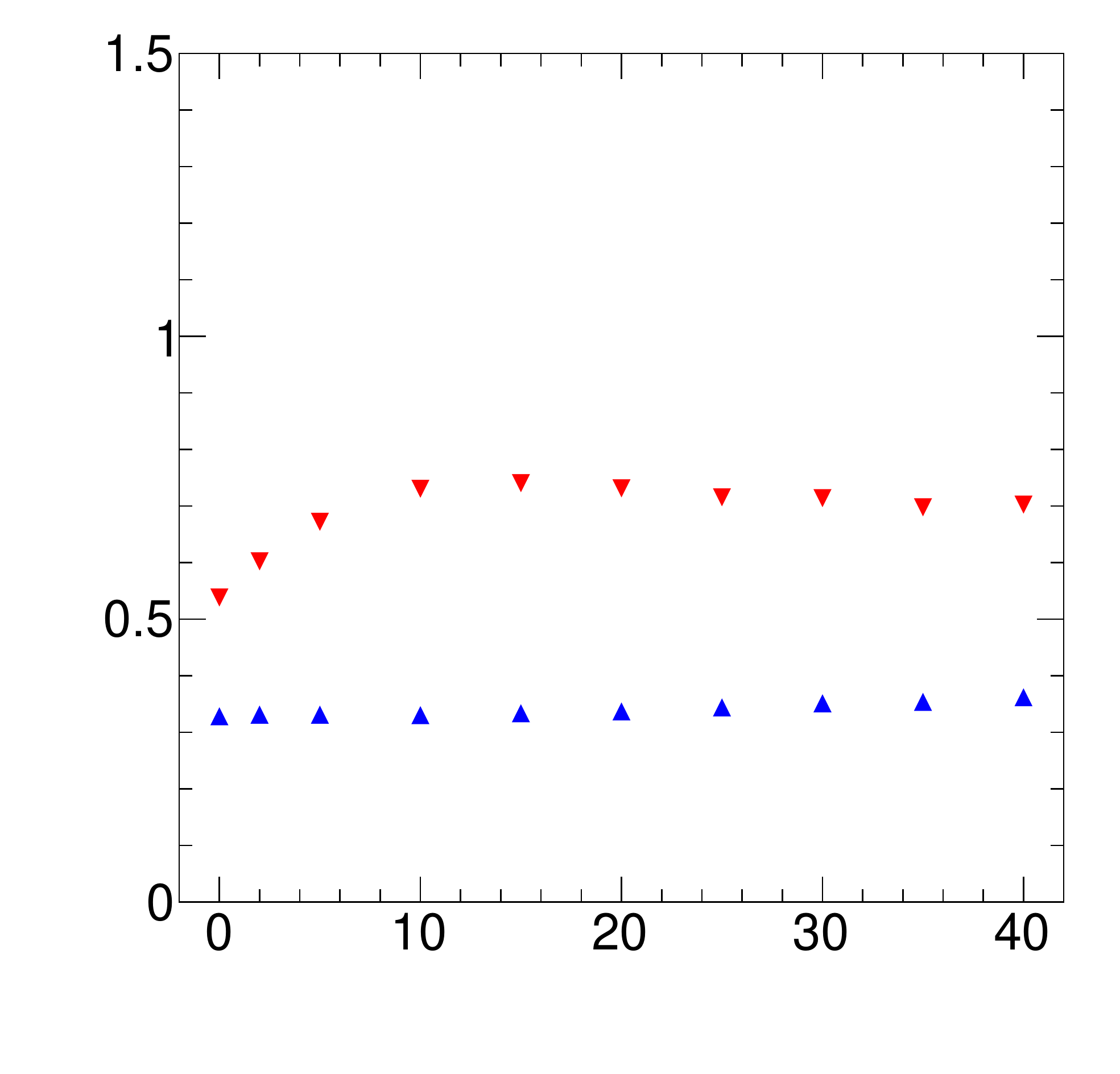}};
  \begin{scope}[x={(image.south east)},y={(image.north west)}]
    % legend
    \draw[fill=red,draw=none] (0.625,0.883)--(0.645,0.883)--(0.635,0.863)--cycle;
    \draw[fill=blue,draw=none] (0.625,0.793)--(0.645,0.793)--(0.635,0.813)--cycle;
    \node[draw=none, anchor=west] at (0.77, 1) {\bfseries \sffamily \large QCD};
    \node[draw=none, anchor=west] at (0.65, 0.875) {\bfseries \sffamily
      \footnotesize Uncorrected};
    \node[draw=none, anchor=west] at (0.65, 0.805) {\bfseries \sffamily
      \footnotesize Corrected};

    % labels
    \node[draw=none] at (0.56,0.065) {\bfseries \small $n_{\text{PU}}$};
    \node[draw=none, rotate=90] at (0.05, 0.55){\bfseries \small Mean
      of $\sigma$};
    \node[draw=none, anchor=west] at (0.17,0.88) {\bfseries \sffamily
      \Large \emph{Pythia 8}};
    \node[draw=none, anchor=west] at (0.17,0.80) {\bfseries \sffamily
      $\sqrt{s} = 8 \text{ TeV}$};
  \end{scope}
\end{tikzpicture} \\
\begin{tikzpicture}
  \node[anchor=south west,inner sep=0] (image) at (0,0)
  {\includegraphics[width=0.43\textwidth]{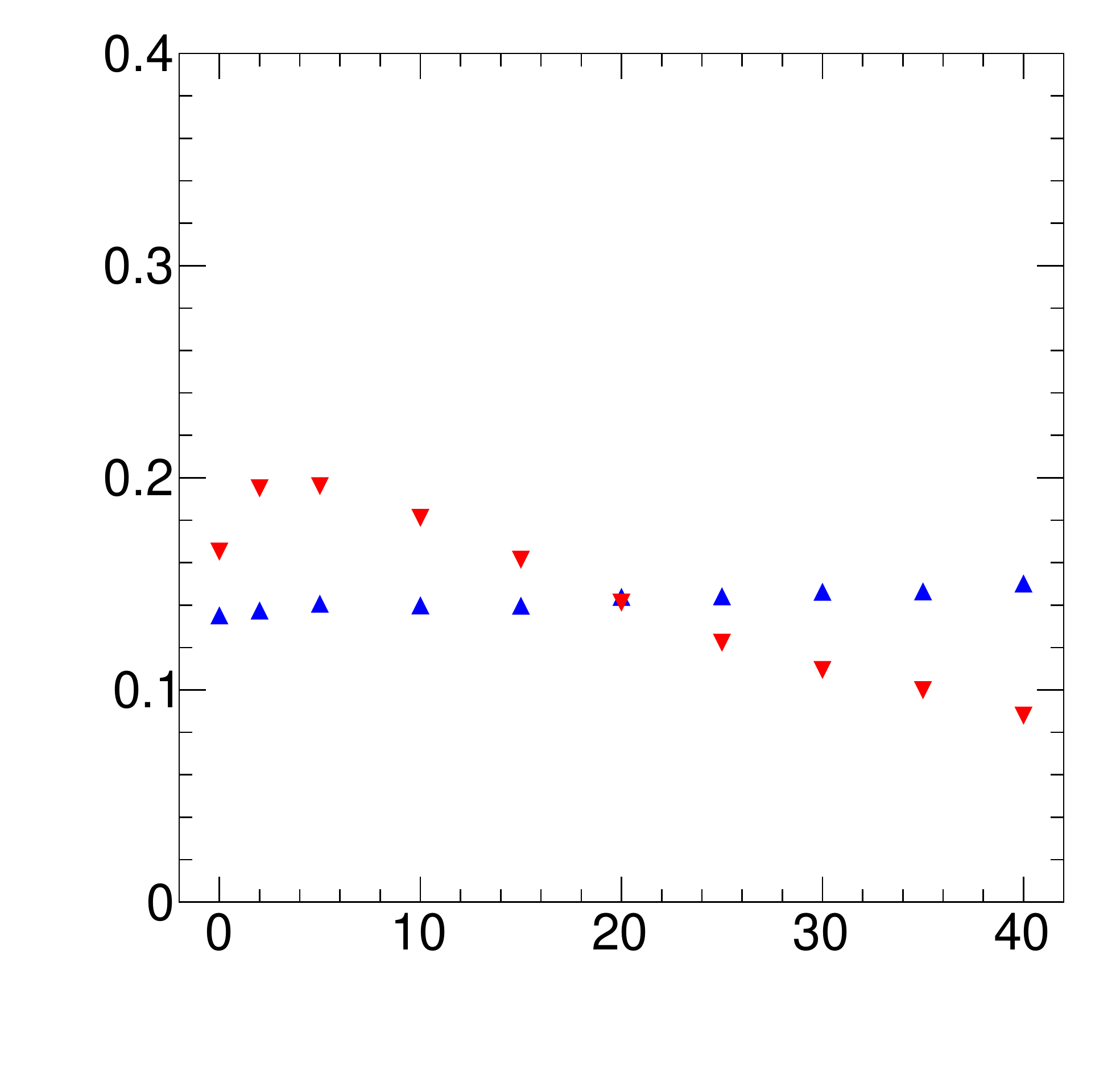}};
  \begin{scope}[x={(image.south east)},y={(image.north west)}]
    % legend
    \draw[fill=red,draw=none] (0.625,0.883)--(0.645,0.883)--(0.635,0.863)--cycle;
    \draw[fill=blue,draw=none] (0.625,0.793)--(0.645,0.793)--(0.635,0.813)--cycle;
    \node[draw=none, anchor=west] at (0.71, 1) {\bfseries \sffamily \large $\text{Z'}
    \rightarrow \text{t}\bar{\text{t}}$};
    \node[draw=none, anchor=west] at (0.65, 0.875) {\bfseries \sffamily
      \footnotesize Uncorrected};
    \node[draw=none, anchor=west] at (0.65, 0.805) {\bfseries \sffamily
      \footnotesize Corrected};

    % labels
    \node[draw=none] at (0.56,0.065) {\bfseries \small $n_{\text{PU}}$};
    \node[draw=none, rotate=90] at (0.05, 0.55){\bfseries \small
      Standard Deviation of $\sigma$};
    \node[draw=none, anchor=west] at (0.17,0.88) {\bfseries \sffamily
      \Large \emph{Pythia 8}};
    \node[draw=none, anchor=west] at (0.17,0.80) {\bfseries \sffamily
      $\sqrt{s} = 8 \text{ TeV}$};
  \end{scope}
\end{tikzpicture} &
\begin{tikzpicture}
  \node[anchor=south west,inner sep=0] (image) at (0,0)
  {\includegraphics[width=0.43\textwidth]{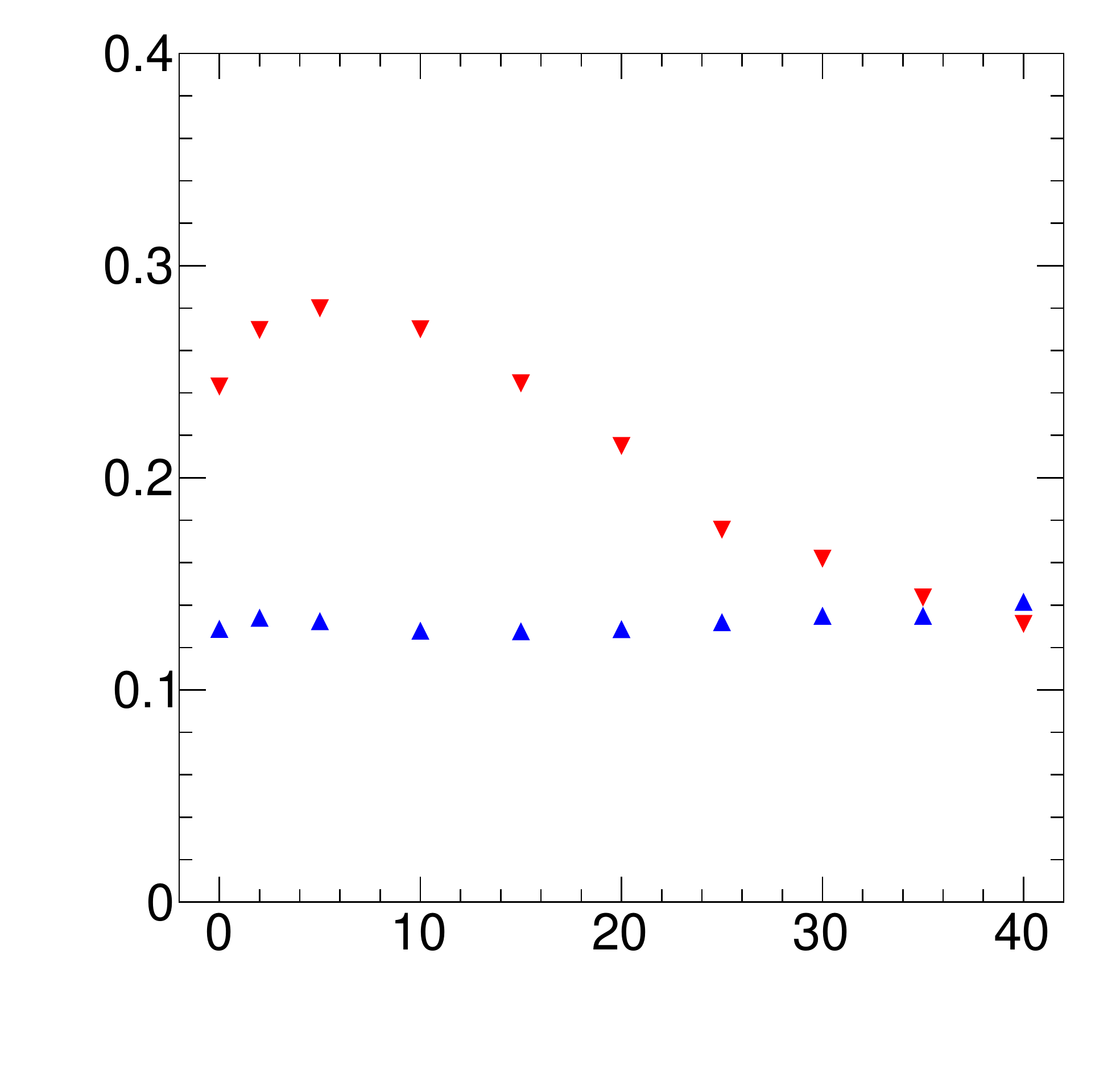}};
  \begin{scope}[x={(image.south east)},y={(image.north west)}]
    % legend
    \draw[fill=red,draw=none] (0.625,0.883)--(0.645,0.883)--(0.635,0.863)--cycle;
    \draw[fill=blue,draw=none] (0.625,0.793)--(0.645,0.793)--(0.635,0.813)--cycle;
    \node[draw=none, anchor=west] at (0.77, 1) {\bfseries \sffamily \large QCD};
    \node[draw=none, anchor=west] at (0.65, 0.875) {\bfseries \sffamily
      \footnotesize Uncorrected};
    \node[draw=none, anchor=west] at (0.65, 0.805) {\bfseries \sffamily
      \footnotesize Corrected};

    % labels
    \node[draw=none] at (0.56,0.065) {\bfseries \small $n_{\text{PU}}$};
    \node[draw=none, rotate=90] at (0.05, 0.55){\bfseries \small
      Standard Deviation of $\sigma$};
    \node[draw=none, anchor=west] at (0.17,0.88) {\bfseries \sffamily
      \Large \emph{Pythia 8}};
    \node[draw=none, anchor=west] at (0.17,0.80) {\bfseries \sffamily
      $\sqrt{s} = 8 \text{ TeV}$};
  \end{scope}
\end{tikzpicture} \\
\end{tabular}
\caption{For both QCD and $Z'\rightarrow t\bar{t}$ samples, using the
  pileup correction (blue triangles) via the event jet and tower
  subtraction stabilizes the mean relative to the uncorrected samples
  (red inverted triangles), and prevents widening of the $\sigma$
  distribution in pileup conditions somewhat worse than during Run 1
  at the LHC.}
\label{fig:corr_mean_var}
\end{center}
\end{figure}
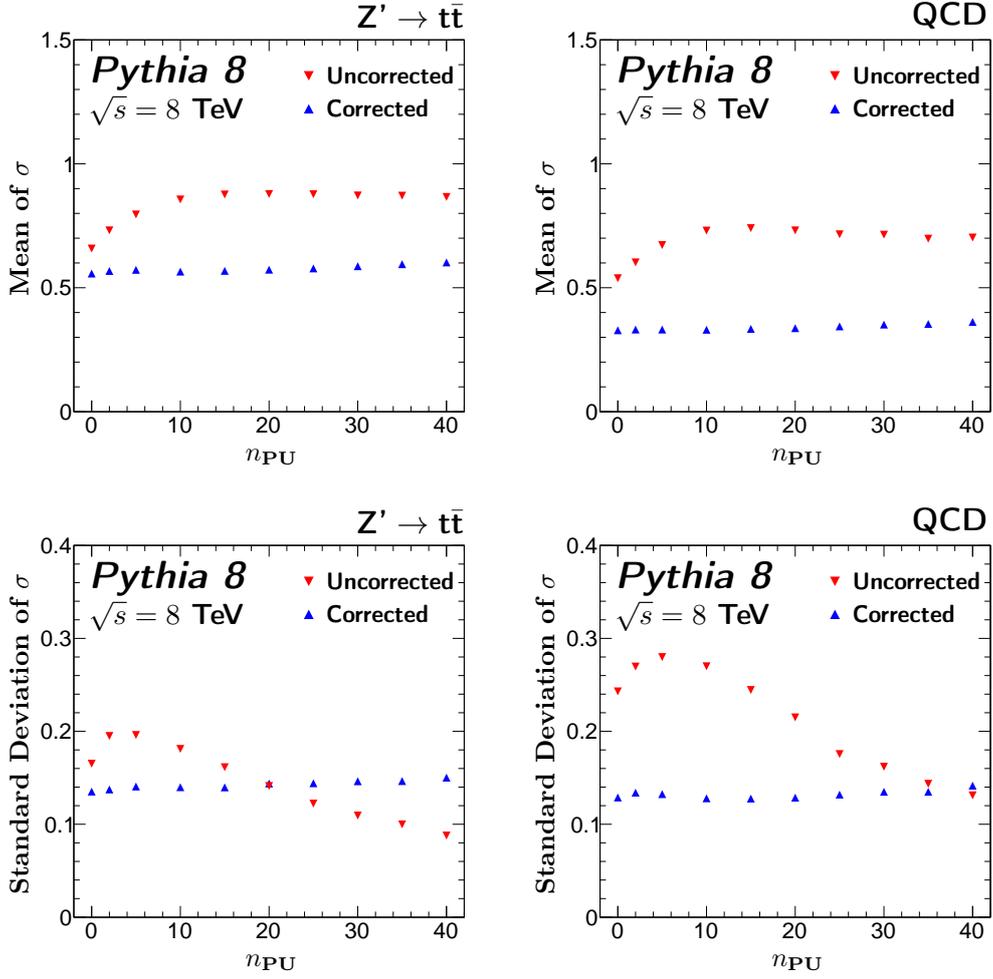

A quantitative study of the pileup mitigation suggested qualitatively by Figure~\ref{fig:pileup_corr} requires an ensemble of events. Figure~\ref{fig:corr_mean_var} shows how the mean and standard deviation of learned $\sigma$ evolve with $n_\text{PU}$.  The uncorrected $\sigma$ is shown in red downward pointing triangles while the tower subtraction and event jet corrections are shown in blue upward pointing triangles.  For both $Z'\rightarrow t\bar{t}$ and QCD, the pileup dependence is dramatically reduced with the tower subtraction and the event jet.  The uncorrected mean $\sigma$ increases as a function of $n_\text{PU}$ as all of the fuzzy jets become the same size.  The standard deviation of the uncorrected $\sigma$ actually decreases beyond $n_\text{PU}\sim 5$ as all of the fuzzy jets become the same size.  For modest levels of pileup, tower subtraction and event and the event jet maintain the mean and standard deviation of the $\sigma$ distribution.

\clearpage
\newpage

\section{Conclusions}
\label{sec:conclusions}

The modified mixture model algorithms provide a new way of looking at whole event
structure.  In contrast to the usual uses of hierarchical-agglomerative algorithms like anti-$k_t$, the number of seeds is fixed ahead of time and their properties are learned during the clustering process.  The learned parameters provide a new set of handles for distinguishing jets of different types.  Even simple variables
constructed out of the learned parameters of a mixture of isotropic
Gaussian jets, like $\sigma$, offer complementary information to the
$n$-subjettiness variables $\tau_{21}$ and $\tau_{32}$ for tagging $W$
boson and top quark jets.  Even though the variable $\sigma$ is sensitive to changes in pileup conditions, small modifications to the fuzzy jets algorithm -- correcting jet inputs and adding a pileup likelihood -- can mitigate the impact of pileup.

Fuzzy jets are new paradigm for jet clustering in high energy physics.  These IRC safe likelihood-based clustering schemes set the stage for many possibilities for future studies related to jet tagging, probabilistic clustering, and pileup suppression.

\section{Acknowledgments}

We would like to thank Jesse Thaler for useful discussions and helpful feedback on the manuscript.  In addition, we thank Gavin Salam for useful comments on the algorithm description.  This work is supported by the US Department of Energy (DOE) Early Career Research Program and grant DE-AC02-76SF00515. BN is supported by the NSF Graduate Research Fellowship under Grant No. DGE-4747 and by the Stanford Graduate Fellowship.

\clearpage
\newpage

\appendix

\section{Wrapped Gaussian}
\label{sec:wrapped}

\begin{figure}[h!]
\vspace{1cm}
\begin{center}
\begin{tabular}{cc}
\begin{overpic}[width=0.43\textwidth]{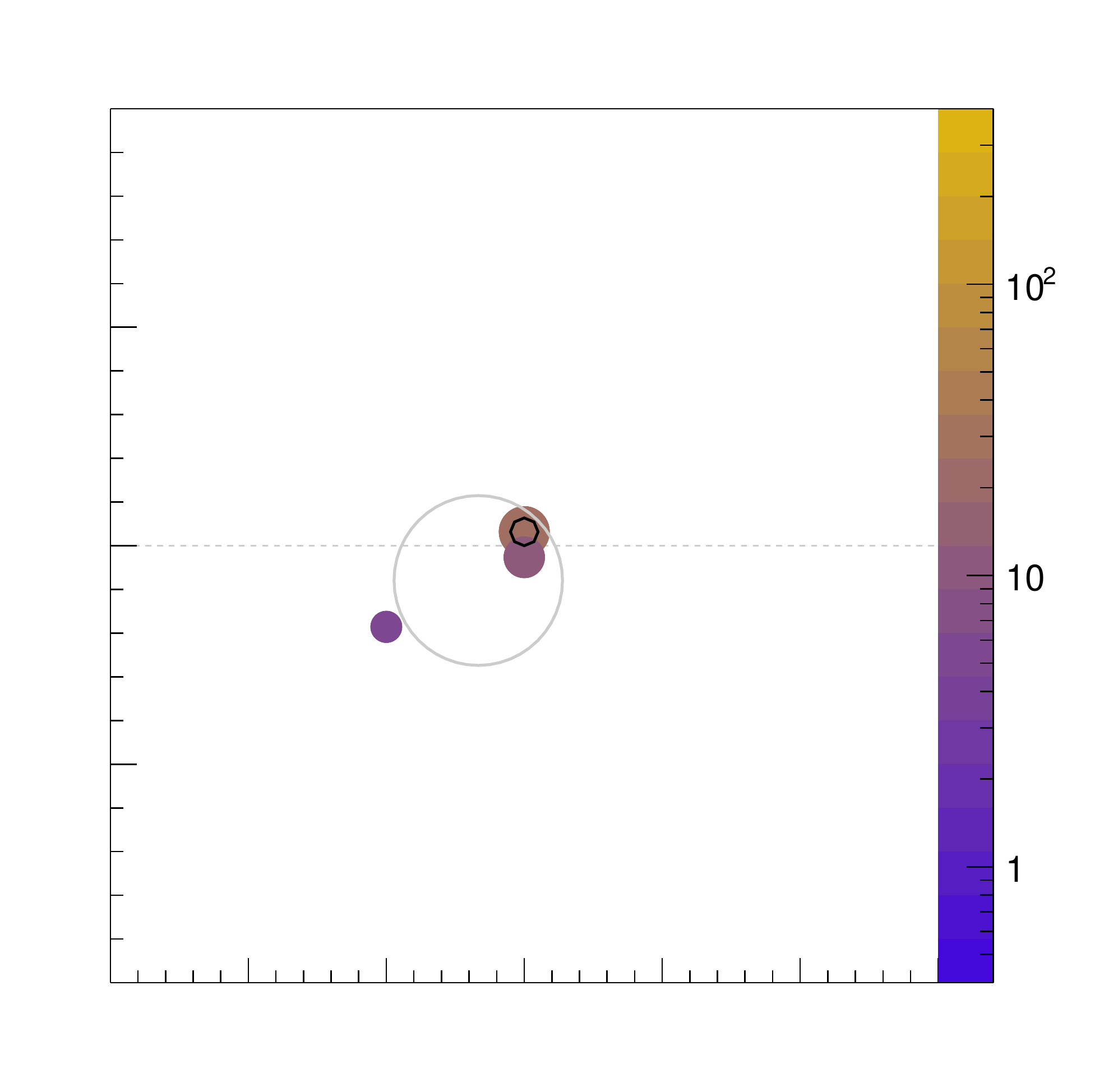}
% general labels
\put(11, 91){\bfseries \sffamily \small Naive Gaussian Density}
\put(72.2, 50.8){\bfseries \sffamily \tiny \textcolor[rgb]{0.8, 0.8, 0.8}{Internal}}
\put(70, 46.2){\bfseries \sffamily \tiny \textcolor[rgb]{0.8, 0.8, 0.8}{boundary}}
% axes labels
\put(24, -4){\bfseries \small Pseudorapidity ($\eta$)}
\put(-4, 10){\rotatebox{90}{\bfseries \small Rotated Azimuthal Angle
    ($\phi$)}}
\put(95, 37){\rotatebox{90}{\bfseries \small $p_T \text{ [GeV]}$}}

% axes tick labels
\put(5, 8){\bfseries \small \sffamily $\pi$}
\put(4, 28.3){\bfseries \small \sffamily $\frac{3\pi}{2}$}
\put(4, 48.3){\bfseries \small \sffamily $2\pi$}
\put(4, 68){\bfseries \small \sffamily $\frac{5\pi}{2}$}
\put(4, 88){\bfseries \small \sffamily $3\pi$}

\put(4,  4){\bfseries \small \sffamily $-3$}
\put(17, 4){\bfseries \small \sffamily $-2$}
\put(29.6, 4){\bfseries \small \sffamily $-1$}
\put(46.2, 4){\bfseries \small \sffamily $0$}
\put(58.8, 4){\bfseries \small \sffamily $1$}
\put(71.3, 4){\bfseries \small \sffamily $2$}
\put(83.8, 4){\bfseries \small \sffamily $3$}
\end{overpic} &
\begin{overpic}[width=0.43\textwidth]{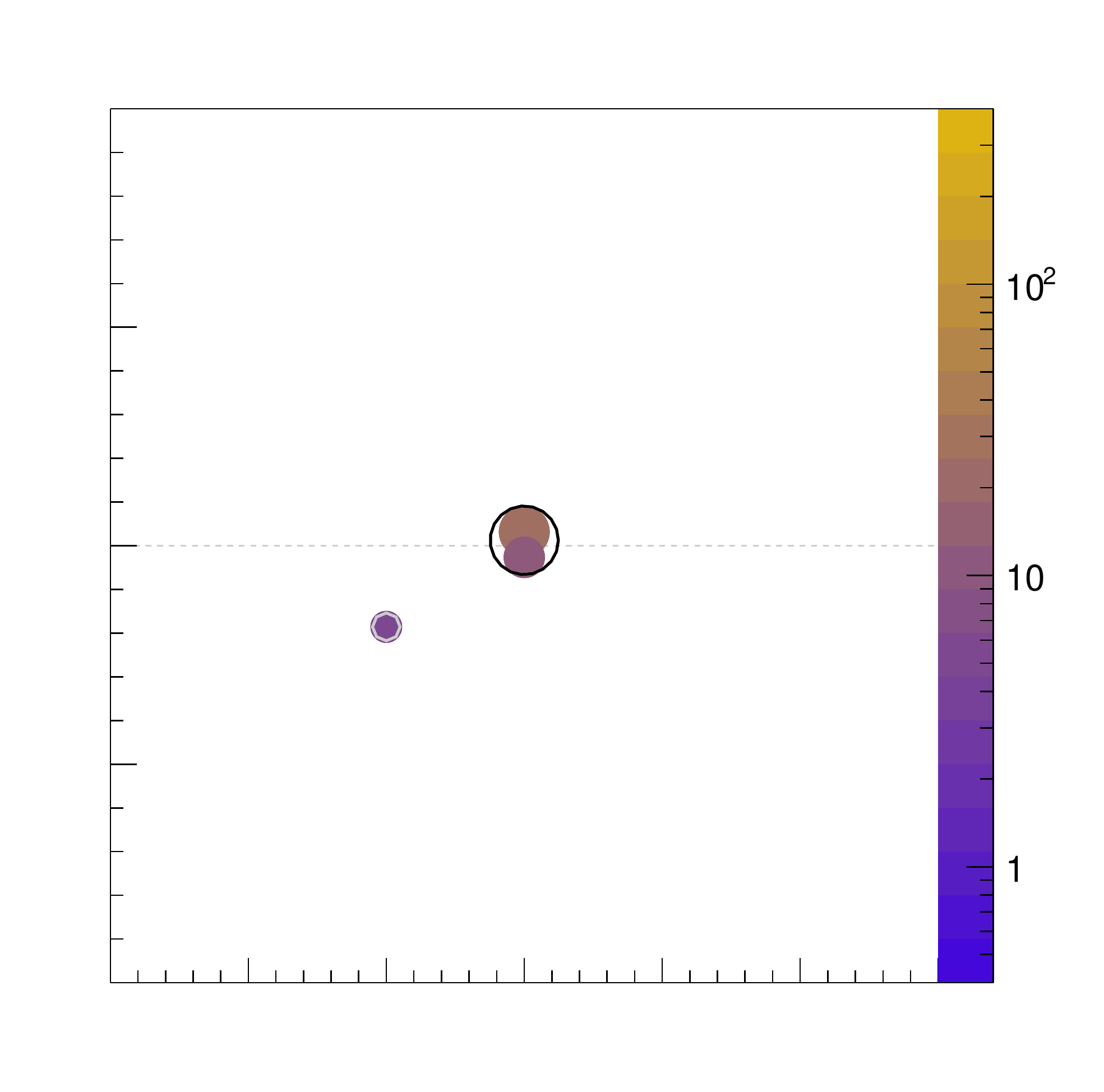}
% general labels
\put(11, 91){\bfseries \sffamily \small Wrapped Approximation}
\put(72.2, 50.8){\bfseries \sffamily \tiny \textcolor[rgb]{0.8, 0.8, 0.8}{Internal}}
\put(70, 46.2){\bfseries \sffamily \tiny \textcolor[rgb]{0.8, 0.8, 0.8}{boundary}}
% axes labels
\put(24, -4){\bfseries \small Pseudorapidity ($\eta$)}
\put(-4, 10){\rotatebox{90}{\bfseries \small Rotated Azimuthal Angle
    ($\phi$)}}
\put(95, 37){\rotatebox{90}{\bfseries \small $p_T \text{ [GeV]}$}}

% axes tick labels
\put(5, 8){\bfseries \small \sffamily $\pi$}
\put(4, 28.3){\bfseries \small \sffamily $\frac{3\pi}{2}$}
\put(4, 48.3){\bfseries \small \sffamily $2\pi$}
\put(4, 68){\bfseries \small \sffamily $\frac{5\pi}{2}$}
\put(4, 88){\bfseries \small \sffamily $3\pi$}

\put(4,  4){\bfseries \small \sffamily $-3$}
\put(17, 4){\bfseries \small \sffamily $-2$}
\put(29.6, 4){\bfseries \small \sffamily $-1$}
\put(46.2, 4){\bfseries \small \sffamily $0$}
\put(58.8, 4){\bfseries \small \sffamily $1$}
\put(71.3, 4){\bfseries \small \sffamily $2$}
\put(83.8, 4){\bfseries \small \sffamily $3$}
\end{overpic} \\
\end{tabular}
\end{center}
\caption{A three-particle event display illustrating the results of fuzzy jet clustering using a Gaussian density for $\phi$ (left) and a wrapped Gaussian density approximation for $\phi$ (right).}  
\label{fig:wrappedgaussian}
\end{figure}
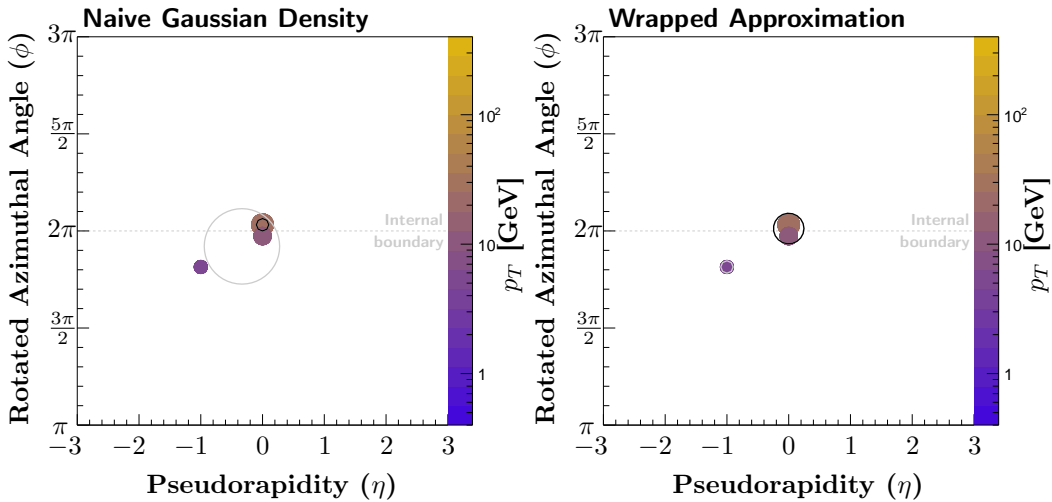

In the EM algorithm described in Sec.~\ref{sec:EMalgorithm}, there are explicit (and implicit) dependencies on the topology.  For instance, if a Gaussian density is used to model $\phi$, then, in the E step, a particle with $\phi_i$ near $2\pi$ will be deemed far from a cluster with location $\phi_j$ near $0$.  To avoid this undesirable behavior and enforce the equivalence of the angles $0$ and $2\pi$, we associate $\phi$ with a {\it wrapped Gaussian density} and $y$ with a standard Gaussian density:

\begin{align}
\label{eq:wrap}
\Phi(y,\phi | \mu_\phi,\mu_y,\sigma^2)=\Phi_y(y|\mu_y,\sigma^2)\frac{1}{\sqrt{2\pi\sigma^2}}\sum_{I=-\infty}^\infty\exp\left[\frac{-(\phi-\mu_\phi(I))^2}{2\sigma^2}\right],
\end{align}

\noindent where $\Phi_y$ is a normal distribution and $\mu_\phi(I)=\mu_\phi+2\pi I$.  In order to approximate the sum in Eq.~(\ref{eq:wrap}), we take only the leading contribution by choosing $\mu_\phi(I^*)$ for $I^*= \text{argmin}_{I'}|\phi-\mu_\phi+2\pi I'|$.  As other contributions are exponentially suppressed, this is a good approximation and recovers continuity near $0$ and $2\pi$.  Figure~\ref{fig:wrappedgaussian} illustrates the improved clustering behavior that results when $\phi$ is modeled using the wrapped Gaussian approximation in place of the standard Gaussian density.

\clearpage
\newpage

\section{The EM algorithm}
\label{sec:emalgo}

This appendix contains two derivations: the modified EM algorithm updates in Eq.~(\ref{eq:emupdates}) and the proof that the modified EM algorithm generically improves the original modified log likelihood Eq.~(\ref{eq:mm2}) with every iteration.  Recall the expected modified complete log likelihood (mmCLL) from Eq.~(\ref{eq:cll}):

\begin{align*}
\sum_{i=1}^n\sum_{j=1}^kp_{Ti}^\alpha\left(q_{ij}\log\Phi(\vec{\rho}_i;\vec{\mu}_j,\Sigma_j)+q_{ij}\log\pi_j\right).
\end{align*}

\noindent Viewing the mCLL as a function of $\vec{\mu},\Sigma$ and $\pi$ for fixed $\lambda$ and $\vec{\rho}$ we can maximize.  For $\pi$, we optimize

\begin{align*}
\sum_{i=1}^n\sum_{j=1}^kp_{Ti}^\alpha\left(q_{ij}\log\pi_j\right)+\lambda\left(\sum_{j=1}^k \pi_j-1\right),
\end{align*}

\noindent where the last term is needed so that the optimal $\pi^*$ is a probability.  The derivative of this expression with respect to $\pi_j$ is

\begin{align*}
\pi_j=-\frac{1}{\lambda}\sum_{i=1}^np_{Ti}^\alpha q_{ij},
\end{align*}

\noindent and then summing the equation over $j$ and using $\sum_{j=1}^kq_{ij}=1$ and the constraint equation $\sum_{j=1}^k\pi_j=1$, we find that

\begin{align*}
\pi_j^*=\frac{1}{\sum_{i=1}^np_{Ti}^\alpha}\sum_{i=1}^np_{Ti}^\alpha q_{ij}
\end{align*}

\noindent The updates for $\vec{\mu}$ and $\Sigma$ follow from the standard derivation (by similarly taking derivatives of the mCLL with respect to components of these multi-dimensional objects) by noting that the only difference is that $q_{ij}\mapsto q_{ij}p_{Ti}^\alpha$ and there are no Lagrange multipliers needed unlike for $\pi_j^*$.

Finally, we prove the claim that the modified EM algorithm described in the body of the text monotonically improves the modified log likelihood in Eq.~(\ref{eq:mm2}).  First, we note that we can rewrite the (log) likelihood as

\begin{align*}
p_T^\alpha \log p(\rho|\theta) &= p_T^\alpha \log\left(\sum_{\lambda\in \{1,2,...,k\}} p(\rho,\lambda;\theta)\right)\\
&=p_T^\alpha \log\left(\sum_{\lambda\in \{1,2,...,k\}} \frac{q(\lambda)p(\rho,\lambda;\theta)}{q(\lambda)}\right)\\
&=p_T^\alpha \log \mathbb{E}_q\left[\frac{p(\rho,\lambda;\theta)}{q(\lambda)}\right]\\
&\geq  \mathbb{E}_q\left[p_T^\alpha\log\left(\frac{p(\rho,\lambda;\theta)}{q(\lambda)}\right)\right]\equiv \mathcal{L}(q,\theta),
\end{align*}

\noindent where the inequality in the last line follows from Jensen's inequality.  Now, we are ready to prove the claim that $p_T^\alpha p(\rho|\theta^{(t}))$ improves monotonically with $t$, the index for the iteration of the EM algorithm.  First, note that

\begin{align*}
\mathcal{L}(q,\theta)&= \mathbb{E}_q\left[p_T^\alpha\log\left(\frac{p(\rho,\lambda;\theta)}{q(\lambda)}\right)\right]\\
&= \mathbb{E}_q\left[p_T^\alpha\log\left(p(\rho,\lambda;\theta)\right)\right]- \mathbb{E}_q\left[p_T^\alpha\log\left(q(\lambda)\right)\right],
\end{align*}

\noindent where the first term is the mCLL and the second term has no $\theta$ dependance and so maximize $\mathcal{L}(q,\theta)$ over $\theta$ is equivalent to maximize the mCLL over $\theta$.  Therefore, $\mathcal{L}(q^{(t+1)},\theta^{(t)})\leq \mathcal{L}(q^{(t+1)},\theta^{(t+1)})$.  By the inequality above, $\mathcal{L}(q^{(t+1)},\theta^{(t+1)})\leq p_T^\alpha p(\rho|\theta^{(t+1)}) $.  The E step can be recast as choosing

\begin{align*}
q^{(t+1)}(\lambda_i=j)=q_{ij}(\theta^{(t)})=\mathbb{E}_{\theta^{(t)}}[q_{ij}]=p(\lambda|\rho,\theta^{(t)}).
\end{align*}

\noindent This enforces:

\begin{align*}
\mathcal{L}(p(\lambda|\rho,\theta^{(t)}),\theta^{(t)})&=\mathbb{E}_{p(\lambda|\rho,\theta^{(t)})}\left[p_T^\alpha\log\left(\frac{p(\rho,\lambda;\theta^{(t)})}{p(\lambda|\rho,\theta^{(t)})}\right)\right]\\
&=\mathbb{E}_{p(\lambda|\rho,\theta^{(t)})}\left[p_T^\alpha\log\left(p(\rho;\theta^{(t)})\right)\right]\\
&=p_T^\alpha\log\left(p(\rho;\theta^{(t)})\right)
\end{align*}

\noindent Putting this together with the bounds from the M step, we arrive at the desired result: $p_T^\alpha p(\rho|\theta^{(t)})\leq p_T^\alpha p(\rho|\theta^{(t+1)})$, i.e., every step of the modified EM algorithm improves or leaves the same the original likelihood.

\clearpage
\newpage

\section{Controlling Jet Multiplicity with $p_T$}
\label{sec:pt_multiplicity}

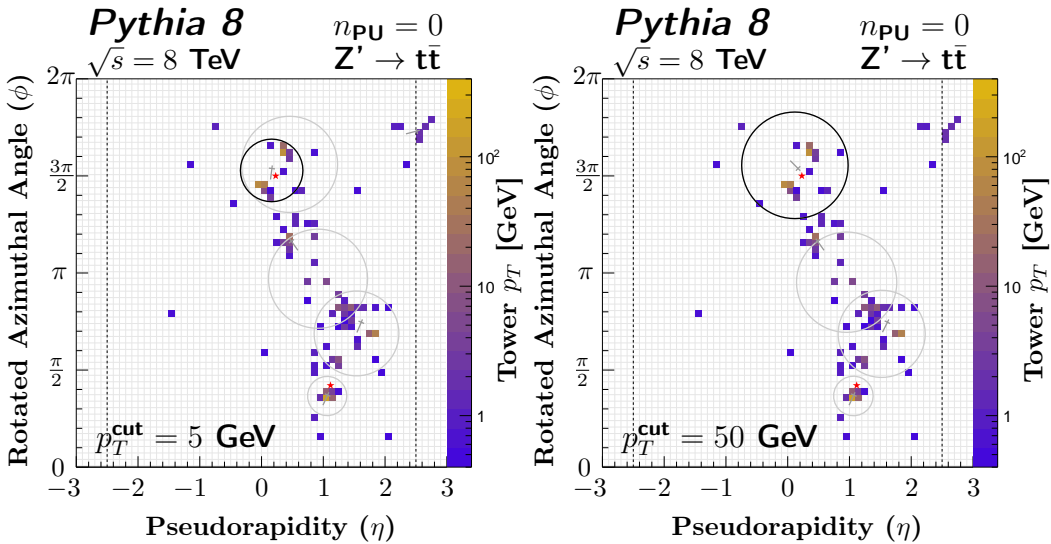
\begin{figure}[h!]
\vspace{1cm}
\begin{center}
\begin{tabular}{cc}
\begin{overpic}[width=0.43\textwidth]{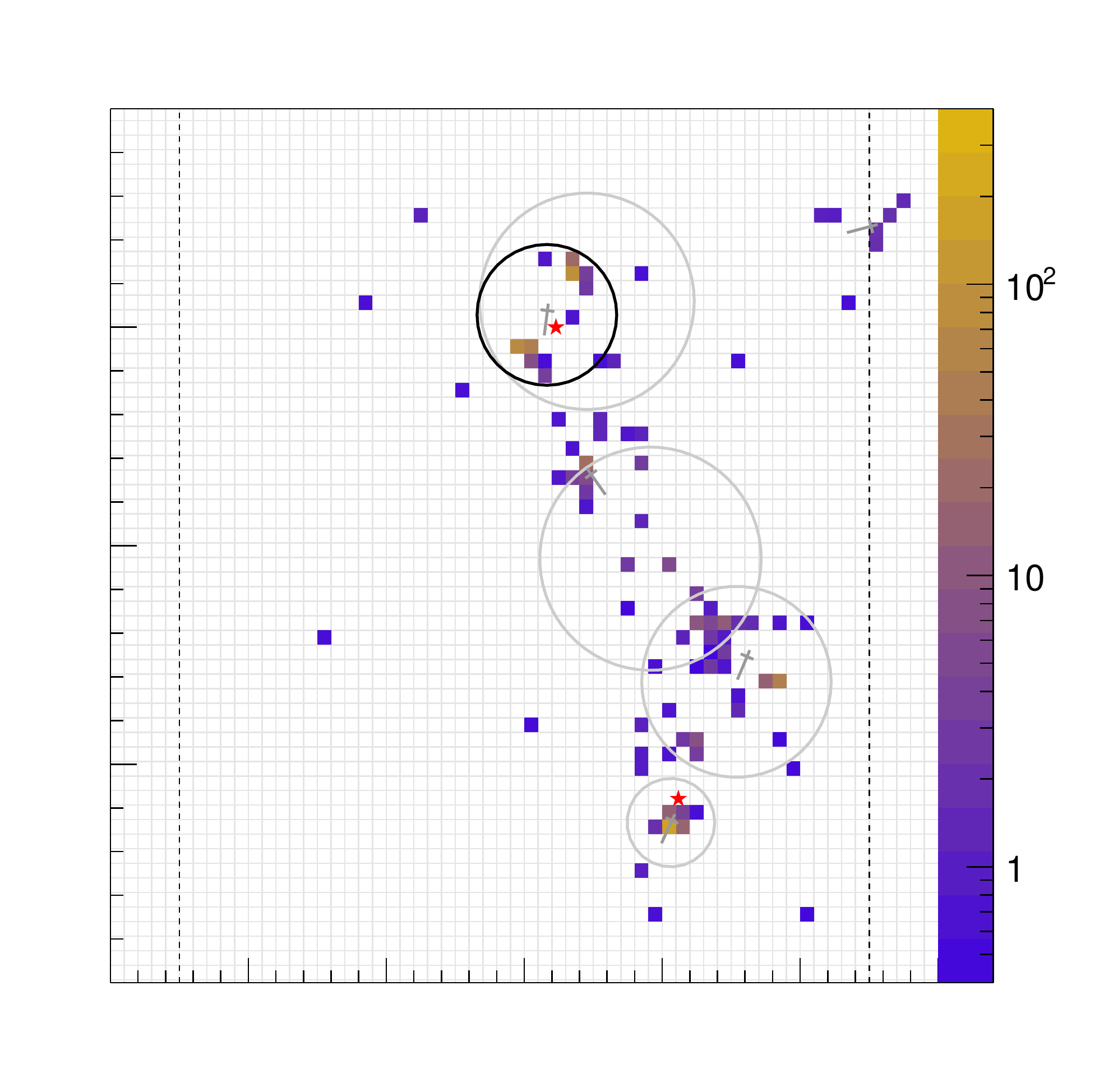}
% general labels
\put(12, 98){\bfseries \sffamily \Large \emph{Pythia 8}}
\put(12, 91){\bfseries \sffamily $\sqrt{s} = 8 \text{ TeV}$}
\put(62, 91){\bfseries \sffamily \large $\text{Z'}
    \rightarrow \text{t}\bar{\text{t}}$}
\put(62, 98){\bfseries \sffamily \large $n_{\text{PU}} = 0$}
\put(14, 14){\bfseries \sffamily \large $p_T^{\text{cut}} = 5 \text{ GeV}$}

% axes labels
\put(24, -4){\bfseries \small Pseudorapidity ($\eta$)}
\put(-4, 10){\rotatebox{90}{\bfseries \small Rotated Azimuthal Angle
    ($\phi$)}}
\put(95, 27){\rotatebox{90}{\bfseries \small Tower $p_T \text{ [GeV]}$}}

% axes tick labels
\put(5, 8){\bfseries \small \sffamily $0$}
\put(4, 28.3){\bfseries \small \sffamily $\frac{\pi}{2}$}
\put(5, 48.3){\bfseries \small \sffamily $\pi$}
\put(4, 68){\bfseries \small \sffamily $\frac{3\pi}{2}$}
\put(4, 88){\bfseries \small \sffamily $2\pi$}

\put(4,  4){\bfseries \small \sffamily $-3$}
\put(17, 4){\bfseries \small \sffamily $-2$}
\put(29.6, 4){\bfseries \small \sffamily $-1$}
\put(46.2, 4){\bfseries \small \sffamily $0$}
\put(58.8, 4){\bfseries \small \sffamily $1$}
\put(71.3, 4){\bfseries \small \sffamily $2$}
\put(83.8, 4){\bfseries \small \sffamily $3$}
\end{overpic} &
\begin{overpic}[width=0.43\textwidth]{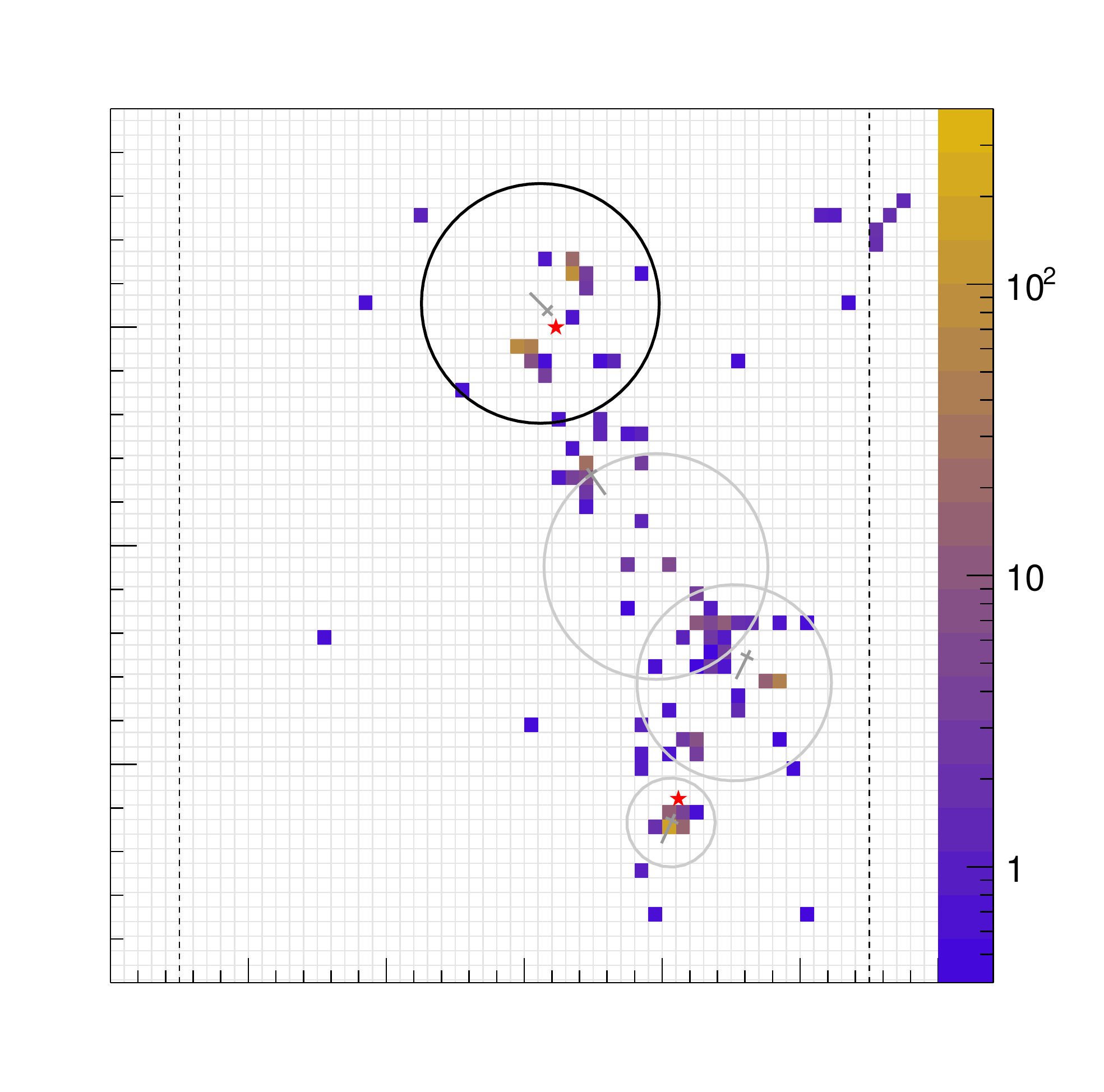}
% general labels
\put(12, 98){\bfseries \sffamily \Large \emph{Pythia 8}}
\put(12, 91){\bfseries \sffamily $\sqrt{s} = 8 \text{ TeV}$}
\put(62, 91){\bfseries \sffamily \large $\text{Z'}
    \rightarrow \text{t}\bar{\text{t}}$}
\put(62, 98){\bfseries \sffamily \large $n_{\text{PU}} = 0$}
\put(14, 14){\bfseries \sffamily \large $p_T^{\text{cut}} = 50 \text{
    GeV}$}

% axes labels
\put(24, -4){\bfseries \small Pseudorapidity ($\eta$)}
\put(-4, 10){\rotatebox{90}{\bfseries \small Rotated Azimuthal Angle
    ($\phi$)}}
\put(95, 27){\rotatebox{90}{\bfseries \small Tower $p_T \text{ [GeV]}$}}

% axes tick labels
\put(5, 8){\bfseries \small \sffamily $0$}
\put(4, 28.3){\bfseries \small \sffamily $\frac{\pi}{2}$}
\put(5, 48.3){\bfseries \small \sffamily $\pi$}
\put(4, 68){\bfseries \small \sffamily $\frac{3\pi}{2}$}
\put(4, 88){\bfseries \small \sffamily $2\pi$}

\put(4,  4){\bfseries \small \sffamily $-3$}
\put(17, 4){\bfseries \small \sffamily $-2$}
\put(29.6, 4){\bfseries \small \sffamily $-1$}
\put(46.2, 4){\bfseries \small \sffamily $0$}
\put(58.8, 4){\bfseries \small \sffamily $1$}
\put(71.3, 4){\bfseries \small \sffamily $2$}
\put(83.8, 4){\bfseries \small \sffamily $3$}
\end{overpic} \\
\end{tabular}
\end{center}
\caption{Changing the choice of the $p_T$ cut used to select seeds can
  make a vast difference in the values of the constructed variables,
  like $\sigma$. In this event, clustered on the left with a cut of $5
  \text{ GeV}$ resulting in five jets, and on the right with a cut of $50
  \text{ GeV}$ resulting four jets. Fewer degrees of freedom in
  the four jet case means a much larger learned value for the $\sigma$ variable.}
\label{fig:pt_cut_ed}
\end{figure}

In contrast to most uses of hierarchical-agglomerative clustering algorithms, the number of fuzzy jets is fixed before clustering
begins. Whereas a single traditional jet can reasonably be considered to
correspond to a parton in appropriate cases, mGMM jets should not be,
as several mGMM jets can together express structure of what would be
one or several jets according to another algorithm.  The choice of the number of jets used in mGMM jet clustering therefore
controls the expressive power of the algorithm to look at the event
structure. In practice, choosing too many jets does not greatly affect
the value of the leading learned $\sigma$ variable, because the
additional jets learn finer features of the event structure. On the
other hand, choosing too few jets is often problematic as can be seen
in Figure~\ref{fig:pt_cut_ed} - the fuzzy jets need to grow in order to cover the full energy distribution in the event.  Using anti-$k_t$ jets as seeds for fuzzy jets has the feature that the number of fuzzy jets change dynamically with the complexity of the event.  The algorithm is not very sensitive to the exact locations of the anti-$k_t$ jets - studies
which randomly perturbed the initial jet locations inside a disc of
radius $1.0$ found that $\sigma$ was robust to such
fluctuations, even on an event by event basis.   However, the $p_T$ threshold for the seed anti-$k_t$ jets can have a significant impact on the fuzzy jets as this alters the number of seeds.  The $p_T$ threshold for the anti-$k_t$ seeds is typically lower than the $p_T$ threshold one would use to consider anti-$k_t$ jets alone because the fuzzy jets algorithm needs enough seeds to populate the low energy regions of the detector.  One way of mitigating the impact of the $p_T$ cut on the fuzzy jet clustering is to introduce an {\it event jet}, described in Section~\ref{sec:event_jet}.

\section{A Leading Order Description of Fuzzy Jet $\sigma$}
\label{sec:theory}

We have seen in Sec.~\ref{sec:tagging} that the fuzzy jet $\sigma$ is correlated with $\rho=m/p_T$.  We can build some intuition for this relationship by considering a leading order QCD calculation of $\sigma$.  Consider an isolated quark jet with energy $E$ which radiates a gluon with angle $\theta\ll 1$ from the jet axis and with energy fraction $z\ll1$.  Without loss of generality, suppose the quark is moving in the $\phi=0$ direction and the splitting happens in the $\phi=\pi/2$ direction so that the four vector of the quark is $q^\mu=E(1-z)(1,0,0,1)$, and the gluon four-vector is $g^\mu=Ez(1,\theta,0,1)$, to leading order.  To this order, the jet mass is simply $m=Ez\theta^2$.  What is $\sigma$?  Consider $k=1$ and something like the event-jet applied so that we can treat this jet in isolation from other hadronic activity in the event.  Since $k=1$, the soft memberships are all one, i.e., $q_{i1}=1$ and there is only one step of the EM algorithm.  The anti-$k_t$ jet has $(y,\phi)$ coordinates $(0,\theta)$, which could be used for the seed, but since $k=1$, the seed is not used.  The quark has coordinates $(0,0)$, and the gluon has coordinates $(0,\theta)$.  We can compute the fuzzy jet coordinates in the (single) M step:

\begin{align}
\mu_y&=0\\
\mu_\phi&=\frac{0\times E(1-z)+\theta\times Ez}{E(1-z)+Ez}=z\theta\\
\sigma^2&=\frac{(0-z\theta)^2\times E(1-z)+(\theta-z\theta)^2\times Ez}{2(E(1-z)+Ez)}\\
&=z\theta^2+\mathcal{O}(\theta^2z^2).
\end{align}

\noindent Therefore, to leading order and $k=1$, the learned $\sigma$ is the jet mass.  For $k=2$, there are enough degrees of freedom to resolve the substructure of the hard splitting and so the relationship between the jet mass and $\sigma$ breaks down.

\clearpage
\newpage

\bibliographystyle{JHEP-2}
\bibliography{myrefs.bib}{}

\end{document}